\documentclass[onecolumn,notitlepage,floatfix,amsmath,amssymb, longbibliography]{revtex4-1}

\usepackage{graphicx}
\usepackage{dcolumn}
\usepackage{color}
\def\AD#1{{#1}}     
  		
\usepackage{amssymb}
\usepackage[toc,page]{appendix}

\newcommand{\be}{\begin{equation}}
\newcommand{\ee}{\end{equation}}

\usepackage[font=small]{caption}
\usepackage[labelformat = empty,position=top]{subcaption}
\usepackage{graphicx}
\begin{document}
\title{Vertical dispersion of Lagrangian tracers in fully developed
  stably stratified turbulence}

\author{N.E. Sujovolsky and P.D. Mininni}
\affiliation{
  Universidad de Buenos Aires, Facultad de Ciencias 
  Exactas y Naturales, Departamento de F\'\i sica, \& IFIBA, CONICET, 
  Ciudad Universitaria, Buenos Aires 1428, Argentina.}

\begin{abstract}
We study the effect of different forcing functions and of the local
gradient Richardson number $\mathrm{Ri_{g}}$ on the vertical transport of
Lagrangian tracers in stably stratified turbulence under the
Boussinesq approximation, and present a wave and continuous-time
random walk model for single- and two-particle vertical
dispersion. The model consists of a random superposition of linear
waves with their amplitude based on the observed Lagrangian spectrum
of vertical velocity, and a random walk process to capture overturning
that depends on the statistics of $\mathrm{Ri_{g}}$ among other Eulerian
quantities. The model is in good agreement with direct numerical
simulations of stratified turbulence, where single- and two-particle
dispersion differs from the homogeneous and isotropic case. Moreover,
the model gives insight into the mixture of linear and non-linear
physics in the problem, as well as on the different processes
responsible for vertical turbulent dispersion.

\end{abstract}
\maketitle

\section{Introduction} \label{sec:INT}

Stably stratified turbulence (SST) is common in geophysical flows,
as the ocean and the atmosphere are usually in a turbulent state and
affected by stratification (and rotation at the largest scales), making it of
fundamental importance in the study of dispersion of pollutants,
transport of nutrients, and turbulent mixing in a wide range of scales
\cite{wyngaard_atmospheric_1992, dasaro_lagrangian_2000,
  watanabe_2017, amir_2017}. As stably stratified turbulence is
anisotropic, it is also inherently different from homogeneous
isotropic turbulence (HIT) \cite{lindborg_vertical_2008, waite_2011,
  marino_large-scale_2014, maffioli_2017}. In SST, the stratification
reduces the vertical velocity, confining the flow into a
quasi-horizontal layered motion, and generating vertically sheared
horizontal winds (VSHWs) with high vertical variability
\cite{smith_generation_2002}. The stratification also results in a
restoring force, allowing for the excitation of waves that can coexist
with the turbulence.

As a result, vertical and horizontal turbulent transports in SST are
fundamentally different. It has been speculated that horizontal transport
could be more efficient than in HIT due to the presence of VSHWs 
\cite{smith_generation_2002,clark_di_leoni_absorption_2015}. Indeed,
horizontal dispersion is dominated by the VSHWs, as shown in direct
numerical simulations and by our recent model for horizontal particle
dispersion in \cite{sujovolsky_single-particle_2017}. \AD{This study
  also showed that the characteristic time scales of dispersion in the 
  vertical and horizontal directions are very different. This
  difference is used in the present work to study vertical
  dispersion separately of its horizontal counterpart.} For vertical
transport and dispersion, stratification has some obvious and some
not-so-obvious implications \cite{kimura_diffusion_1996,
  kaneda_suppression_2000, liechtenstein_role_2006}. On the one hand,
the vertical velocity in SST is intermittent, implying that arguments
based solely on mean values of the vertical velocity or its power
spectrum could be misleading due to the spontaneous occurrence of
extreme values \cite{rorai_turbulence_2014, feraco_vertical_2018}. On
the other hand, while it is well understood that in stratified
turbulence as the stratification is increased the mean vertical
velocity is quenched, vertical gradients also increase with increasing
stratification, possibly balancing the vertical transport
\cite{billant_self-similarity_2001, deBruyn_2004_reynolds,
  lindborg_vertical_2008}.

Mixing in stratified turbulence has been largely studied from an
Eulerian point of view \cite{bauer_dispersion_1974,
  fernando_turbulent_1991, polzin_spatial_1997, wunsch_vertical_2004,
  ivey_density_2008, klein_oceanic_2009}, but Lagrangian measurements
with floaters are also common nowadays, specially in oceanic
measurements of waves and turbulence \cite{dasaro_lagrangian_2000}, 
where they are relevant to understand the transport of nutrients with
applications for the fishing industry. Vertical dispersion is also
important in the atmosphere \cite{mingari_2017}, and particle
dispersion has also been studied recently in atmospheric flows for
forecasting purposes using Lagrangian models
\cite{jones_2007}. In spite of this, there are few studies of
stratified turbulence from the Lagrangian point of view
\cite{godeferd_eulerian_1996, nicolleau_turbulent_2000, 
  aartrijk_single-particle_2008}, where linear theories of SST predict
the bounding of particles in vertical layers and the saturation for
long times of single- and two-particle vertical dispersion
\cite{nicolleau_turbulent_2000}. However, these linear models cannot
capture the effect of overturning, or of thermal diffusion, that can
be relevant at intermediate times and dominate the dynamics of the
vertical transport at very long times
\cite{aartrijk_single-particle_2008}. It is also worth noting that 
inertial particles with density different from that of the fluid are
also relevant to study transport, and have received substantial
attention in HIT (see, e.g., \cite{bec_heavy_2007, sumbekova_2017}),
with a special emphasis on the mechanisms leading to its spatial
distribution and clustering. However, transport and distribution of
inertial particles in SST have also been studied only recently
\cite{van_aartrijk_vertical_2010,sozza_large-scale_2016, amir_2017,
  sozza2018inertial}.

In this paper we present several direct numerical simulations (DNSs)
of the Boussinesq equations with Reynolds buoyancy numbers
$\mathrm{Rb}>1$ in two different domains: one cubic, and
the other anisotropic (an elongated domain with the horizontal sides
longer than the vertical), and using two different mechanical forcing
functions. We applied random forcing (RND), or a Taylor-Green
forcing (TG) which generates a coherent large-scale flow at the
largest available scales thus affecting vertical transport. The
Boussinesq Eulerian flow is evolved together with Lagrangian
particles. We study single- and two-particle vertical dispersion, and
analyze the role of the Froude number, the vertical shear, the
large-scale flow, and the local gradient Richardson number in the
vertical dispersion of particles. We also present a model for single-
and two-particle vertical dispersion that is in good agreement with
the DNS results. 
\AD{In
  a previous work \cite{sujovolsky_single-particle_2017}, we focused
  on the study of horizontal displacements of Lagrangian particles in
  SST and we developed a model for single-particle horizontal
  dispersion; single-particle vertical displacements were considered
  but at moderate values of Rb and in cases dominated by waves. The
  model introduced here for vertical dispersion, together with the
  results in \cite{sujovolsky_single-particle_2017} for horizontal
  dispersion, provide a description of transport of Lagrangian tracers
  in SST in both the horizontal and vertical directions, for a wide
  range of parameters, and for both early and late times in the
  particles'  evolution. In particular,} the superposition of linear
and turbulent effects in the model for vertical dispersion presented
in this work allows us to identify the leading physical effects
resulting in vertical dispersion at early and at late times (compared
with the period of the internal gravity waves). Moreover, as all
parameters in the model can be obtained from large-scale Eulerian 
data, the model could be used autonomously to obtain statistical
predictions of vertical particle dispersion provided a large-scale
flow.

\section{Numerical Simulations} \label{sec:NSIM}

For this study we solved numerically the incompressible Boussinesq
equations for the velocity ${\bf u}$ and for buoyancy (or
``temperature'') fluctuations $\theta$,
\begin{eqnarray}
\label{eq:n-s_strat}
\partial_t {\bf u} +{\bf u}\cdot{\bf \nabla}{\bf
  u} &=& -{\bf \nabla}p - N \theta {\hat z} +\nu \nabla^{2}{\bf u}+{\bf
  f}, \\
\label{eq:theta}
\partial_t \theta+ {\bf u}\cdot{\bf \nabla} \theta
  &=& N {\bf u} \cdot {\hat z}  + \kappa \nabla^{2} \theta, \\
\label{eq:incomp}
 {\bf \nabla}\cdot{\bf u}&=&0,
\end{eqnarray}
where $p$ is the correction to the hydrostatic pressure, $\nu$ is the 
kinematic viscosity, ${\bf f}$ is an external mechanical forcing, $N$
is the Brunt-V\"{a}is\"{a}l\"{a} frequency (which sets the
stratification), and $\kappa$ is the diffusivity. In terms of the
density fluctuations $\rho$, the Brunt-V\"{a}is\"{a}l\"{a} frequency
is $N^2=-(g/\rho_0) (d\bar \rho /dz)$, with $d\bar \rho /dz$ the
imposed (linear) background density stratification, and $\rho_0$ the
mean density. We write the buoyancy field $\theta$ in units of
velocity by defining $\theta=g\rho/(\rho_0N)$.  All quantities are
then made dimensionless using a characteristic length $L_{0}$ and a
characteristic velocity $U_{0}$. All runs in this paper have a Prandtl
number $\textrm{Pr} = \nu/\kappa = 1$.

The Boussinesq equations were solved in a three-dimensional periodic
domain, using a parallelized and fully dealiased pseudospectral
method, and a second-order Runge-Kutta scheme for time integration 
\cite{mininni_hybrid_2011}. In the turbulent steady state of each
simulation we also injected ${\cal O}(10^6)$ Lagrangian particles, and
integrated their trajectories in time using 
\begin{equation}
{\bf v}_{i}=\dfrac{d {\bf x}_{i}}{d t} = {\bf u}({\bf x}_{i},t),
\end{equation}
where the subindex $i$ labels each particle. Here and in the
following, the velocity of Lagrangian particles and its Cartesian
components are represented as ${\bf v} = (v_x,v_y,v_z)$, while the
Eulerian fluid velocity is given by 
${\bf u} = (u_x,u_y,u_z)$. Integration of particles' trajectories was
done using a second-order Runge-Kutta method in time, 
and a three-dimensional cubic spline spatial interpolation to estimate
Lagrangian velocities at the particles positions ${\bf x}_i$ 
from the velocity ${\bf u}$ in the regular Eulerian grid
\cite{yeung_1988}. All simulations were done using the GHOST code
({\it Geophysical High-Order Suite for Turbulence}), recently 
extended to work with non-cubic boxes \cite{mininni_generation_2017}.

Equations (\ref{eq:n-s_strat}) and (\ref{eq:theta}) have two
controlling dimensionless parameters, the Reynolds and the Froude
numbers, respectively given by
\begin{equation}
 \textrm{Re} = \frac{LU}{\nu}, \,\,\,\,\,\, 
 \textrm{Fr} = \frac{U}{LN},
 \label{eq:Re_Fr}
\end{equation}
where $L$ and $U$ are respectively the characteristic Eulerian
integral length and r.m.s.~velocity of the flow. From
Eq.~(\ref{eq:Re_Fr}) we can also define the buoyancy Reynolds number
\begin{equation}
 \textrm{Rb}  =   \textrm{Re} \, \textrm{Fr}^{2},
\end{equation}
which gives an estimation of how turbulent the flow is at the buoyancy
scale $L_{b}=U/N$, and as a result can be expected to play an
important role in turbulent transport. In the following we will consider
simulations with $\textrm{Rb}>1$. The Ozmidov scale,
$L_{oz}=2\pi/k_{oz}$ (with $k_{oz}=\sqrt{N^3/\epsilon}$ and $\epsilon$ the
energy injection rate), will also
play an important role in the following discussions, as for scales
sufficiently small when compared with $L_{oz}$ the flow is expected
to recover isotropy. When $\textrm{Rb}>1$ the Ozmidov scale is larger
than the Kolmogorov dissipation scale $\eta$, and quasi-isotropic
turbulent transport can thus be expected to take place at small
scales. Another parameter that will be useful to quantify small scale
turbulence and transport is the local gradient Richardson number
\begin{equation}
 \textrm{Ri}_{g} =
  \dfrac{N(N-\partial_{z}\theta)}{(\partial_{z}u_{\perp})^{2}},
\label{eq:RIG}
\end{equation}
where $u_{\perp}$ is the horizontal velocity. When
$\textrm{Ri}_{g}<1/4$ the flow can develop shear instabilities
\cite{billant_theoretical_2000}, while for $\textrm{Ri}_{g}<0$ local
overturning can take place.

\AD{A relevant time scale for the tracers is the Lagrangian turnover 
  time (or the Lagrangian time), 
  \begin{equation}
    T_{L}=\int_{0}^{\infty}\left< {\bf v}_{i}(t-\tau) \cdot 
      {\bf v}_{i}(t) \right> d\tau / \left< v_i^{2} \right>
  \label{eq:LAG_TIME}
  \end{equation}
  where the averages are done over the time $t$ and over all particles
  (i.e., over the subindex $i$). $T_L$ quantifies the time over which
  particles' velocities are auto-correlated.} 
Other relevant parameters for the next sections are the Eulerian
turnover time at the Ozmidov scale $\tau_{oz}=L_{oz}/U_z$ (with $U_z$
the characteristic Eulerian vertical velocity), and the
energy-containing (or {\it integral}) isotropic and parallel length
scales 
\begin{equation}
 L=2\pi \dfrac{\int E_V(k)k^{-1}dk}{\int E_V(k) dk},
\end{equation}
\begin{equation}
 L_{\parallel}=2\pi \dfrac{\int
   E_V(k_{\parallel})k_{\parallel}^{-1}dk_{\parallel}}{\int E_V(k) dk},
\end{equation}
where $E_V(k)$ and $E_V(k_\parallel)$ are respectively the isotropic
and parallel kinetic energy spectra. From these lengths we can also
define an energy-containing isotropic wavenumber as $K=2\pi/L$,
and an energy-containing parallel wavenumber as 
$K_{\parallel}=2\pi/L_{\parallel}$.

The numerical simulations were performed in three-dimensional periodic
domains with different aspect ratios. A first set of runs has a cubic
box with domain lengths $L_{x}=L_{y}=L_{z}$ (equal to  $2\pi$ in
dimensionless units) and isotropic linear resolution
$n_{x}=n_{y}=n_{z}$, and therefore with an aspect ratio of the 
vertical to horizontal lengths of $1:1$. Another set of simulations
was done in elongated boxes with sizes $L_{x}=L_{y}=\alpha L_{z}$ and
resolution $n_{x}=n_{y}=\alpha n_{z}$. Thus, the aspect ratio of the
domain is  $1:\alpha$, and we will consider in the following
$\alpha=4$ or 8. Note that in all cases the spatial resolution is
isotropic, i.e., the distance between grid points is the same in the
three directions, $\Delta x=\Delta y=\Delta z$, and thus isotropy can
in principle be recovered by the flow at the smallest scales.

In each domain, simulations were done using two different forcing
functions. Some simulations were forced with Taylor-Green (TG)
forcing (see, e.g., \cite{riley_2003, clark_di_leoni_absorption_2015,
  mininni_generation_2017}), which only excites directly the two
horizontal components of the velocity field, and has vertical
shear. The geometry of the large-scale flow generated by this forcing
is that of pairs of counter-rotating horizontal vortices at large
scales, and the expression of the forcing is
\begin{equation}
 {\bf f_{TG}} = f_{0} \left( \sin (x) \cos (y) \cos (\alpha z) , 
  - \cos (x) \sin (y) \cos (\alpha z) , 0 \right) .
\end{equation}
The effectively forced wavenumber is then
$k_f=(2+\alpha^2)^{1/2}$. Note that changing the aspect ratio of the
domain modifies the factor $\alpha = L_{x}/L_{z}$ and thus, the
strength of vertical gradients in the flow. For $\alpha=1$ (cubic
domain) $k_f \approx 1.7$, while for $\alpha=4$ or 8 we obtain
respectively $k_f \approx 4.2$ or $8.1$. The flow generated by these
forces (for $\alpha \neq 1$) still has a large-scale circulation at
$k_x=k_y=1$, while developing stronger shear in the vertical direction
as $\alpha$ is increased (see \cite{mininni_generation_2017} for more
details).

Other simulations were done using a random isotropic three-dimensional
forcing (RND), with a correlation time $\tau_{corr}$ of half an eddy
turn-over time. A forcing with random phases in the Fourier shell 
$k_{f} = \alpha$ is computed at a given time as
\begin{equation}
{\bf f_1} = f_{0} \sum_{|{\bf k}| \in [k_{f}, k_{f}+1)} \Re \left[
  i{\bf k} \times {\bf \hat{u}_k} e^{i ({\bf k} \cdot {\bf r} +
  \varphi_{\bf k})} \right],
\end{equation}
where $\Re$ stands for the real part, ${\bf \hat{u}_k}$ is a
unit vector, and $\varphi_{\bf k}$ are uniformly distributed random
phases. The actual forcing ${\bf f_{RND}}$ is obtained by slowly
interpolating the forcing from a previous random state ${\bf f_0}$ to
the new random state ${\bf f_1}$, in such a way that 
${\bf f_{RND}} = {\bf f_1}$ after $\tau_{corr}$. The process is then
repeated to obtain a slowly evolving random forcing. As the forcing
wavenumber depends on the aspect ratio, in the cubic box
$k_{f}=1$ while in the elongated domains $k_{f}=4$ or $8$, similarly
to the Taylor-Green case. However, note that in this case the choice
$k_{f} = \alpha$ to maintain the forcing isotropic for all aspect
ratios also implies that, as the aspect ratio $1:\alpha$ is decreased
and the forcing is applied (isotropically) at smaller scales, the
Reynolds number (based on the energy containing scale) will also
decrease.

\begin{table}
\centering
\setlength{\tabcolsep}{.3 cm}
\begin{tabular}{p{1.0cm} c c c c c c c c c c c c}
\hline \hline
Run & Aspect ratio & $n_{x}=n_{y}$ & $n_{z}$ & Forcing & $N$ & $Re$ &
    $Fr$ & $R_{b}$ & $R$  & $2 \pi / N$ & $L_{oz}$ &
    $\tau_{oz}$ \\  
\hline
TG$_{1}4$  & 1:1 & 512  & 512  &  TG  &  $4$ & $ 7000 $ & 0.04 & $11$
                   & 0.12  & 1.57 & 0.28 & 1.2 \\
TG$_{1}8$  & 1:1 & 512  & 512  &  TG  &  $8$ & $ 8000 $ & 0.02 & $3$
                   & 0.03  & 0.79 & 0.1  & 0.8 \\
TG$_{4}4$  & 1:4 & 768  & 192  &  TG  &  $4$ & $ 10000$ & 0.05 & $25$
                   & 0.25  & 1.57 & 0.36 & 1.4 \\
TG$_{4}8$  & 1:4 & 768  & 192  &  TG  &  $8$ & $ 14000$ & 0.03 & $13$
                   & 0.09  & 0.79 & 0.14 & 0.7 \\
TG$_{4}12$ & 1:4 & 768  & 192  &  TG  & $12$ & $ 15000$ & 0.02 & $4$
                   & 0.03  & 0.52 & 0.07 & 1.0 \\
TG$_{8}8$  & 1:8 & 2048 & 256  &  TG  &  $8$ & $ 35000$ & 0.03 & $30$
                   & 0.30  & 0.79 & 0.18 & 0.8 \\
RND$_{1}4$ & 1:1 & 512  & 512  &  RND &  $4$ & $ 6000 $ & 0.07 & $29$
                   & 0.06  & 1.57 & 0.24 & 0.9 \\
RND$_{1}8$ & 1:1 & 512  & 512  &  RND &  $8$ & $ 8000 $ & 0.03 & $7$
                   & 0.02  & 0.79 & 0.07 & 0.3 \\
RND$_{4}8$ & 1:4 & 768 & 192  &  RND &  $8$ & $ 3000 $ & 0.11 & $36$
                   & 0.16  & 0.79 & 0.17 & 0.7 \\
RND$_{4}8$B & 1:4 & 512 & 128  &  RND &  $8$ & $ 2000 $ & 0.10 & $20$
                   & 0.07  & 0.79 & 0.14 & 0.6 \\
RND$_{4}8$C & 1:4 & 256 & 64   &  RND &  $8$ & $ 800  $ & 0.17 & $23$
                   & 0.02  & 0.79 & 0.16 & 0.8 \\
RND$_{4}8$D & 1:4 & 128 & 32   &  RND &  $8$ & $ 300  $ & 0.20 & $12$
                   & 0.25  & 0.79 & 0.15 & 0.2 \\
\hline
\end{tabular}
\caption{Relevant parameters of the simulations. The aspect ratio
  gives the vertical to horizontal aspect ratio of the spatial domain
  $1:\alpha$, $n_{x}$, $n_{y}$, and $n_{z}$ are the grid points in
  each spatial direction, forcing indicates the forcing function,
  $\textrm{Re}$ is the Reynolds number, $\textrm{Fr}$ is the Froude
  number, $\textrm{Rb}$ is the buoyancy Reynolds number, $R$ is the
  fraction of particles with $\mathrm{Ri_{g}}<0$, $L_{oz}$ is the Ozmidov
  length scale, and $\tau_{oz}=L_{oz}/U_{z}$ is the Eulerian
  turnover time at the Ozmidov scale.}
\label{tab:parameters}
\end{table}

All flows were evolved from ${\bf u}=\theta=0$, and once they reached
the turbulent steady state Lagrangian particles were injected and
integrated in time together with the flow. The list of all runs with
their respective relevant parameters is presented in table
\ref{tab:parameters}. Runs are labeled using the forcing (TG or RND),
a subindex for the inverse aspect ratio ($\alpha = 1$, 4, or 8), and a
number indicating the value of the Brunt-V\"{a}is\"{a}l\"{a} frequency
($N=4$, 8, or 12). As mentioned above, note
that run RND$_{4}8$ has a lower $\textrm{Re}$ than, e.g., run
TG$_{4}8$ (although it has the same spatial resolution and kinematic
viscosity), as the isotropic forcing at $k_{f}=\alpha=4$ in the
RND$_{4}8$ run results in a smaller integral length scale when 
compared to the TG$_{4}8$ run, which has a large-scale flow at
horizontal scales (with $k_\perp \approx 1$) with shear at smaller
vertical scales (with $k_\parallel = 4$). However, note run 
RND$_{4}8$ also has a larger $\textrm{Fr}$, thus resulting in a
larger $\textrm{Rb}$. To explore the effect of varying $\textrm{Re}$
and $\textrm{Rb}$, while keeping the forcing and aspect ratio fixed
and $\textrm{Fr}$ approximately the same, runs RND$_{4}8$ and
RND$_{4}8$B to RND$_{4}8$D were done at decreasing spatial resolution
and at increasing values of $\nu=\kappa$.

\begin{figure}
\centering
\includegraphics[width=8cm]{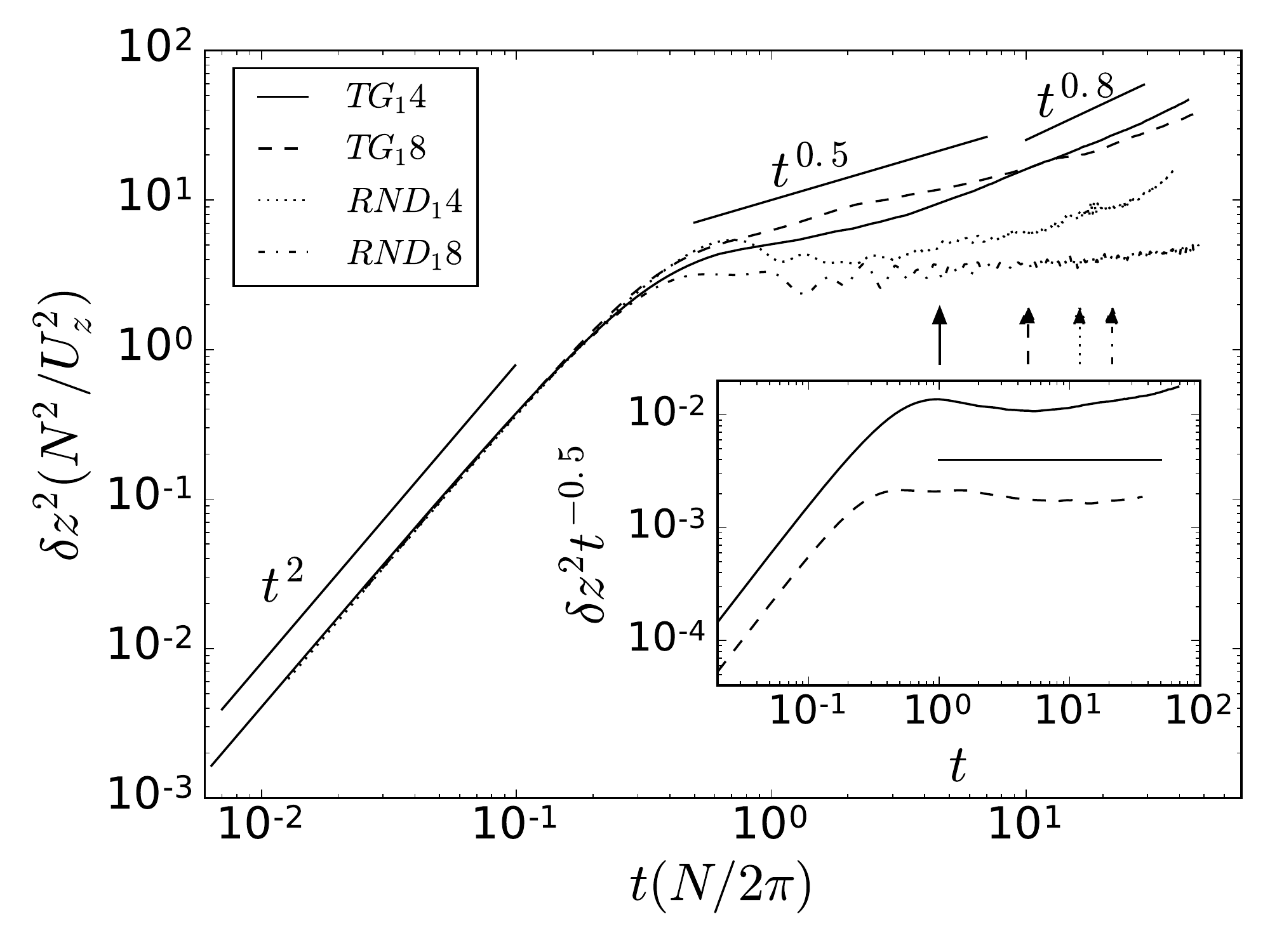}
\includegraphics[width=8cm]{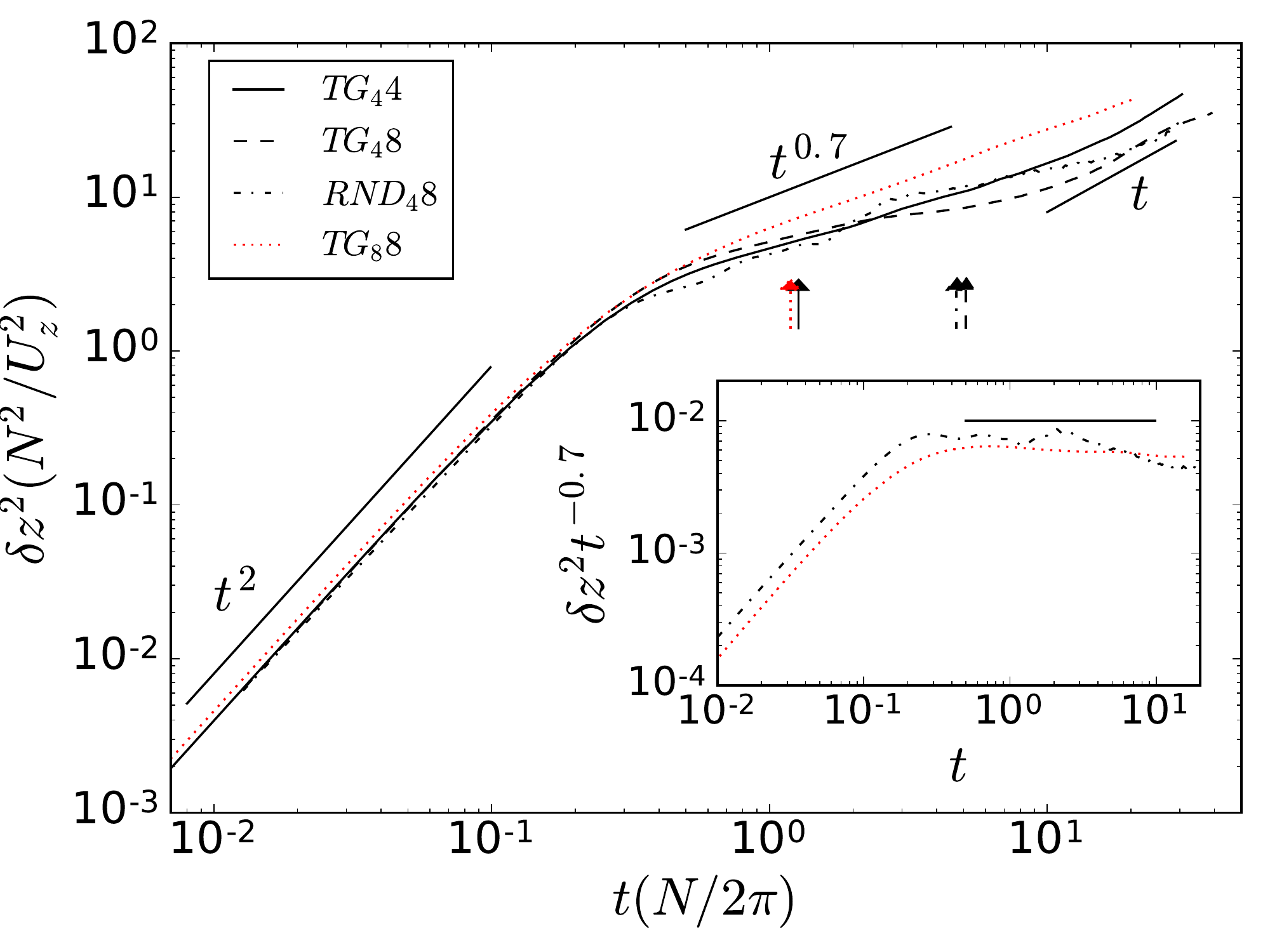}
\caption{({\it Color online)} Mean vertical dispersion 
  $\delta z^2$ for ({\it left}) runs in cubic domains (boxes with
  aspect ratio 1:1), and ({\it right}) runs in elongated domains with
  aspect ratio 1:4, in both figures with TG and RND forcing and with
  different Brunt-V\"{a}is\"{a}l\"{a} frequencies. The dispersion is
  normalized by $U_{z}^{2}/N^2$, the ratio of the squared mean vertical
  velocity to the Brunt-V\"{a}is\"{a}l\"{a} frequency, and time is
  normalized by the Brunt-V\"{a}is\"{a}l\"{a} period. Power laws are
  indicated as references. \AD{In each figure, the Lagrangian time of
    each run is indicated by a vertical arrow with the same line style
    as the corresponding run, while the insets show the mean 
    vertical dispersion of some of the simulations, not normalized,
    and compensated by the power laws indicated in the main panels for
    intermediate times.}}
\label{f:DZ2}
\end{figure}

\section{Single-particle vertical dispersion in stably stratified
  turbulence} \label{sec:SIN}

Particle dispersion in SST is inherently different from HIT as
stratification suppresses vertical dispersion. As mentioned in the
Introduction, linear models of SST predict the saturation of the
vertical dispersion for $t \approx 2\pi / N$, as the displacement of
particles is in practice vertically bounded by the stratification,
resulting in an oscillatory motion of the
particles \cite{nicolleau_turbulent_2000}. This is confirmed by
numerical simulations at moderate buoyancy Reynolds number
\cite{kimura_diffusion_1996,lindborg_vertical_2008}.

\begin{figure}
\centering
\includegraphics[width=8cm]{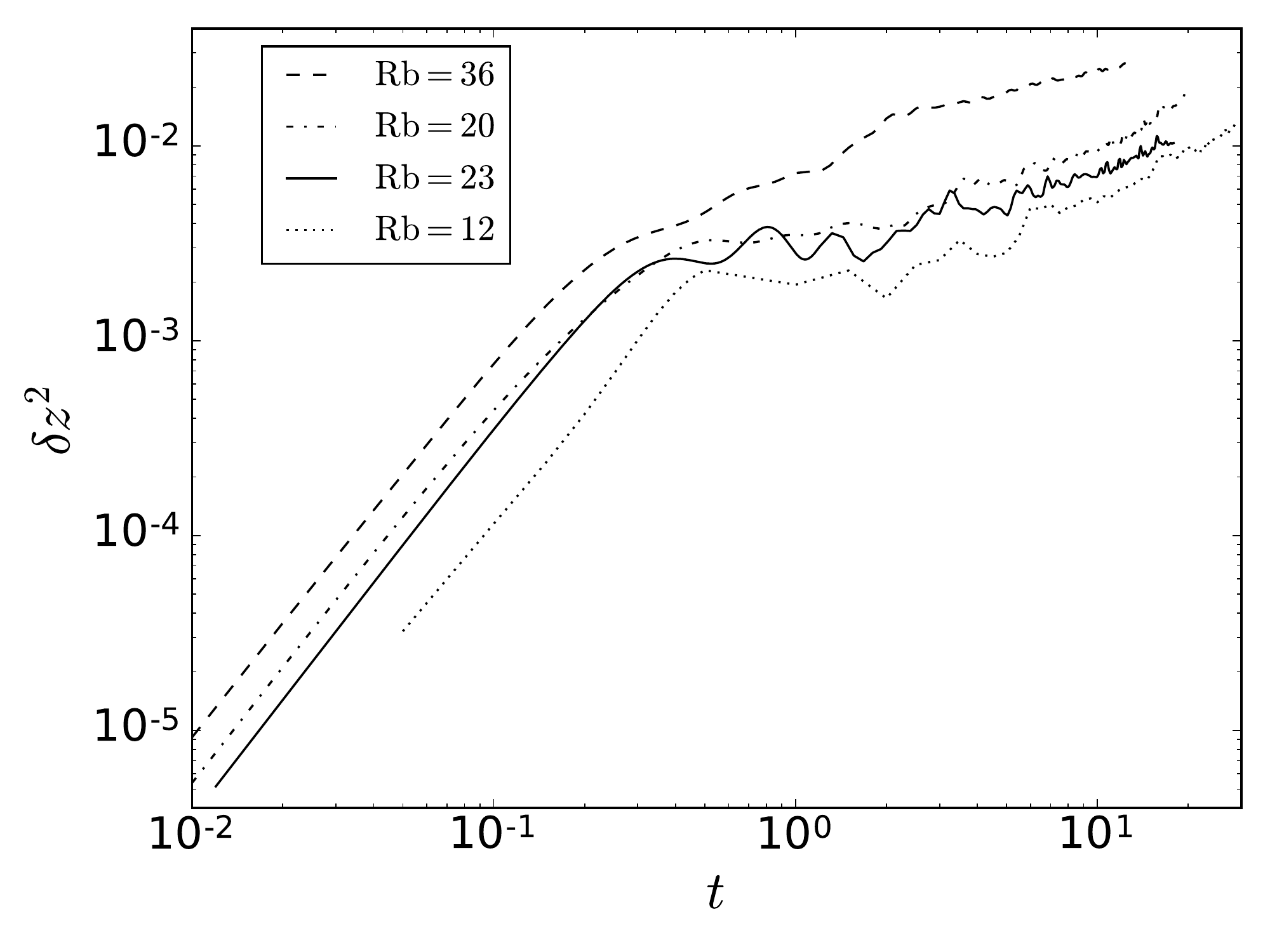}
\includegraphics[width=8cm]{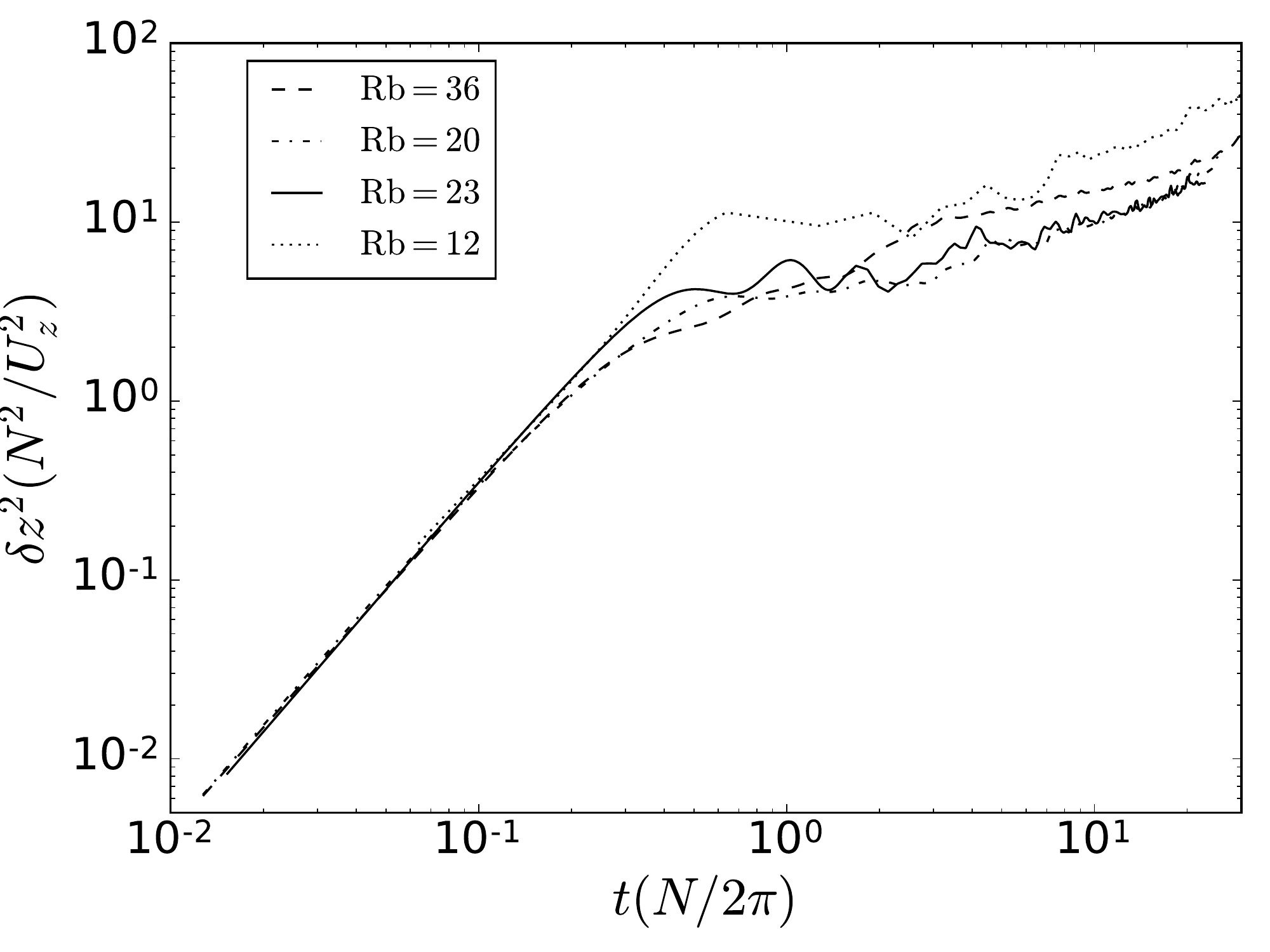}
\caption{Mean squared vertical displacement $\delta z^2$ for RND runs
  in domains with aspect ratio 1:4 and with $N=8$ (run RND$_{4}8$ and
  runs RND$_{4}8$B to RND$_{4}8$D, from higher to lower $\textrm{Re}$
  and $\textrm{Rb}$. {\it Left:} $\delta z^2(t)$ without
  normalization. {\it Right:} same, normalized by the ratio $U_{z}^{2}/N^2$,
  and with time normalized by the Brunt-V\"{a}is\"{a}l\"{a}
  period. } 
\label{f:DZ2_RE}
\end{figure}

We computed the single-particle vertical dispersion as 
$\delta z^2=\left<[z_{i}(t)-z_{i}(0)]^{2}\right>_{i}$, where $i$ is
the particle label, and the average is computed over all particles. 
Figure~\ref{f:DZ2} shows the resulting mean vertical dispersion
in our simulations, for TG and RND forcing, different
aspect ratios, and different Brunt-V\"{a}is\"{a}l\"{a} frequencies
(and thus, different Froude numbers). Time is normalized by 
$2 \pi / N$ (the Brunt-V\"{a}is\"{a}l\"{a} period), while $\delta z^2$
is normalized by $(U_{z}/N)^{2} $ (following the normalization used
in \cite{aartrijk_single-particle_2008}) \AD{where $U_{z}$ is, as
  already mentioned, the (Eulerian) r.m.s.~vertical velocity in the
  turbulent steady state when the particles were injected in the flow
  (note that while $U_{z}$ does not change significantly in time and
  has similar values for both forcing functions, the pointwise value
  of $u_{z}$ changes significantly in space in the runs with TG
  forcing, as will be discussed in detail in
  Sec.~\ref{sec:spatialTG}).} With this normalization all curves
collapse from $t=0$ until $t\approx2\pi/N$, in a time range when they
display ballistic behavior $\delta z^2 \sim t^2$. \AD{In
  Fig.~\ref{f:DZ2} we also indicate with arrows the Lagrangian time
  $T_L$ of each simulation. These times are very different for each
  run, and are also different from the time at which the ballistic
  behavior ends. The end of the ballistic regime at the time of the
  wave period $2\pi/N$, instead of at the Lagrangian time $T_L$,
  indicates that the early-time fast vertical dispersion is dominated
  by the waves,} in good agreement with previous studies of SST
\cite{aartrijk_single-particle_2008, sujovolsky_single-particle_2017}:
particles are first displaced ballistically by the internal gravity 
waves, which for $\textrm{Fr}<1$ are faster than the large-scale
turbulent eddies.

The ballistic behavior observed in Fig.~\ref{f:DZ2} finishes after one
Brunt-V\"{a}is\"{a}l\"{a} period, resulting in a change in the growth
of $\delta z^2(t)$. In some cases (see runs RND$_{1}4$ and RND$_{1}8$
in Fig.~\ref{f:DZ2}) $\delta z^2(t)$ grows very slowly or even
saturates at late times, displaying a plateau. The saturation was
reported before in simulations of SST at moderate $\textrm{Re}$ and 
$\textrm{Rb}$ numbers \cite{aartrijk_single-particle_2008,
  sujovolsky_single-particle_2017}, where a very slow growth at late
times was attributed to the effect of molecular diffusion. However,
some of our runs (all TG runs even at moderate $\textrm{Rb}$, and
simulations with RND forcing at higher $\textrm{Rb}$ in elongated
domains) display a more efficient transport (i.e., a faster growth of
$\delta z^2(t)$ for $t>2\pi/N$) when compared with the runs that
display the plateau. The enhanced dispersion after $t>2\pi/N$ seems 
to be controlled, at least for RND forcing, by $\textrm{Rb}$,
suggesting it may be caused by turbulence generated by shear
instabilities or by overturning events. \AD{As a reference, some power
  laws are shown in Fig.~\ref{f:DZ2} in this late time regime, and the
  mean vertical dispersions of some of the simulations,
  compensated by these power laws, are shown in insets. However, note
  that as will be discussed in Sec.~\ref{sec:MOD}, the enhanced
  dispersion seen in some of these runs is a combination of both the
  effects of the waves and of the turbulent eddies, and as a result it
  is not captured by a unique power law, and tends in some of the runs
  (at sufficiently high $\textrm{Rb}$ or for sufficiently strong
  overturning events) to a $\delta z^2 \sim t$ behavior for
  sufficiently long times.}

To further illustrate the effect of varying $\textrm{Rb}$,
Fig.~\ref{f:DZ2_RE} shows the single-particle vertical dispersion for
several runs with RND forcing, and with the same parameters as
RND$_{4}8$ (runs RND$_{4}8$B to RND$_{4}8$D), but with different
spatial resolution and values of $\textrm{Rb}$ (by decreasing
$\textrm{Re}$). Run RND$_{4}8$D, with the lowest values of
$\textrm{Re} \approx 300$ and of $\textrm{Rb} \approx 12$, displays
a saturation in $\delta z^2(t)$ at $tN/(2\pi) \approx 7\times 10^{-1}$,
a plateau until $tN/(2\pi) \approx 5$, and then a slow growth. As
$\textrm{Rb}$ increases the plateau shortens, until it completely
disappears for run RND$_{4}8$ (with $\textrm{Re} \approx 3000$ and 
$\textrm{Rb} \approx 36$). \AD{Note that while one panel in
  Fig.~\ref{f:DZ2_RE} shows $\delta z^2$ without any normalization
  (and thus the increase in its amplitude with increasing
  $\textrm{Rb}$ can be easily appreciated), the other panel shows 
  $\delta z^{2} (N^2/U_z^{2})$. In this latter case, the change in the
  amplitude seen at late times is thus associated with the fact that
  $U_z^{2}$ also increases with increasing Rb, resulting in a
  net decrease of $\delta z^{2} (N^2/U_z^{2})$ with increasing
  $\textrm{Rb}$. However, this normalization (together with 
  the normalization of the time $t$ by the Brunt-V\"{a}is\"{a}l\"{a}
  period) makes all curves collapse again at early times, further
  showing that the ballistic behavior is independent of
  $\textrm{Rb}$ and thus of the strength of the small-scale
  turbulence.}

The case of TG forcing is different, as the plateau in $\delta z^2(t)$
at intermediate times is not present even in runs at moderate 
$\textrm{Rb}$. Although turbulence plays an important role in the
dispersion at high $\textrm{Rb}$, the TG forcing function generates a
coherent large-scale flow which creates strong fronts and helps
instabilities to develop \cite{mininni_generation_2017}, enhancing
vertical dispersion even at values of $\textrm{Rb}$ which are low when
compared to the RND case. In the next section we study the gradient
Richardson number \textrm{Ri}$_{g}$, with a special focus on the
TG simulations, to characterize the features of this flow that result
in differences in the vertical dispersion.

\begin{figure}
\centering
\includegraphics[width=8cm]{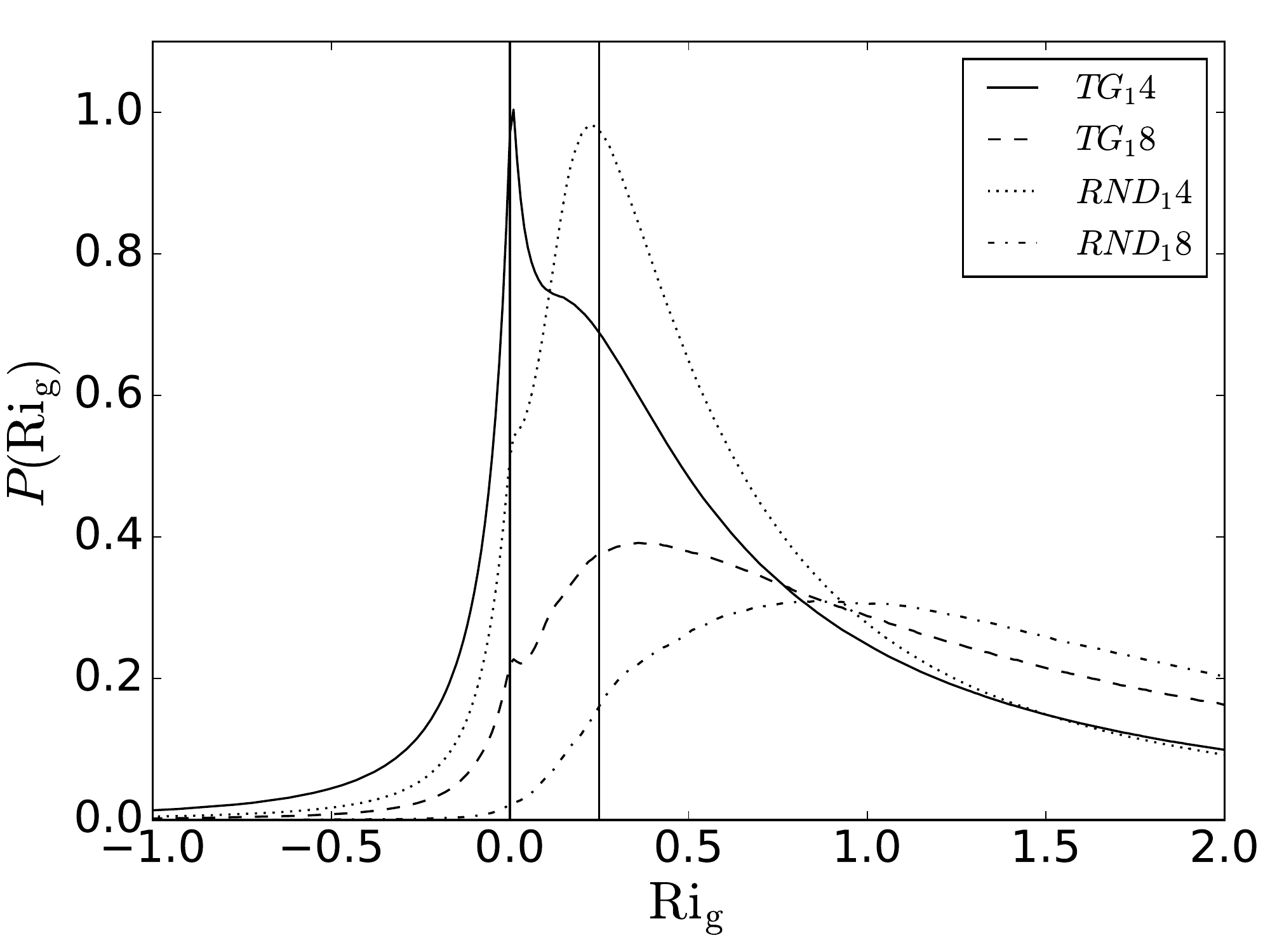}   
\caption{PDFs of the Eulerial local gradient Richardson number
  \textrm{Ri}$_{g}$, for all runs in cubic domains (runs TG$_{1}4$
  and TG$_{1}8$ with TG forcing and $N=4$ and 8 respectively, and runs
  RND$_{1}4$ and RND$_{1}8$ with RND forcing and $N=4$ and 8
  respectively). Vertical solid lines at $\textrm{Ri}_{g}=0$ and $1/4$
  are shown as references. }
\label{f:RIG_BOX}
\end{figure}

\section{The local gradient Richardson number} \label{sec:RIG}

\subsection{General properties of the local gradient 
  Richardson number}

The local gradient Richardson number provides a measure of the
vertical stability of stratified flows. When 
$\textrm{Ri}_{g}({\bf r})<1/4$ pointwise, local shear instabilities
can take place \cite{davidson_turbulence_2013}, while if
$\textrm{Ri}_{g}<0$, then $\partial_{z}\theta > N$, and an overturning
instability can develop generating convection locally in the
flow. Figure \ref{f:RIG_BOX} shows the probability density functions
(PDFs) of the Eulerian $\textrm{Ri}_{g}$ for all runs in cubic
domains. The PDFs of runs with $N=4$ (TG$_{1}4$ and RND$_{1}4$)
display larger probabilities of low values of $\textrm{Ri}_{g}$
($<1/4$ and $<0$) than the runs with the same forcing but with $N=8$
(TG$_{1}8$ and RND$_{1}8$). As $N$ is increased (for a given forcing),
the peak of the PDF moves to larger values of $\textrm{Ri}_{g}$. This
indicates, as expected, that as stratification increases vertical
instabilities are inhibited, and as a result we can also expect a less
efficient vertical transport (in agreement with the single-particle
vertical dispersion observed in the previous section). However, when
we compare TG and RND runs with the same value of $N$, we see that TG
runs still shows larger probabilities of $\textrm{Ri}_{g}<1/4$ and of
$\textrm{Ri}_{g}<0$. Indeed, the PDFs of the TG runs are shifted
towards the left relative to the RND set, indicating that this flow is
more vertically unstable and, consequently, can be more efficient at
vertically displacing particles.

\begin{figure}
\centering
\includegraphics[width=7.5cm]{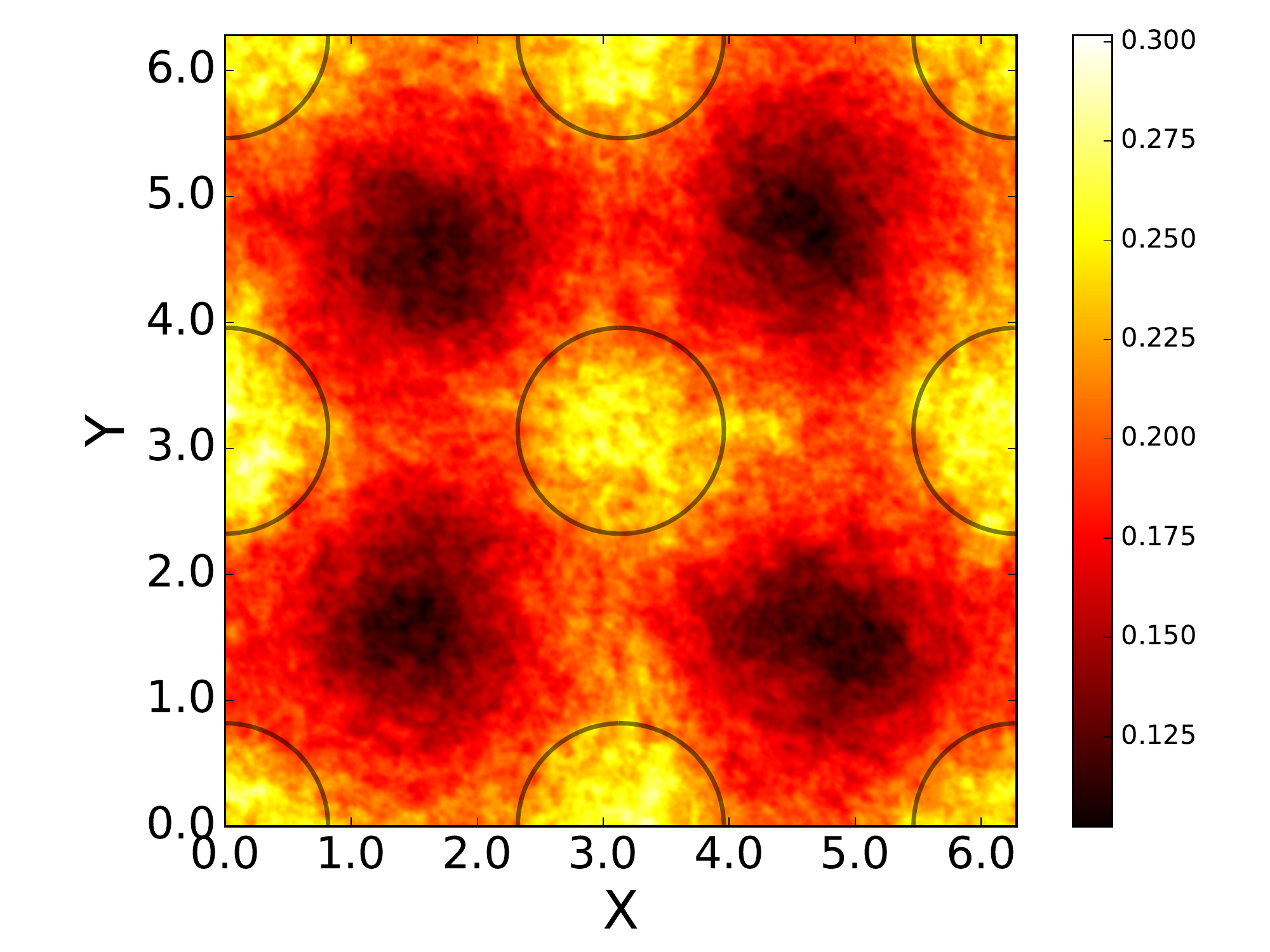} 
\caption{({\it Color online}) Vertically averaged absolute value of
the Eulerian vertical velocity, $\left<|u_{z}|\right>_{z}$, for run
TG$_{4}4$ (TG forcing, $N=4$, and aspect ratio 1:4). Bright regions
correspond to large vertical velocities in absolute value. As the
domain is periodic in both $x$ and $y$ directions, the regions with
large $\left<|u_{z}|\right>_{z}$ can be enclosed by four circles
(cylinders when extended in the $z$ direction), indicated as a
reference by the black solid lines. }
\label{f:CYL}
\end{figure}

\begin{figure}
\centering
\includegraphics[width=8cm]{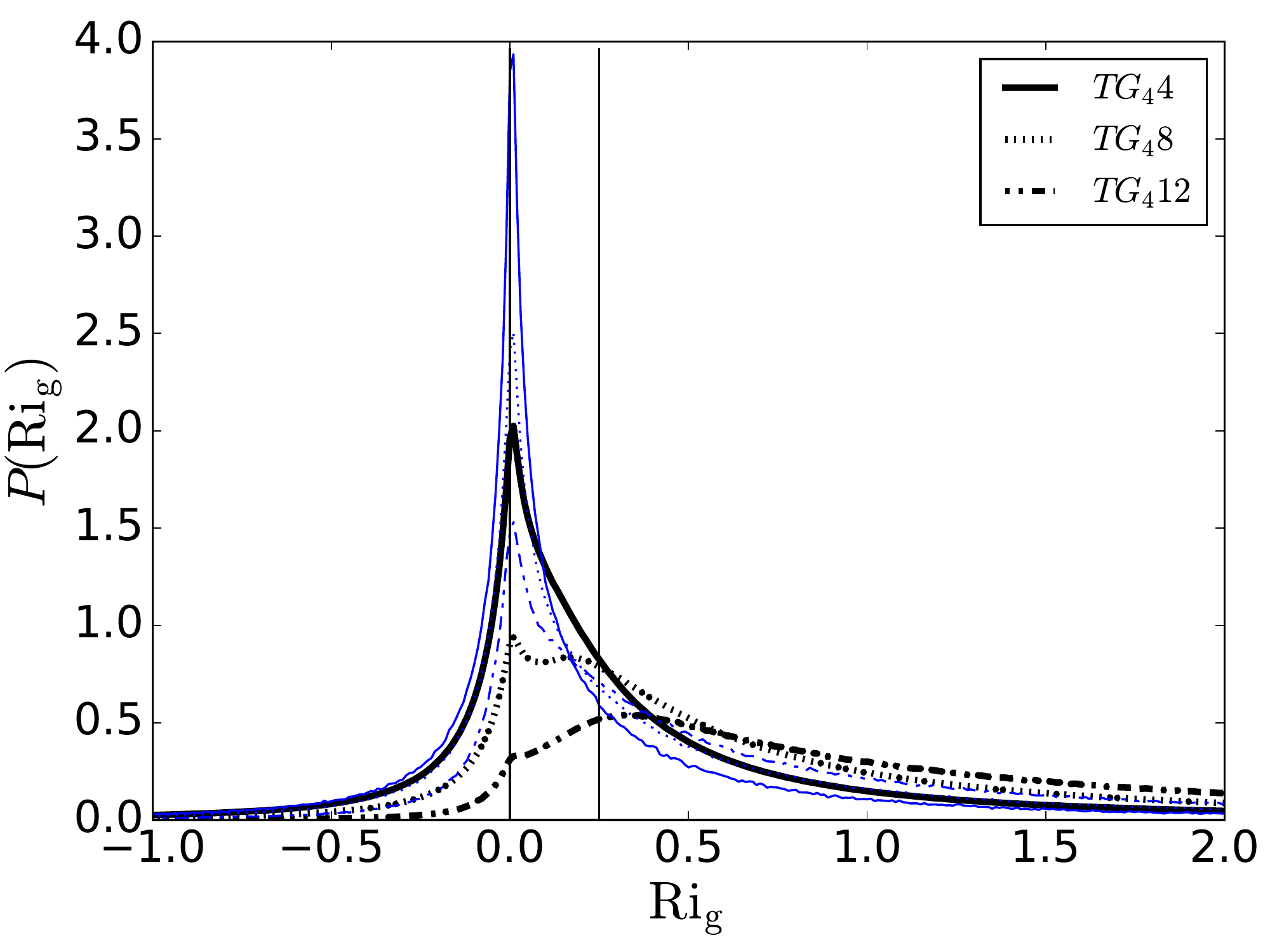}   
\includegraphics[width=8cm]{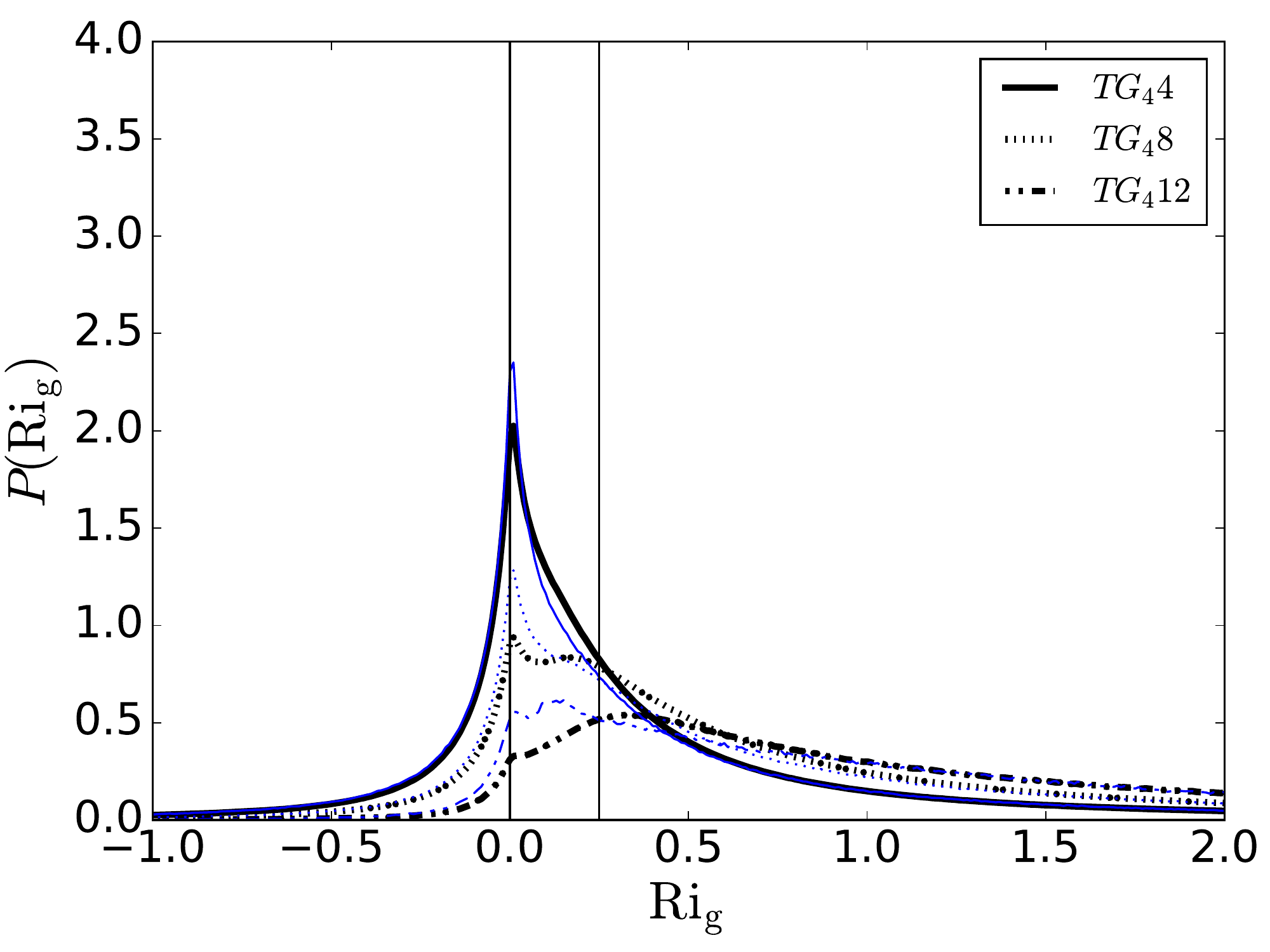}  
\caption{({\it Color online}) Both panels show, in black thick lines,
the Lagrangian PDFs of the gradient Richardson number
\textrm{Ri}$_{g}$ for TG runs in boxes with 1:4 aspect ratio. This
PDF is compared with ({\it left}) the PDFs of $\textrm{Ri}_{g}$
restricted to Lagrangian velocities
$|v_z|> \left<|v_z|\right>+2\sigma_{v_z}$ (where $\sigma_{v_z}$ is
the dispersion in $v_z$) in thin (blue) lines, and ({\it right}) the
PDFs of $\textrm{Ri}_{g}$ restricted to particles in the circular
regions indicated in Fig.~\ref{f:CYL}. Vertical lines at
$\textrm{Ri}_{g}=0$ and $\text{Ri}_{g}=1/4$ are shown as
references.}
\label{f:RIG_R}
\end{figure}

\subsection{The spatial structure of the local gradient Richardson
number in the TG flow \label{sec:spatialTG}}

We are interested in how the TG forcing affects the structure of the
local gradient Richardson number. As the forcing generates a coherent
large-scale flow, which in principle can affect vertical transport, we
show first in Fig.~\ref{f:CYL} the mean vertical value of the absolute
Eulerian vertical velocity $\left<|u_{z}|\right>_{z}$ computed 
for run TG$_44$ (where the subscript $z$ in the brackets indicates the
average was computed along the $z$-coordinate). As explained in
Sec.~\ref{sec:NSIM}, the TG flow consists of pairs of counter-rotating
horizontal vortices, separated vertically by shear layers. Pressure
gradients create a vertical circulation
\cite{mininni_generation_2017}, and as a result the forcing generates 
a coherent structure at the largest scales that organize the flow into
regions of high and low $\left<|u_{z}|\right>$. As a result, some
well-defined spatial regions in the flow display a bias towards larger
values of $|u_{z}|$ (also associated with the generation of front- and
filament-like structures in the flow, as discussed in
\cite{mininni_generation_2017}). \AD{As a comparison, the runs with
  RND forcing do not display such a large-scale structure (not
  shown).} It can thus be expected that Lagrangian particles
approaching these regions in the TG flow will have a tendency to
suffer larger displacements in the vertical direction, thus increasing 
$\delta z^2$ even at moderate $\textrm{Rb}$.

To confirm this effect, Fig.~\ref{f:RIG_R} shows the PDFs of the
Lagrangian \textrm{Ri}$_{g}$ (i.e., now computed using the gradients
as seen by the Lagrangian particles) for runs with TG forcing in the
box with 1:4 aspect ratio. \AD{Gradients (as well as velocity and
  density fluctuations) seen by the Lagrangian tracers are computed
  for each time and at each particle position using the same
  three-dimensional cubic spline interpolation used to integrate the
  particles discussed in Sec.~\ref{sec:NSIM}. From these quantities,
  the PDFs of \textrm{Ri}$_{g}$ are computed. As expected, the
  ``Lagrangian'' PDFs coincide with the Eulerian PDFs, which are
  computed at a fixed time and for all points in the Eulerian spatial
  grid; however, the PDFs from Lagrangian data will allow us next to
  more easily compute statistics restricted to specific conditions
  over the fluid elements. As a result,} as observed before for the
Eulerian statistics, for the complete dataset as the stratification
increases (i.e., for higher $N$) the mean gradient Richardson number
also increases, and the fraction of fluid  elements with 
$\textrm{Ri}_{g}<1/4$ or $\textrm{Ri}_{g}<0$ (i.e., prone to
overturning) decreases. But, \AD{as we just mentioned}, the 
computation of \textrm{Ri}$_{g}$ using the gradients as seen by the
Lagrangian particles also allow us to compute  conditional statistics,
e.g., only for instants when the particles suffer large vertical
velocities, or when the particles are in a specific region in
space. Using the mean of the absolute Lagrangian vertical velocity
$\left<|v_z|\right>$ (averaged over all particles and over time), and
the standard deviation of $v_z$ ($\sigma_{v_z}$), we computed the PDF
of $\textrm{Ri}_{g}$ restricted to particles with absolute vertical
velocity $2\sigma_{v_z}$ larger than $\left<|v_z|\right>$ (see
Fig.~\ref{f:RIG_R}). With this restriction, the fraction of fluid
elements that can suffer overturning instabilities increases (note the
PDFs have a larger peak at $\textrm{Ri}_{g}=0$, display larger values
for $\textrm{Ri}_{g}<0$, and smaller values for $\textrm{Ri}_{g}>0$
when compared with the PDFs at the same $N$ without any
restriction). This indicates that there is a correlation between fluid
elements with $\textrm{Ri}_{g} \le 0$ and large values of $|v_z|$ (and
thus, of particles displacing larger distances in the vertical direction, and
 contributing to $\delta z^2$). We also see that as $N$ is
increased, the probability of finding fluid elements with 
$\textrm{Ri}_{g} \le 0$ decreases even when restricted to parcels with
large $|v_z|$. Finally, Fig.~\ref{f:RIG_R} also shows the PDF of
$\textrm{Ri}_{g}$ restricted to the instants the particles are in the
spatial regions of the large-scale circulation for which the largest
absolute values of $u_{z}$ were observed in Fig.~\ref{f:CYL}. A
similar (albeit weaker) behavior as for the restriction in $v_{z}$ is
found, with the shift in the peak of the PDFs towards smaller values
of $\textrm{Ri}_{g}$, confirming the relevance of the geometry of the
large-scale flow in the TG runs for the vertical transport of
Lagrangian particles.

\begin{figure}
\centering
\includegraphics[width=8cm]{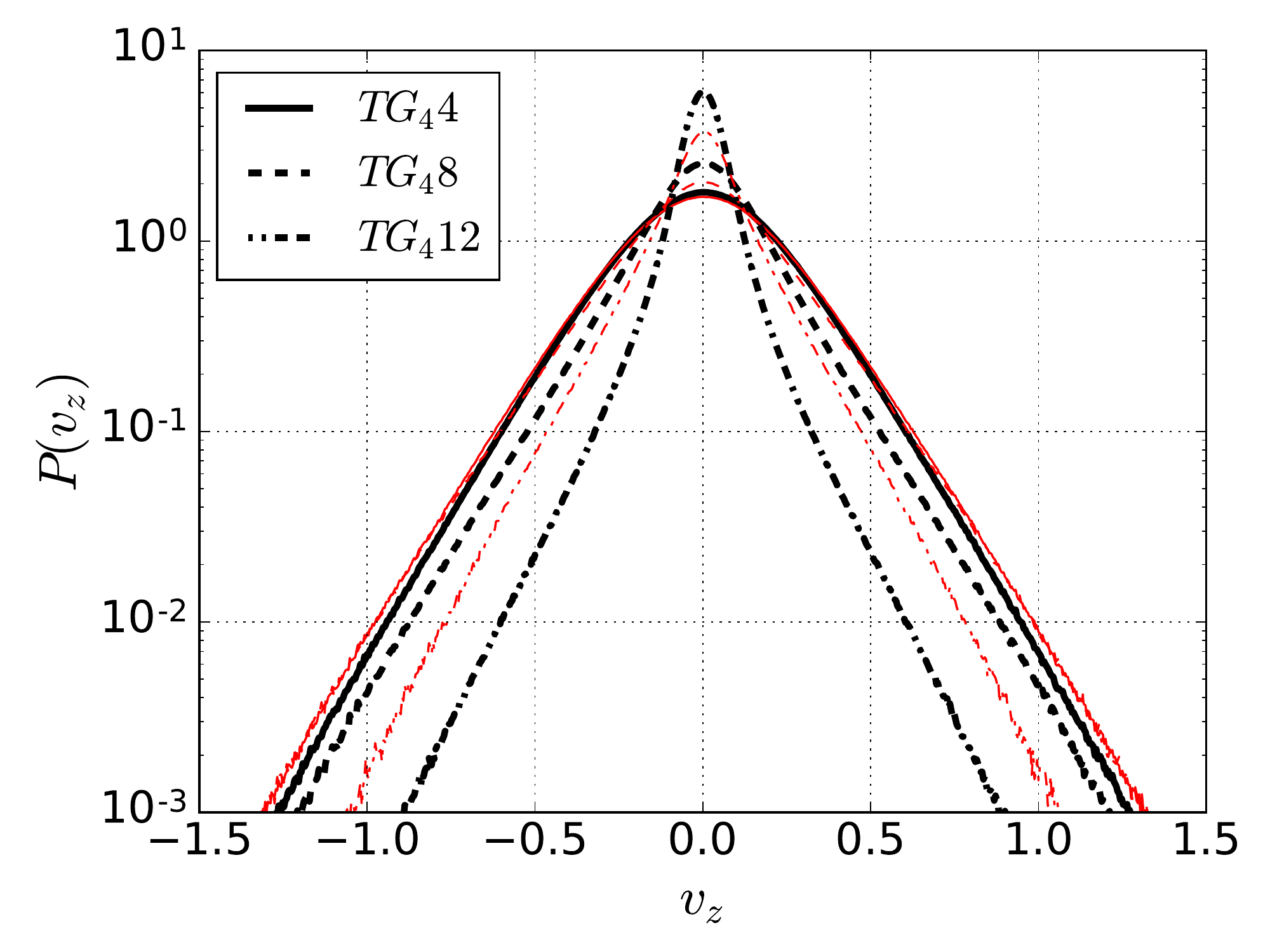}   
\includegraphics[width=8cm]{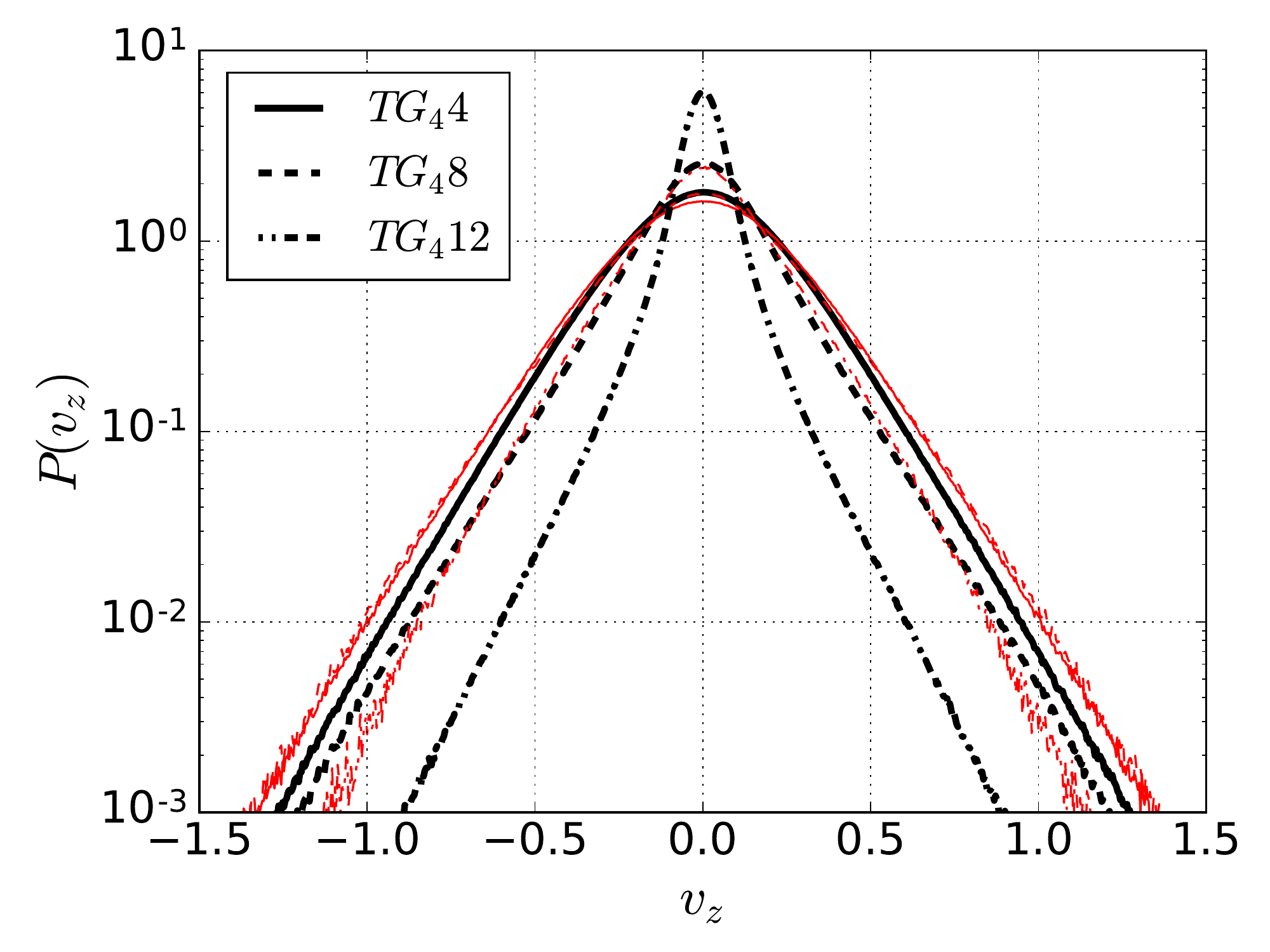}   
\caption{({\it Color online}) Both panels show in thick black lines
the PDFs of the Lagrangian vertical velocity $v_{z}$ for TG runs in
domains with 1:4 aspect ratio and varying $N$. PDFs restricted to
particles in instants with $\textrm{Ri}_{g}<1/4$ ({\it left}), or
with $\textrm{Ri}_{g}<0$ ({\it right}), for the same runs, are shown
in (red) thin lines.}
\label{f:HVZ}
\end{figure}

\begin{figure}
\centering
\includegraphics[width=5.5cm]{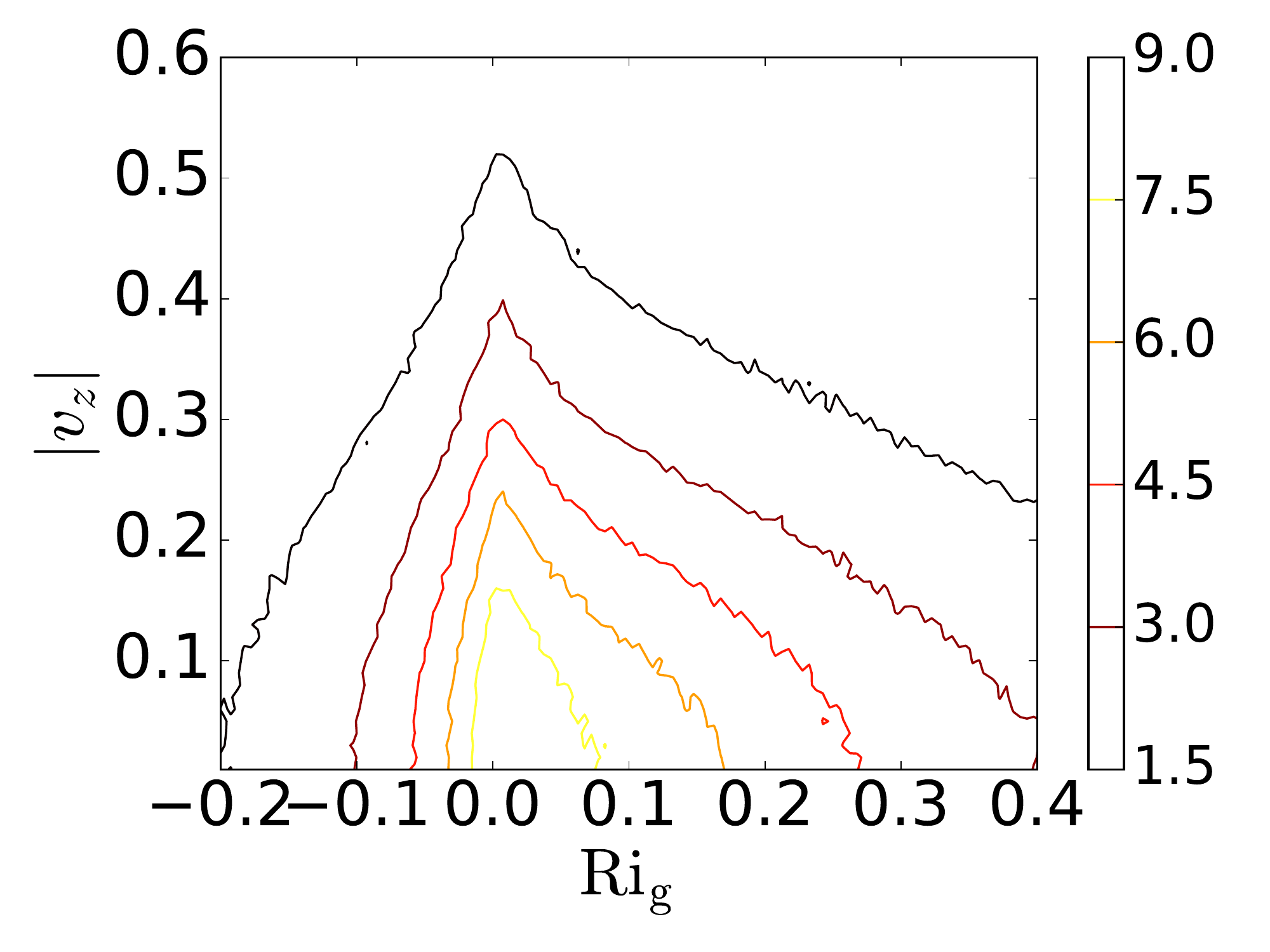}   
\includegraphics[width=5.5cm]{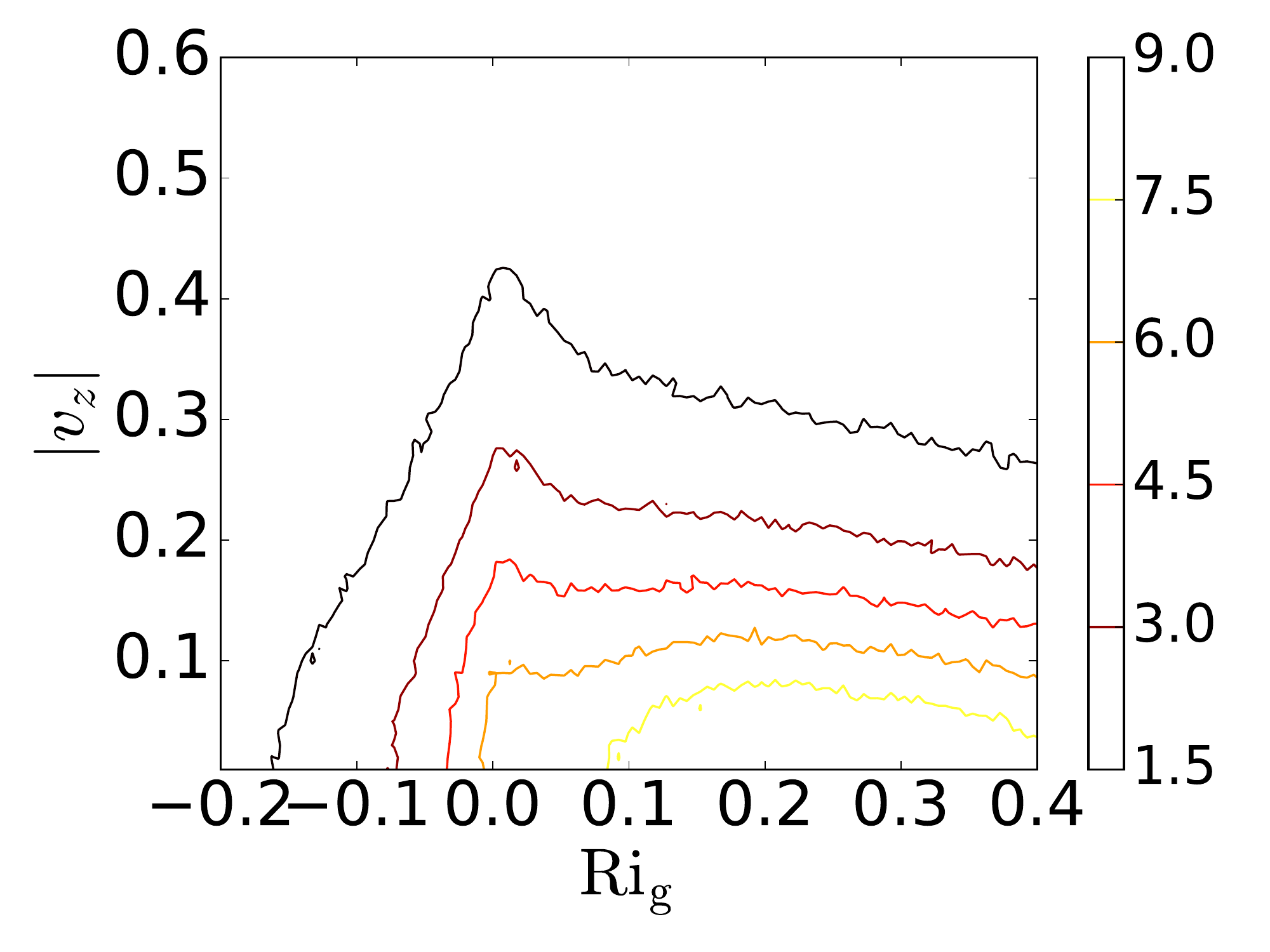}  
\includegraphics[width=5.5cm]{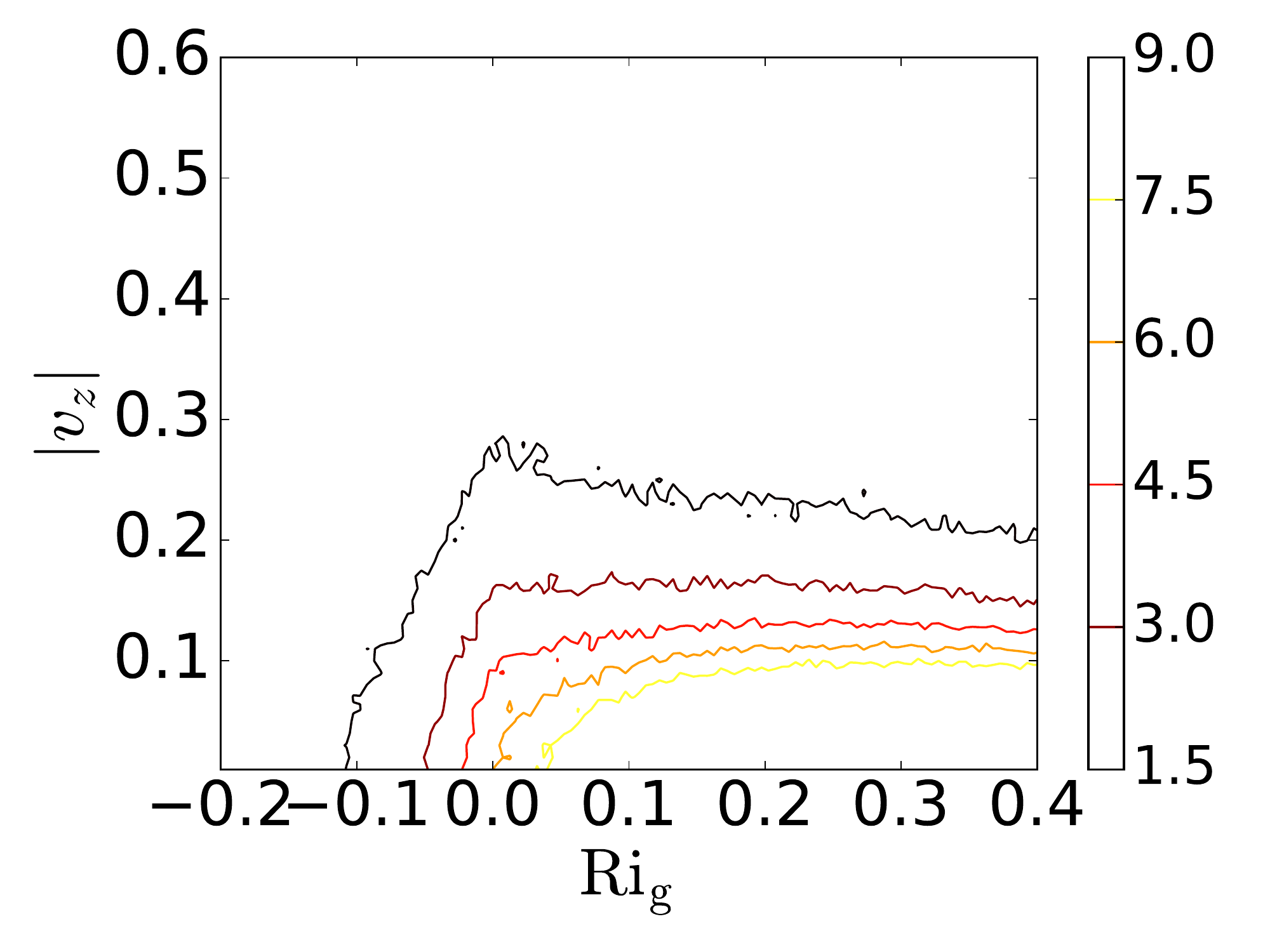}  
\caption{({\it Color online}) Isocontours of the joint probability
distribution function of $\textrm{Ri}_{g}$ and $|v_{z}|$,
$P(\textrm{Ri}_{g},|v_{z}|)$, for runs TG$_{4}4$ ({\it left}),
TG$_{4}8$ ({\it middle}), and TG$_{4}12$ ({\it right}).}
\label{f:h_Ri_vs_Vz}
\end{figure}

To further study the effect of $\textrm{Ri}_{g}$ on the vertical
velocity of the particles, Fig.~\ref{f:HVZ} shows the PDFs of the
Lagrangian vertical velocity for all particles in TG runs with aspect
ratio 1:4 and with varying $N$. As previously reported in
\cite{rorai_turbulence_2014,feraco_vertical_2018}, the vertical
velocity does not follow Gaussian statistics, and display strong tails
(this feature is not exclusively associated with the TG forcing, as
the same behavior was found in simulations with random forcing, see
\cite{rorai_turbulence_2014}). In \cite{feraco_vertical_2018} the
extreme values were shown to be associated to intermittent overturning
instabilities in the flow. Note that the behavior reported in
\cite{feraco_vertical_2018} is non-monotonous in $\textrm{Fr}$,
although for sufficiently small $\textrm{Fr}$ (or sufficiently large
values of $N$) the maximum values of $v_z$ decrease with increasing
stratification (see Fig.~\ref{f:HVZ}). When we compute the PDFs
restricted to particles in instants for which $\textrm{Ri}_{g}<1/4$ or
$\textrm{Ri}_{g}<0$, while for the runs with moderate stratification
($N=4$ and 8) there are only small changes in the 
tails of the PDFs (albeit extreme values of $v_z$ become more
probable), for stronger stratification ($N=12$) the changes are
significantly larger, with stronger tails. This further confirms that
points with $\textrm{Ri}_{g}<1/4$ or $\textrm{Ri}_{g}<0$ are
associated with larger values of $v_z$, and can thus be expected to be
associated with the enhanced dispersion after $t > 2\pi/N$ at least in
the TG runs.

This can be also confirmed in Fig.~\ref{f:h_Ri_vs_Vz}, which shows the
joint probability density function as a function of $\textrm{Ri}_{g}$
and $|v_{z}|$, $P(\textrm{Ri}_{g},|v_{z}|)$, for the TG runs with
aspect ratio 1:4 and with varying $N$. As the stratification
increases, the probability of finding particles with large values of
$|v_{z}|$ decreases, while that of finding larger values of
$\textrm{Ri}_{g}$ increases. For $N=4$ and $N=8$ note the 
correlation between larger absolute values of the vertical velocity
with $\textrm{Ri}_{g}\approx0$ values, which is significantly weaker
in the run with $N=12$.

\begin{figure}
\centering
\includegraphics[width=8cm]{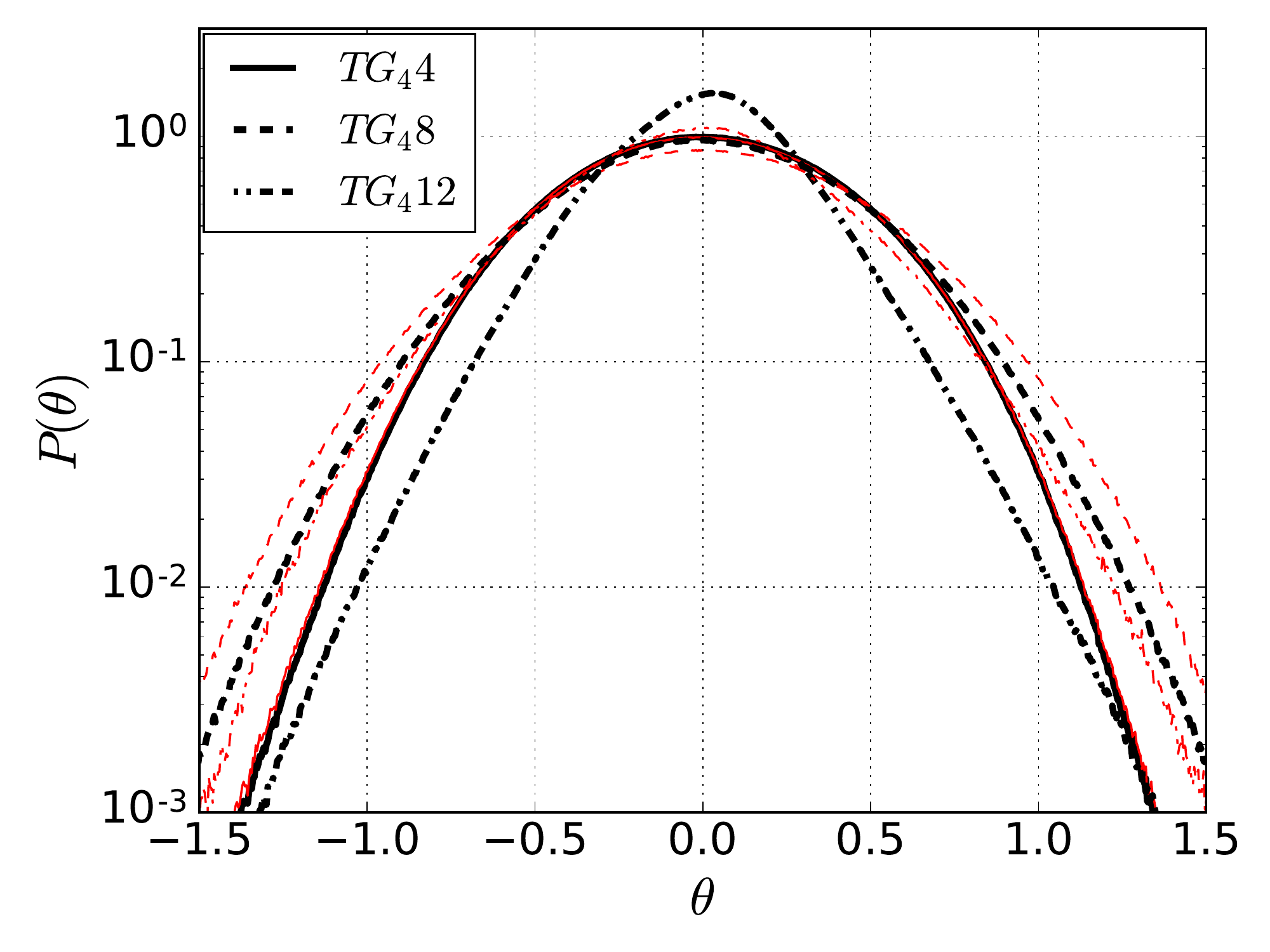}   
\includegraphics[width=8cm]{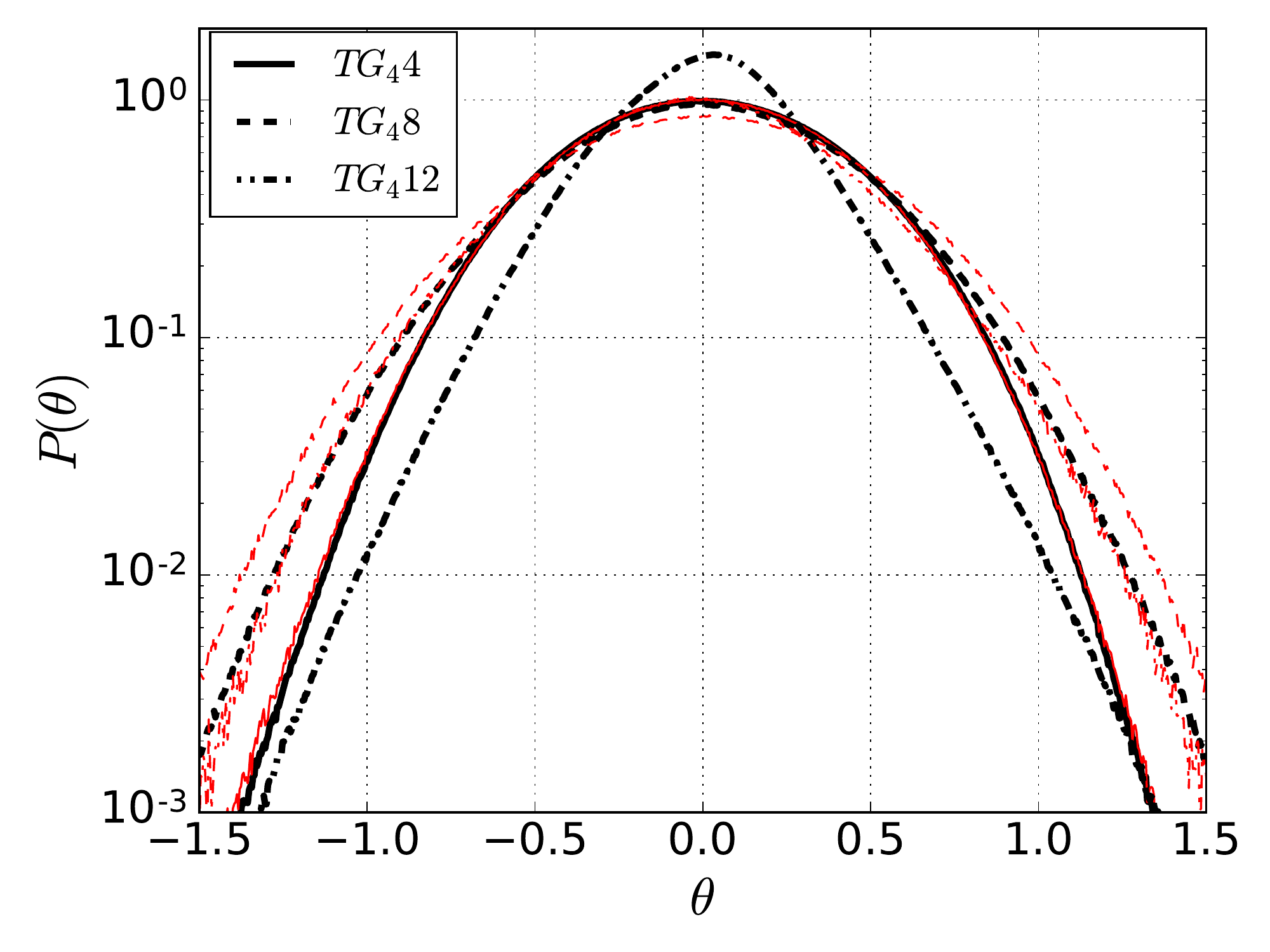}   
\caption{({\it Color online}) PDFs of $\theta$ as seen by the
Lagrangian particles (thick black curves), and the same PDFs
restricted (in thin red curves) to ({\it left}) particles at times
with $\textrm{Ri}_{g}<1/4$, and ({\it right}) particles at times
with $\textrm{Ri}_{g}<0$. Except for the run TG$_{4}12$, all PDFs
are compatible with Gaussian statistics for $\theta$.}
\label{f:HTH}
\end{figure}

\begin{figure}
\centering
\includegraphics[width=8cm]{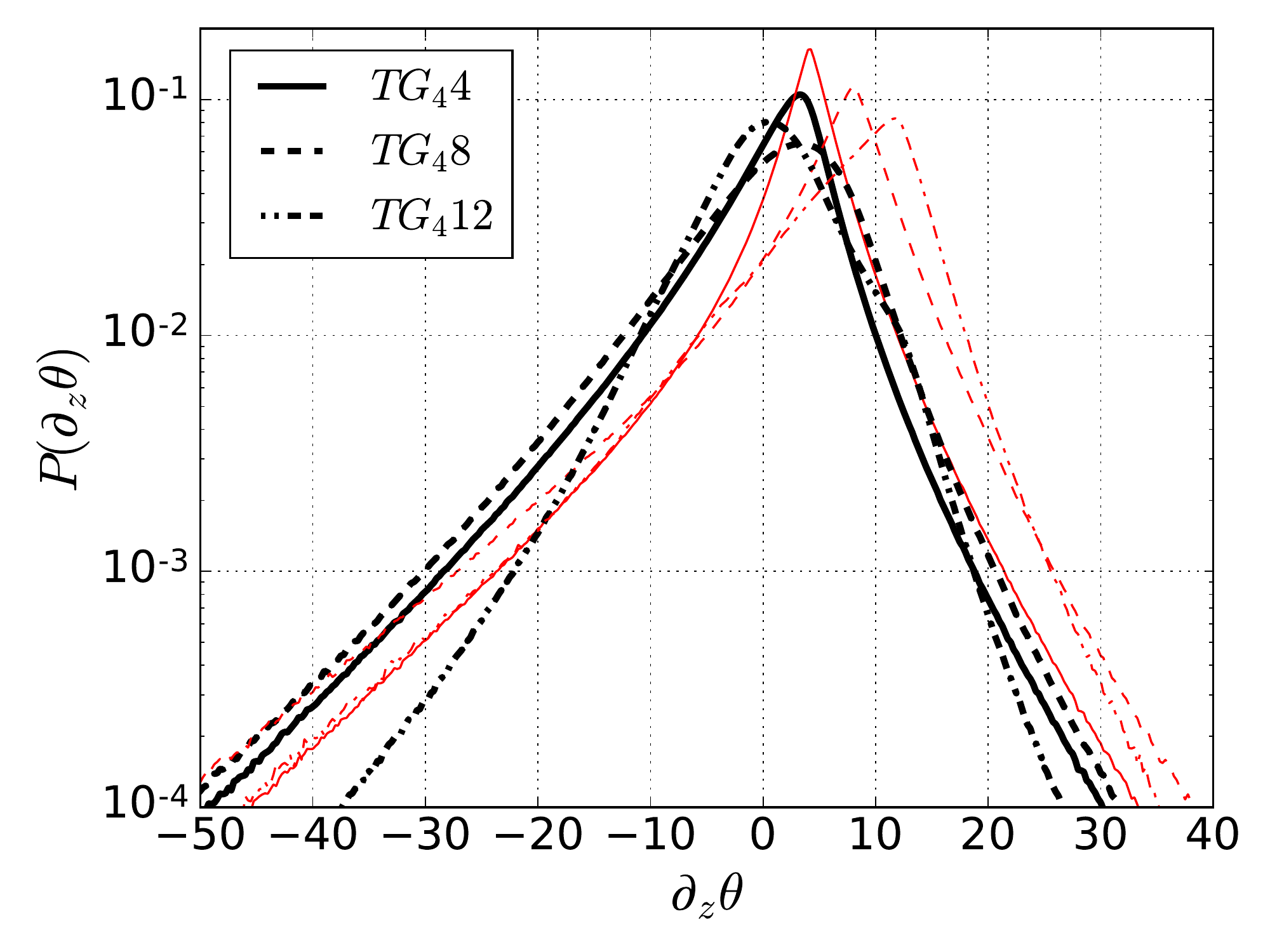}   
\includegraphics[width=8cm]{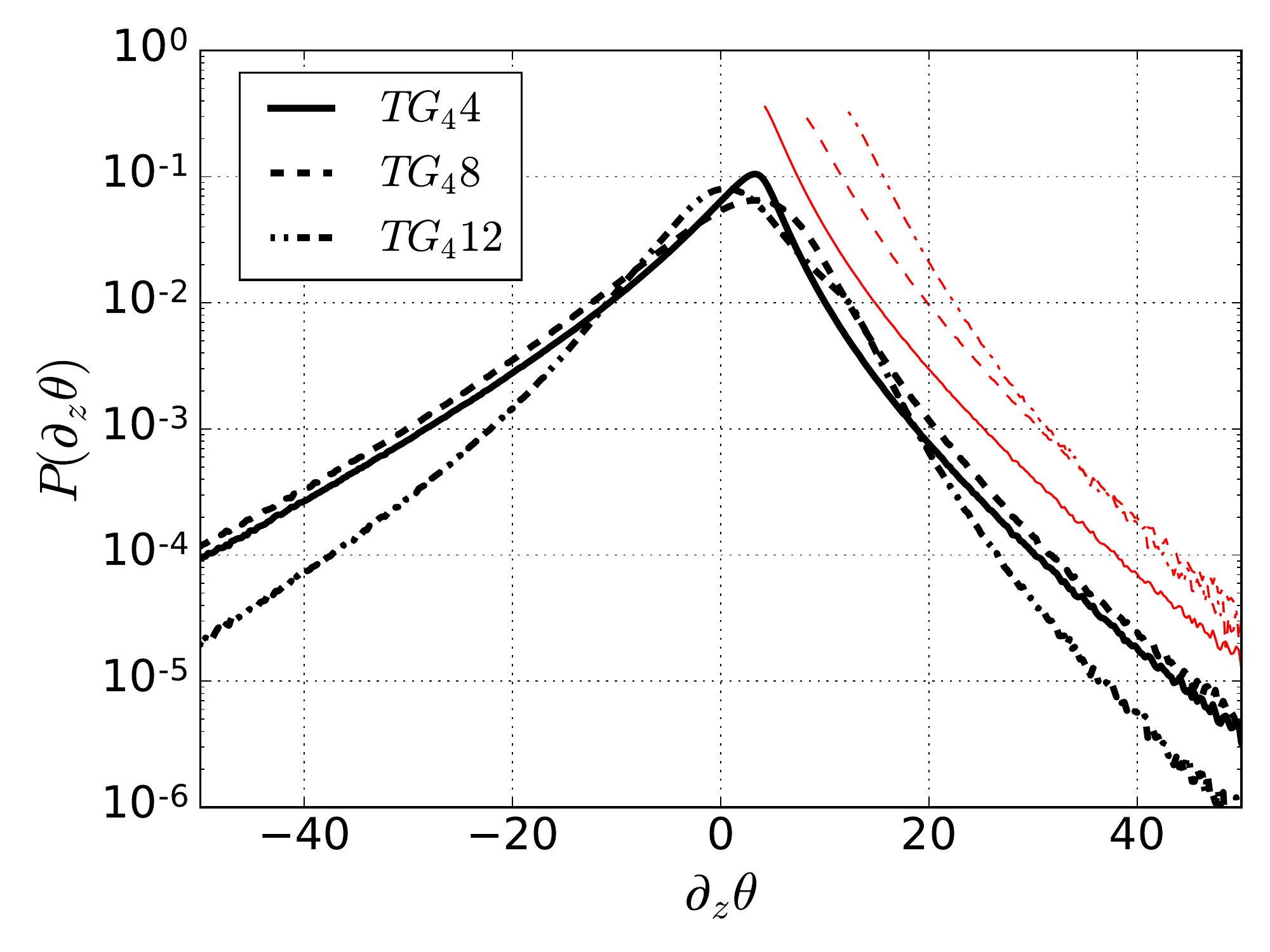}   
\caption{({\it Color online}) PDFs of the Lagrangian vertical
temperature gradients $\partial_{z}\theta$ (thick black curves), and
the same PDFs restricted (in thin red curves) to ({\it left})
$\textrm{Ri}_{g}<1/4$  and ({\it right}) $\textrm{Ri}_{g}<0$.}
\label{f:HDTH}
\end{figure}

Finally, we also studied how the value of $\textrm{Ri}_{g}$ affects
$\theta$ and $\partial_{z}\theta$ with increasing stratification (note
that the local value of $\partial_{z}\theta$ is important for overturning
instabilities, as the gradient of the buoyancy fluctuations can
compete with the background gradient, resulting in local inversion of
the stratification). Figure~\ref{f:HTH} shows the PDFs of $\theta$ as
seen by the Lagrangian particles, and the same PDFs restricted to
instants when $\textrm{Ri}_{g}<1/4$ or  $\textrm{Ri}_{g}<0$, 
in all cases for the $TG_{4}$ runs. For $N=4$ the 
PDFs of $\theta$ are close to Gaussian, and the restriction in the
values of $\textrm{Ri}_{g}$ has a negligible effect in the
statistics. However, for $N=8$, while the PDFs are still close to
Gaussian, the restricted PDFs show a lower probability for
$|\theta|<0.5$ and higher probability for $|\theta|>0.5$, indicating
particles with $\textrm{Ri}_{g}<1/4$ or $\textrm{Ri}_{g}<0$ 
are more likely to be found in points with higher potential energy
density $\sim \theta^{2}$. This behavior is enhanced for $N=12$, for
which the PDFs also display non-Gaussian tails. Finally,
Fig.~\ref{f:HDTH} shows the PDFs of the Lagrangian vertical gradients
of $\theta$, $\partial_{z}\theta$, which are non-Gaussian and
asymmetric. The asymmetry is enhanced when the PDFs are restricted to
instants when $\mathrm{Ri_{g}}<1/4$ or $\mathrm{Ri_{g}}<0$.   While
the non-restricted PDFs have their maximum at 
$\partial_{z}\theta \gtrsim 0$, for the restricted PDFs the maximum is
at $\partial_{z}\theta \approx N$. From the ideal Boussinesq equation
for $\theta$ (Eq.~\ref{eq:theta}, with $\kappa=0$), it can be seen
that ${\boldsymbol \nabla}\theta = \left(0,0,N\right)$ is a fixed
point of both the equations for $\theta$ and for the Lagrangian
evolution of $\partial_{z}\theta$, which could explain the
accumulation of (restricted) particles with $\partial_{z}\theta
\approx N$. Also, at points where $\mathrm{Ri_{g}} < 0$, then
$\partial_z \theta > N $ (for which overturning events can
occur). This is the reason why the PDFs of particles restricted to
$\mathrm{Ri_{g}}<0$ in Fig.~\ref{f:HDTH} only take values of
$\partial_{z} \theta$ greater than $N$. \AD{Finally, note that since
  $\textrm{Ri}_{g}$ depends explicitly on $\partial_{z}\theta$ and not
  on the pointwise value of $\theta$, a restriction on the values of
  $\textrm{Ri}_{g}$ can be expected to affect the PDFs in
  Fig.~\ref{f:HDTH} more strongly than those in Fig.~\ref{f:HTH}, as
  is indeed observed in the figures.}

\begin{figure}
\centering
\includegraphics[width=8cm]{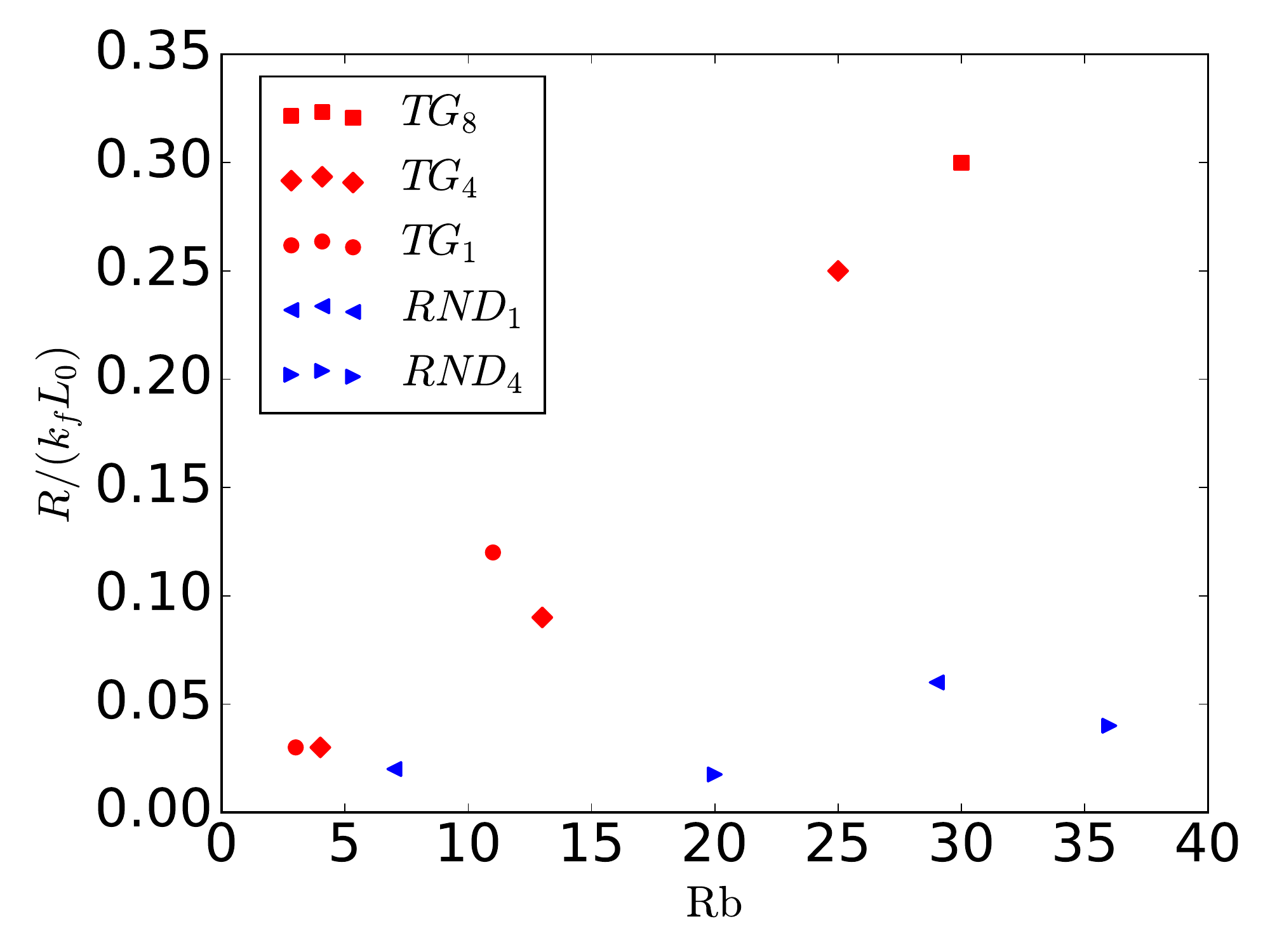}   
\caption{({\it Color online}) Overturning probability normalized by
the forcing wave number and unit length, $R/(k_{f}L_{0})$, as a
function of the Reynolds Buoyancy number $\textrm{Rb}$ for all
simulation in table \ref{tab:parameters}.}
\label{fig:R_VZ_RB}
\end{figure}

\subsection{Overturning probability and the buoyancy Reynolds number}

As already mentioned, the extreme vertical velocities reported in the
previous subsection are not exclusive of the TG flow. In
\cite{rorai_turbulence_2014, feraco_vertical_2018}, non-Gaussian PDFs
of $u_z$, $v_z$, and $\theta$ were reported for RND forcing depending
on the values of $\textrm{Fr}$ and $\textrm{Rb}$. However, it is clear
from the results shown so far that the geometry of the TG flow
facilitates the development of overturning instabilities and the
occurrence of extreme values of the vertical velocity even at moderate
$\textrm{Rb}$.

In the next section we will use these results to build a simple model
for single-particle vertical dispersion, for all cases considered and
independently of the two specific forcing function used. The results
in Sec.~\ref{sec:SIN} suggest that while the ballistic behavior of
$\delta z^2$ for $t<2\pi/N$ is dominated by the waves, the differences
in $\delta z^2$ for $t>2\pi/N$ depend on the strength of the vertical
velocity and of the turbulence. For moderate turbulence (i.e.,
moderate values of $\textrm{Rb}$) and without a large-scale vertical
circulation, $\delta z^2$ is dominated by the waves even at 
late times, resulting in the observed saturation of the
single-particle vertical dispersion. But for larger values of
$\textrm{Rb}$ (as in some of the RND runs), or in the 
presence of a large-scale flow (as in all TG runs), strong vertical
updrafts or downdrafts can enhance vertical transport resulting in the
growth of $\delta z^2$ at late times. We will measure the probability
of this happening by introducing an overturning probability $R$,
defined as the fraction of particles (in the Lagrangian frame) or
the fraction of space volume (in the Eulerian frame) with
$\textrm{Ri}_{g}<0$. Figure \ref{fig:R_VZ_RB} (see also table
\ref{tab:parameters}) gives $R/(k_{f}L_{0})$ as a function of
$\textrm{Rb}$ for all runs, where $R$ is measured as the integral of
the PDF of $\textrm{Ri}_{g}$ for $\textrm{Ri}_{g}<0$. \AD{The
  normalization of $R$ by the product $k_{f}L_{0}$ (where $k_{f}$ is
  the forcing wave number and $L_{0}$ the unit length) makes all
  simulations with a given forcing (either RND or TG) collapse in the
  vicinity of approximate linear relations independently of the
  forcing wave number used. Indeed,} the data follows (for the range
of $\textrm{Rb}$ considered) a linear relation with $\textrm{Rb}$,
with two different slopes for the TG and RND runs (even when the runs
in each set also have different aspect ratios, forcing wavenumbers,
Reynolds, and Froude numbers). As expected, for fixed $\textrm{Rb}$,
the TG runs display larger values of $R$ than the RND runs. 

\begin{figure}
\centering
\includegraphics[width=8cm]{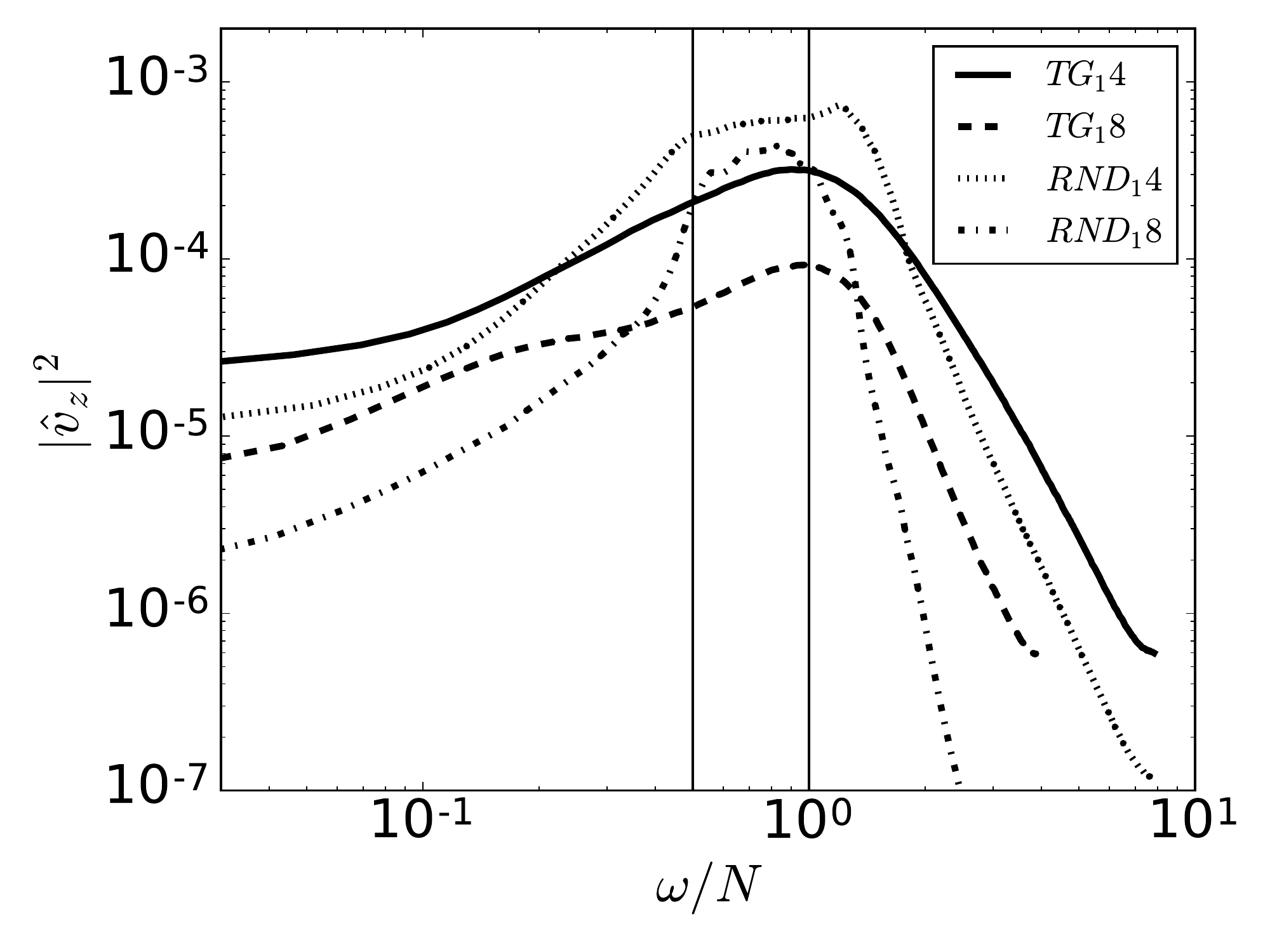}   
\includegraphics[width=8cm]{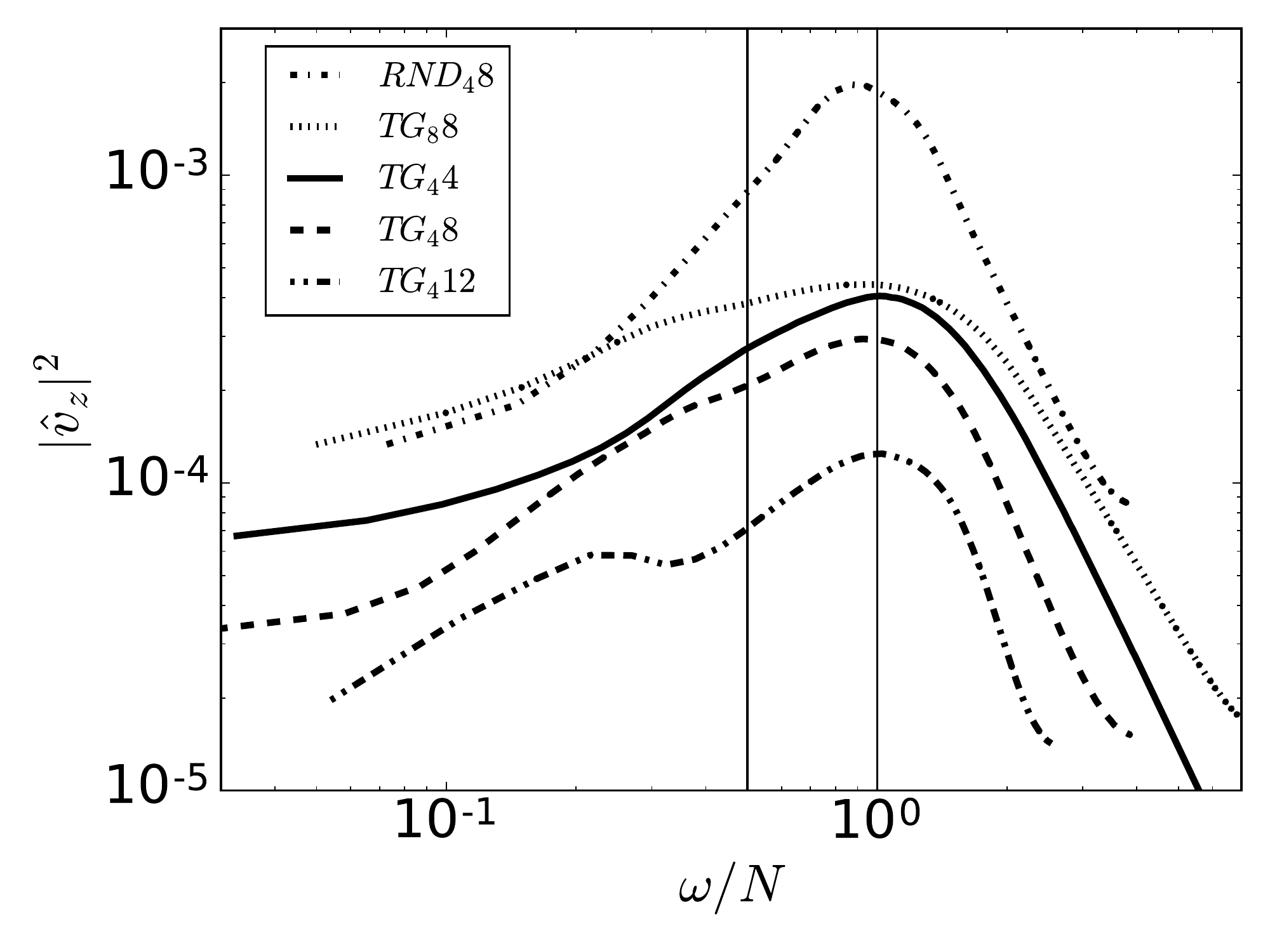}   
\caption{Power spectrum of the Lagrangian vertical velocity for ({\it
left}) runs in cubic domains, and ({\it right}) runs in elongated domains. 
Frequencies have been normalized by the
Brunt-V\"{a}is\"{a}l\"{a} frequency. The solid vertical lines
indicate (from left to right) $\omega=N/2$ and $\omega=N$.}
\label{fig:LPS}
\end{figure}

\section{Single-particle vertical dispersion model} \label{sec:MOD}

To study the vertical dispersion of single-particles observed in the
DNSs of SST in section \ref{sec:SIN}, we now present a stochastic
model that combines a random wave model (to consider the effect of
internal gravity waves) with a CTRW \cite{rast_turbulent_2016} (to
capture the effect of overturning by turbulent or large-scale eddies). 

\AD{Based on the results presented so far (and in particular, on the
  observation that at early times the behavior is dominated by
  waves),} the wave model we present consists of a sum of linear
waves with random phases. \AD{The presence of internal gravity waves
  in these flows, and their dispersion relation, have been studied
  before in spatio-temporal studies (see, e.g.,
  \cite{clark_di_leoni_absorption_2015}), further indicating their
  relevance in the dynamics of SST.} We can thus approximate the
trajectory of a Lagrangian particle moving vertically following these
waves as
\begin{equation}
z_{wav}(t) = \Re\left[ {\sum_{\omega} A_{\omega} e^{i(\omega
t+\phi_{\omega})}} \right] , 
\label{eq:traj}
\end{equation}
where $\phi_{\omega}$ is a random phase \AD{(note that as we are
  interested only in the vertical motion of the particles, the
  possible dependence of travelling waves on $x$ and $y$ can be
  ignored or absorbed into the random phase)}, and where the sum is
performed over $N_\omega$ uniformly distributed frequencies in the
range of frequencies $\omega \in [\omega_{min}, \omega_{max}]$.
The amplitude of the waves satisfies the spectral relation
\begin{equation}
A_\omega = A_{0} \, \omega ^{-1} ,
\label{eq:AW}
\end{equation}
for the same range of frequencies, and where $A_{0}$ is a
normalization factor. The dependence of $A_\omega \sim \omega^{-1}$
follows from observations that the power spectrum of the actual
Lagrangian vertical velocity has a broad maximum with approximately
constant amplitude near the Brunt-V\"{a}is\"{a}l\"{a} frequency. Note
that associating $v_z$ with  
$\dot{z}_{wav} = \Re[ {\sum A_{\omega} \omega e^{i(\omega
t+\phi_{\omega})}} ]$,
a flat power spectrum for $v_z$ implies Eq.~(\ref{eq:AW}) for the
amplitude of the waves. Once $N_{\omega}$ 
is chosen, the normalization factor $A_{0}$ can
then be fixed as $A_{0}=\left( 2 U_{z}^{2}/N_{\omega}\right)^{1/2}$ 
by imposing that for each particle $\left<\dot{z}_{wav}^2\right>_t$
(averaged over time) must be equal to the mean squared Eulerian
vertical velocity $U_{z}^{2}$ (also equal to the mean squared
Lagrangian vertical velocity) using Parseval's theorem.

Note also that a flat Lagrangian spectrum for a range of frequencies
is compatible with oceanic observations of the so-called Garrett-Munk
spectrum, and also with numerical simulations of SST 
\cite{sujovolsky_single-particle_2017}. As an example, 
Fig.~\ref{fig:LPS} shows the power spectrum of the Lagrangian vertical
velocity for all runs in table \ref{tab:parameters}. There is a broad
peak near $\omega=N$, and in several of the runs an approximately
flat spectrum can be observed in its vicinity (as a reference, the 
figure indicates a range of frequencies $[N/2,N]$), with a decay
compatible with a power law for $\omega>N$, and a slowly decaying, or
almost flat, spectrum for $\omega \ll N$). Also, for the runs with the
smallest values of $\textrm{Rb}$ considered in this study (runs
TG$_{1}8$ with $\textrm{Rb}=3.2$, and TG$_{4}12$ with
$\textrm{Rb}=4.3$), a secondary peak at smaller frequencies can be
observed. In these runs turbulence is moderate, and the waves dominate
the dispersion at intermediate times. 

As the dispersion relation of internal gravity waves is
$\omega=Nk_\perp/k \le N$, we set $\omega_{max}=N$, and for
simplicity, from the results in Fig.~\ref{fig:LPS} we set
$\omega_{min}=N/2$ in all cases. It then follows from
Eq.~(\ref{eq:traj}) that the vertical displacement of any particle
following the waves is given by 
\begin{equation}
\delta z_{wav}(t) = z_{wav}(t) - z_{wav}(0) = \sum_{\omega} A_{w} \left[
\cos(\omega t + \phi_{\omega}) - \cos(\phi_{\omega}) \right] .
\label{eq:displacement}
\end{equation}
The square of this expression, when averaged over an ensemble of
particles and waves, can be approximated by (see Appendix \ref{ap:A})
\begin{equation}
\label{eq:dz_GM}
\left< \delta z_{wav}^{2} \right>(t) = \left\{ \begin{array}{lcc}
U_{z}^{2} t^{2} &   \textrm{if}  & t \le N^{-1} , \\
\\ Q(t) & \textrm{if} & N^{-1} < t < 4 N^{-1} , \\
\\ \dfrac{4 U_{z}^{2}}{N^2} & \textrm{if} & t \geq
4 N^{-1} ,
\end{array}
\right.
\end{equation} 
where $Q(t)$ is a third order polynomial function obtained by
imposing $\left< \delta z_{wav}^{2} \right>$ and its time derivative
to be continuous in time (see Appendix \ref{ap:A}). Figure
\ref{fig:dz2_GM} 
shows the mean dispersion for many particles calculated from
a stochastic superposition of waves as in Eq.~(\ref{eq:displacement}),
and from the function in Eq.~(\ref{eq:dz_GM}), in both cases using
values of $A_0$ and $N$ that adjust the vertical r.m.s.~velocity and
the Brunt-V\"{a}is\"{a}l\"{a} frequency of several TG runs. The
function in Eq.~(\ref{eq:dz_GM}) is in good agreement with the sum of
random waves, specially for short and long times. Note also that this
simple model based on a superposition of waves captures the early-time
ballistic behavior of  $\delta z^{2} \sim t^2$ seen for all DNSs in
Fig.~\ref{f:DZ2}, as well as the saturation at late times seen in
Fig.~\ref{f:DZ2} for some of the simulations.

\begin{figure}
\centering
\includegraphics[width=8cm]{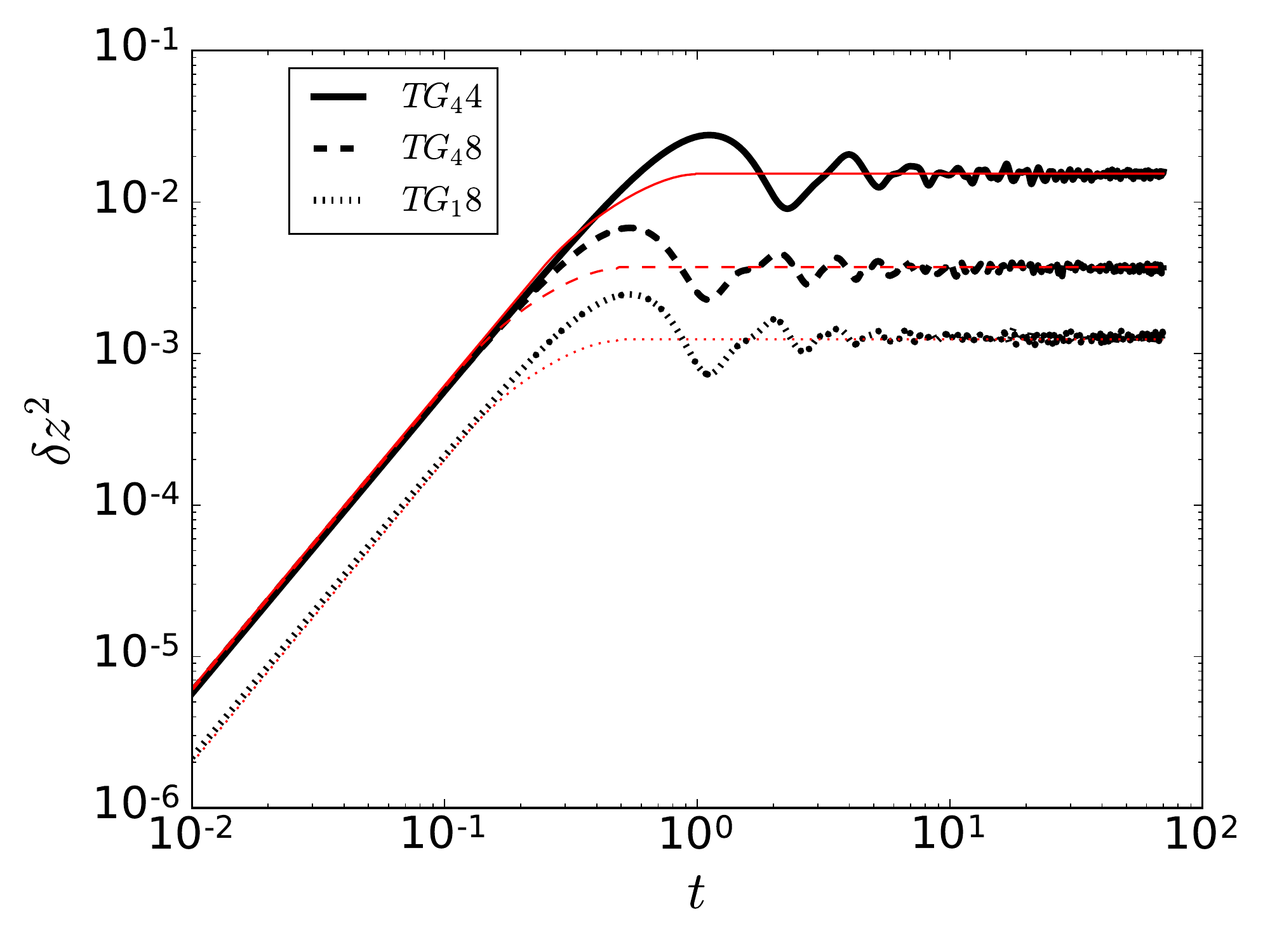}
\caption{({\it Color online}) Mean dispersion obtained from a random
superposition of waves as in Eq.~(\ref{eq:displacement}) (in thick
black lines), and from Eq.~(\ref{eq:dz_GM}) (in thin red lines),
with the parameter $A_0$ to match the Eulerian vertical 
r.m.s.~velocity and $N$ the Brunt-V\"{a}is\"{a}l\"{a} frequency
for some of the TG runs (see labels in the inset).}
\label{fig:dz2_GM}
\end{figure}

\begin{figure}
\centering
\includegraphics[width=8cm]{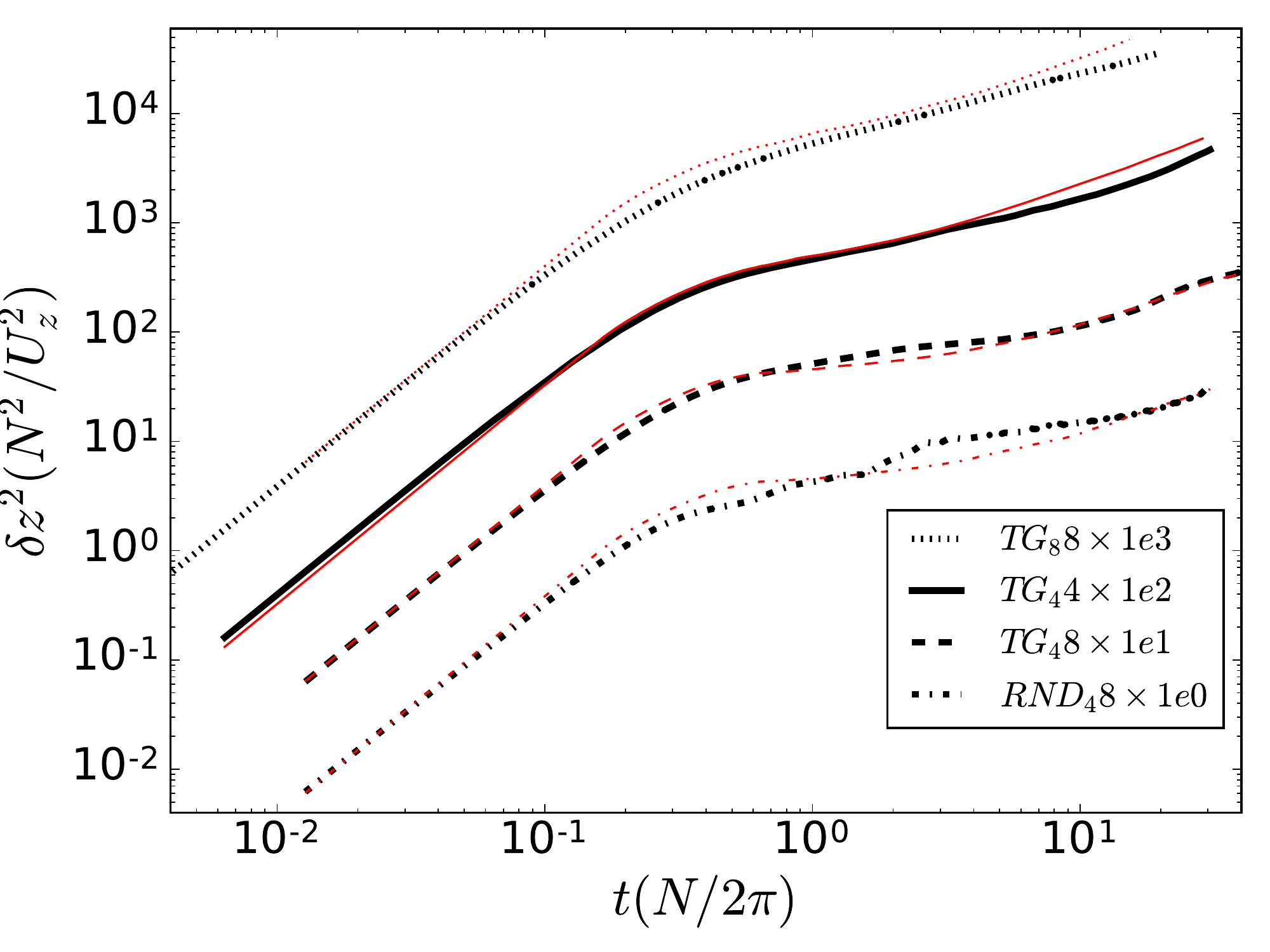}
\includegraphics[width=8cm]{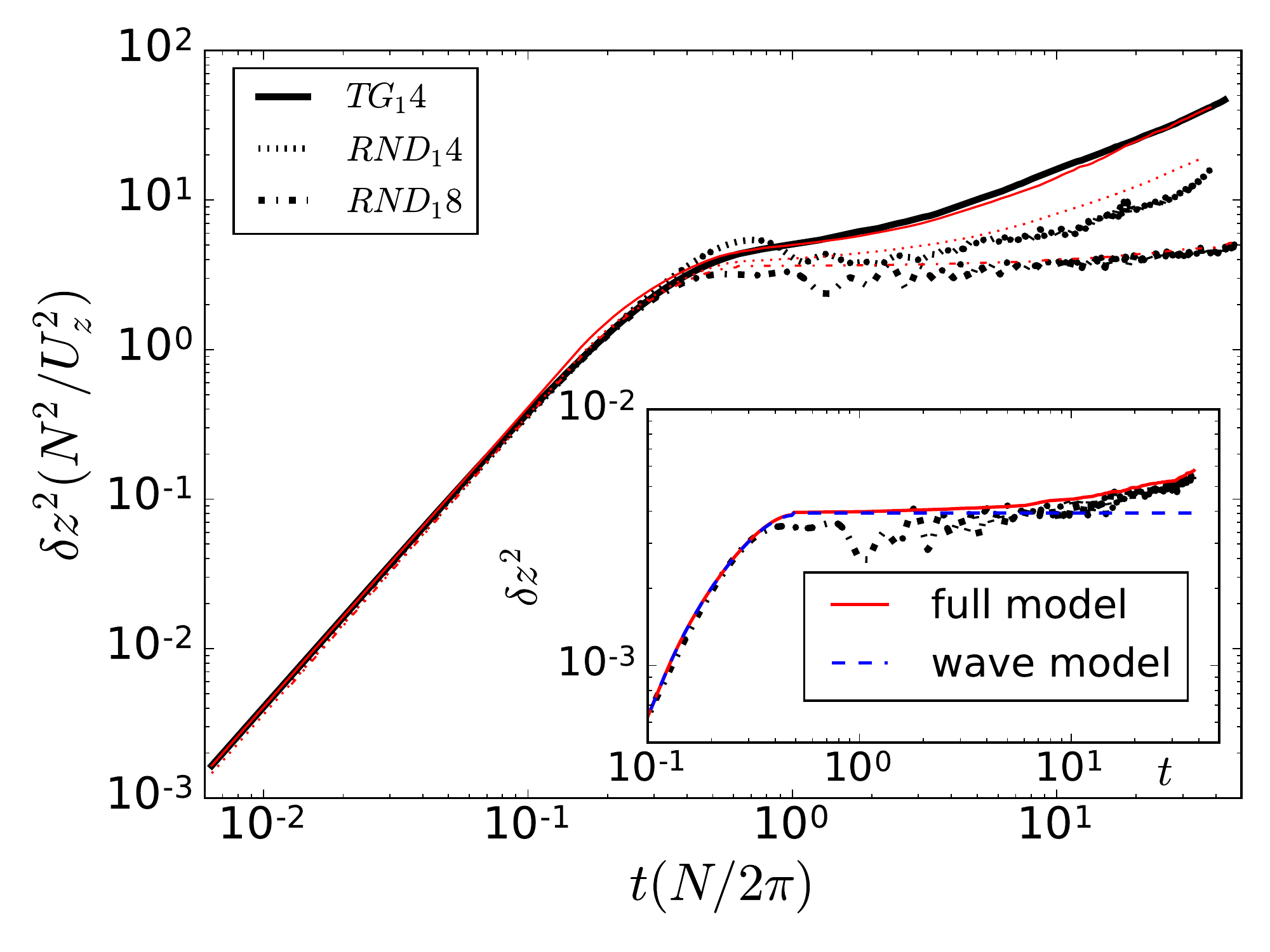}
\caption{({\it Color online}) Mean vertical dispersion $\delta z^2$
for ({\it left}) runs in elongated domains with amplitudes rescaled
for better visualization (see the inset), and ({\it right}) runs in
cubic domains, for RND and TG forcing, and with different values
of $N$. Normalizations of $\delta z^2$ and of time are the same as
in Fig.~\ref{f:DZ2}. In both figures the thick (black) lines show the
results from the DNSs, and the thin (red) lines show the results
obtained from the single-particle dispersion model. \AD{The inset
  on the right shows a detail of $\delta z^2$ for run RND$_18$ (thick
  dash-dotted curve), together with the mean squared vertical
  displacement predicted by the full model, and by the superposition
  of random waves alone.}}
\label{f:DZ2_MOD}
\end{figure}

The behavior shown in Fig.~\ref{fig:dz2_GM} is similar to the
vertical dispersion predicted for SST by other models based on a
linear superposition of waves \cite{nicolleau_turbulent_2000}, and is
reminiscent of the vertical dispersion observed in previous DNSs of
SST at moderate values of $\textrm{Rb}$
\cite{aartrijk_single-particle_2008,
sujovolsky_single-particle_2017}. However, this wave model fails 
to reproduce the dispersion observed at long times in some of our
runs. To introduce an enhanced vertical dispersion by turbulent
overturning, we add a CTRW model that mimics the trapping of particles
by eddies, resulting in vertical displacements when the flow becomes
unstable such that the total vertical dispersion will be
$\delta z(t) = \delta z_{wav}(t) + \delta z_{CTRW}(t)$.

To compute $\delta z_{CTRW}(t)$, in each step $t$ of the random walk
process we assume a particle has a probability $R$ of being trapped 
for a time $t_{t}$ inside an eddy of radius $r_{t}$ with velocity
$U_{t}$. As in the previous section, $R$ is the probability of finding
particles with $\textrm{Ri}_{g}<0$. \AD{Whether at a given step $t$
  the particle is trapped or not is a binary decision, and thus the
  probability of the particle not being trapped is $1-R$ (in which
  case the particle does not move following eddies).} When trapped,
the probability of being advected by an eddy of radius $r_{t}$ is
given by a Kolmogorov distribution $P(r_{t}) \sim r_{t}^{4/3}$ for 
$r < L_{oz}$, compatible with an isotropic energy spectrum 
$\sim k^{-5/3}$ for wavenumbers $k > k_{oz}$; in other words, we
assume that the eddies responsible for the vertical dispersion at late
times are associated with overturning instabilities in the turbulent
inertial range for scales 
equal to or smaller than the Ozmidov scale. The distribution of
trapping times $P(t_{t})$ is continuous and uniform between $t_{t}=0$
and the Eulerian turnover time at the Ozmidov scale
$\tau_{oz}$. Finally, \AD{at each step $t$ and if the particle is 
  trapped, the particle velocity (or equivalently, the characteristic
  velocity of the eddy trapping the particle) is $U_{t}$, given by a
  probability distribution $P(U_t)$ that is obtained from} the
observed PDF of the absolute value of the Eulerian vertical velocity
(which, in practice, can be very well approximated by assuming that it
follows a Rayleigh distribution, so knowledge of the r.m.s.~value of
$u_{z}$, $U_z$, is sufficient to estimate $P(U_t)$).

\AD{As mentioned above,} in each step of the CTRW if a particle is not
trapped by an eddy, $\delta z_{CTRW}(t)$ will remain constant (i.e.,
the particle will not move as a result of eddy trapping). If it gets
trapped, it will be displaced along a circle as the result of the
trapping, with a vertical displacement of $r_{t} \sin(\theta_{t})$
which is just the projection of the circular trajectory of radius
$r_{t}$ in the $z$  direction, and where  $\theta_{t} =
U_{t}t_{r}/r_{t} $ is the angle of the arc traveled during the time
$t_{r}$. Thus, the random walk process just mimics in a simplified way
the eventual presence of vertical eddies that can result in upward or
downward transport of the Lagrangian particles. As described above the
model has no free parameters; all parameters are obtained from
Eulerian characteristic lengths and time scales of the flow. 

\begin{figure}
\centering
\includegraphics[width=4.4cm]{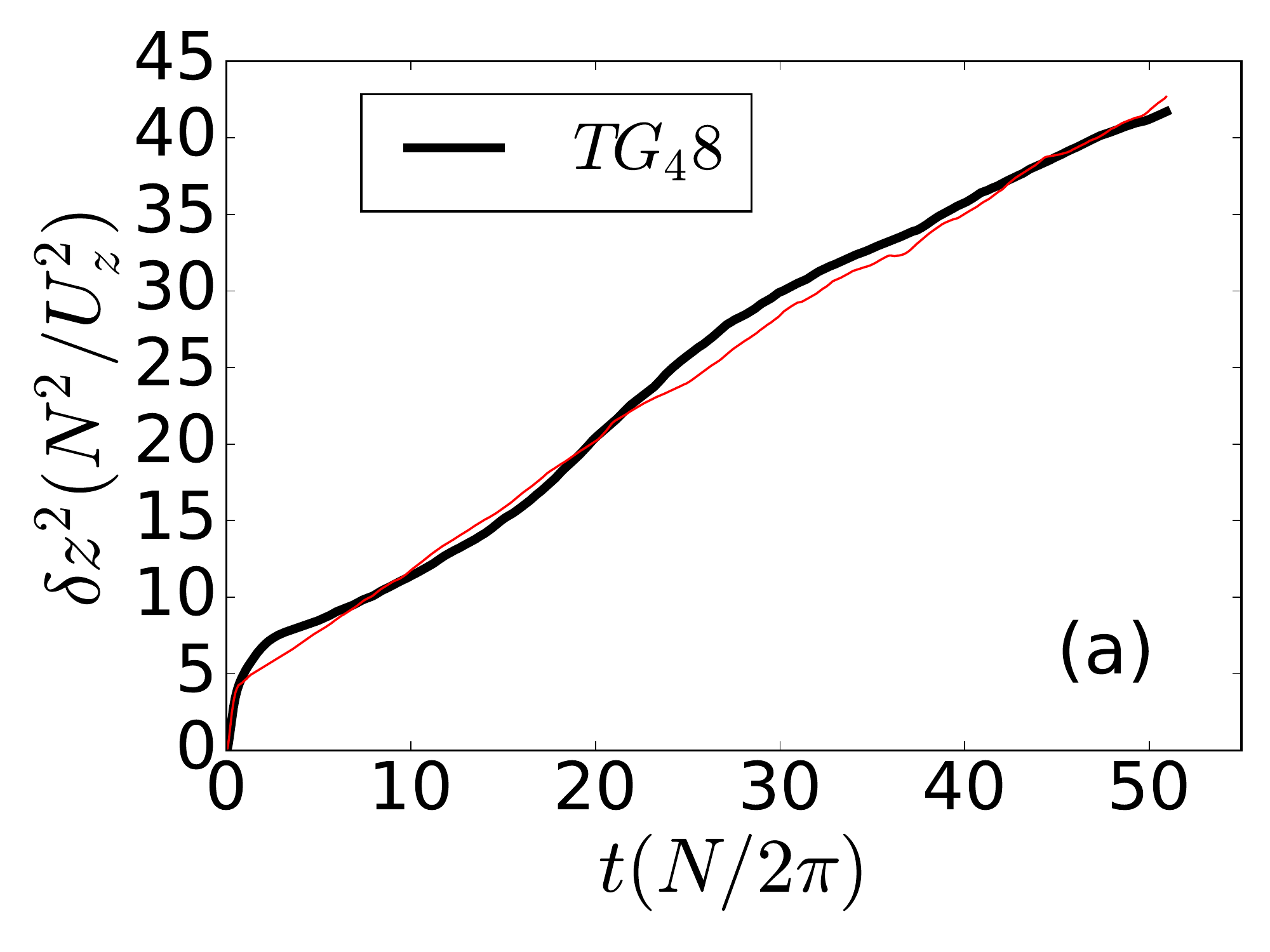}   
\includegraphics[width=4.4cm]{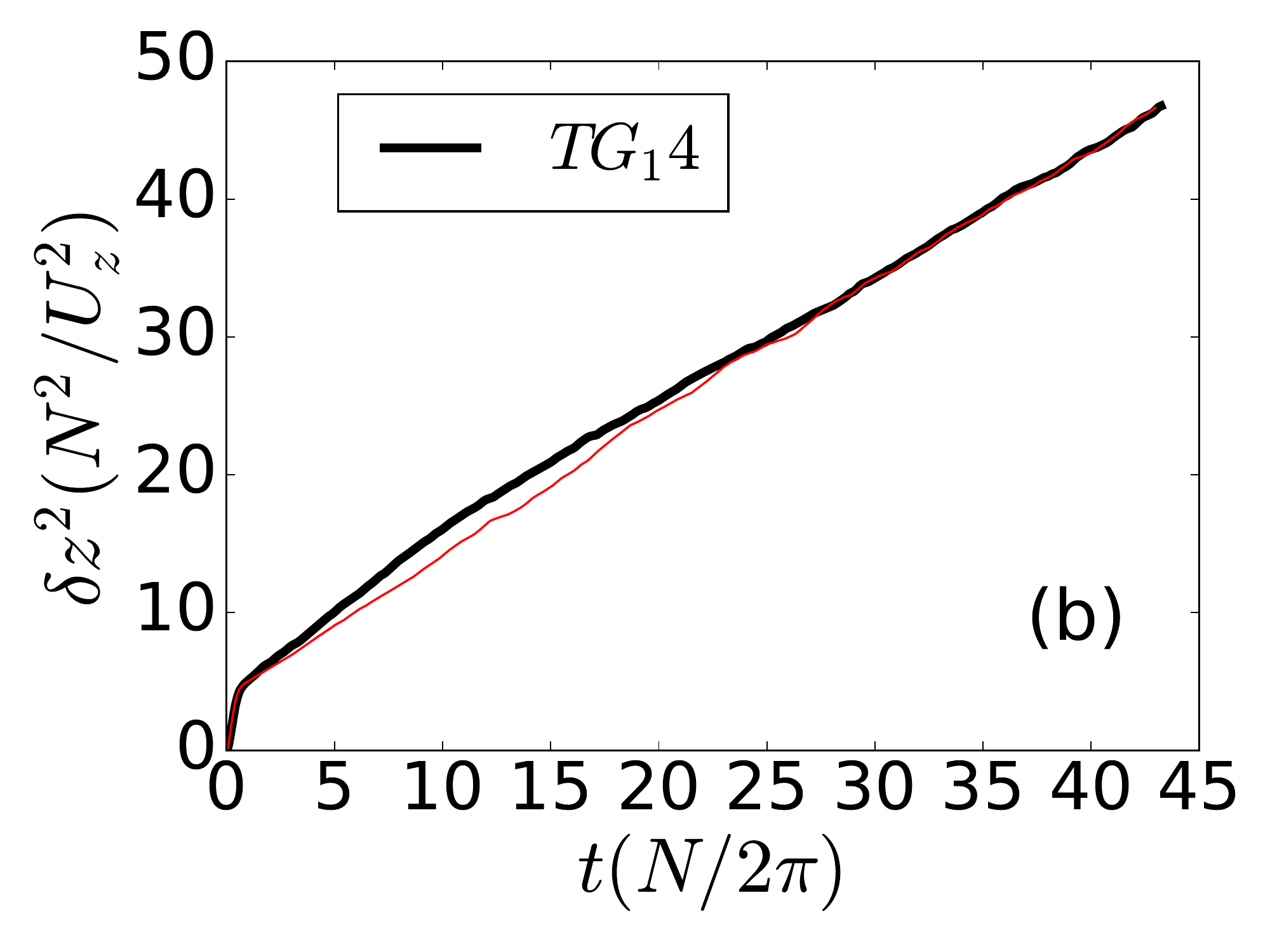}   
\includegraphics[width=4.4cm]{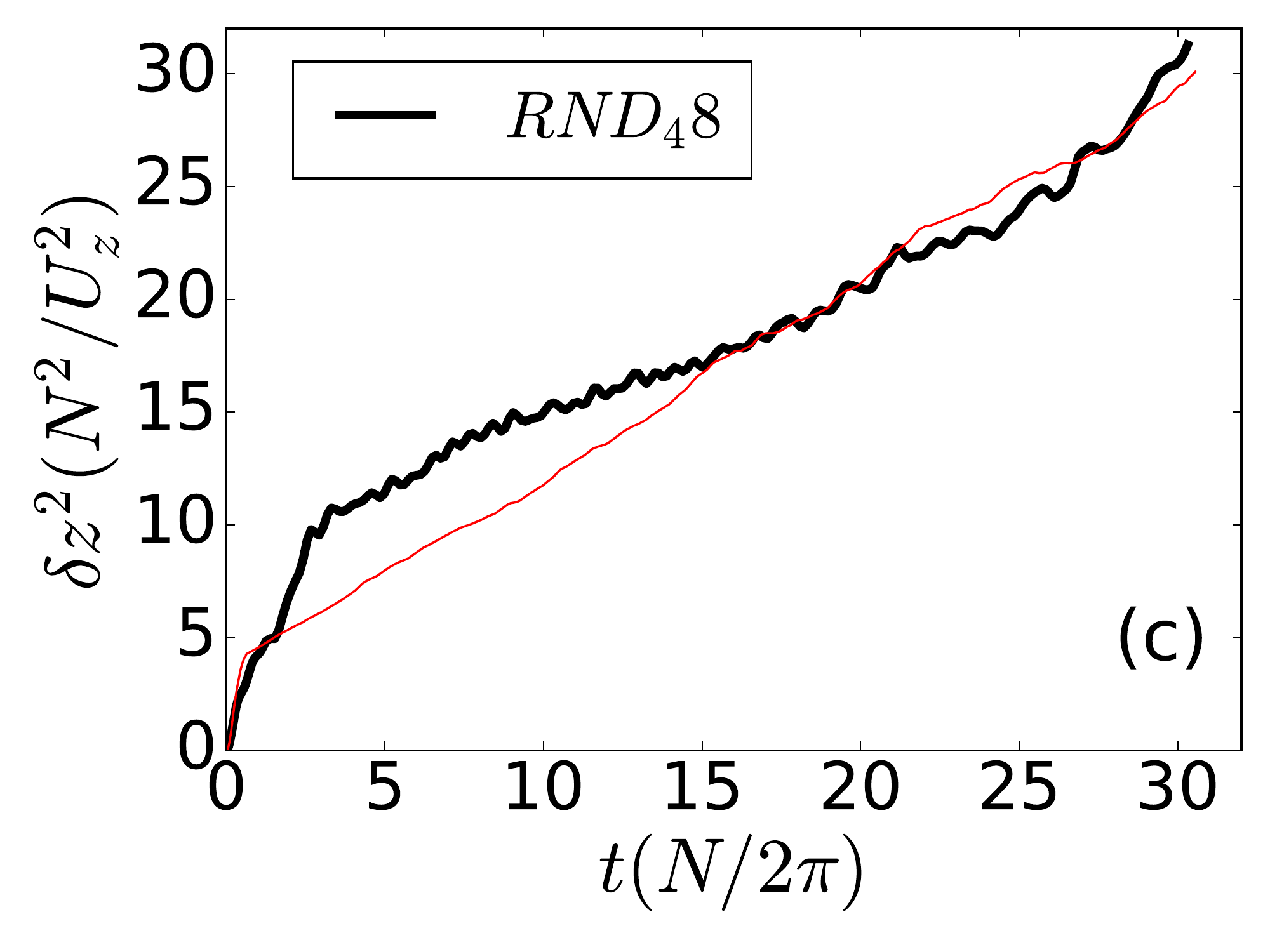}   
\includegraphics[width=4.4cm]{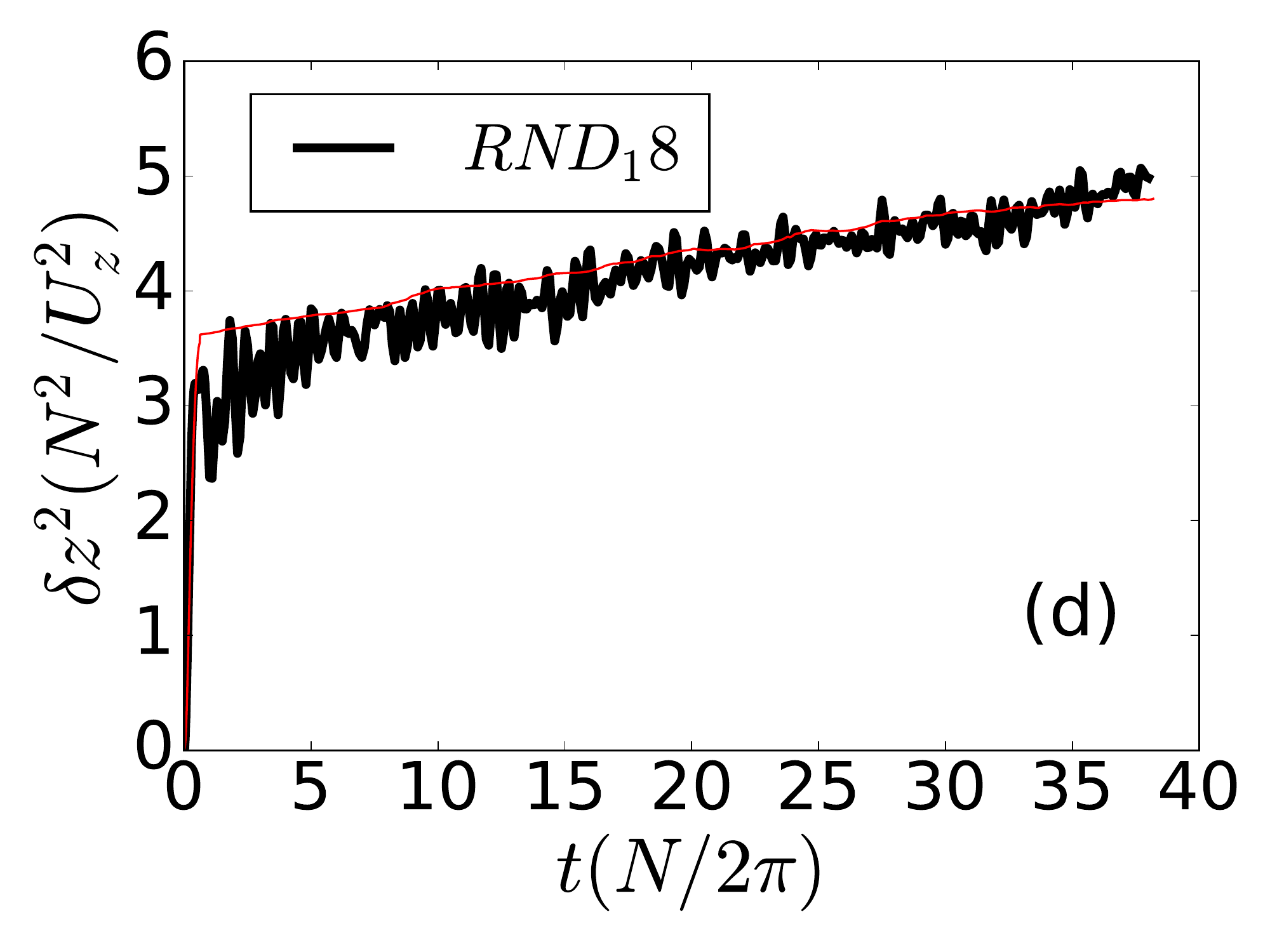}   
\caption{\AD{({\it Color online}) Mean squared vertical displacement
  $\delta z^2$ in linear scale, for four cases: (a) TG$_{4}8$,
  (b) TG$_{1}4$, (c) RND$_{4}8$, and (d) RND$_{1}8$. The four
  cases are illustrative of runs with TG or RND forcing, with
  different levels of stratification, and with different domain aspect
  ratios. The thick (black) curves show the results from the DNSs,
  while the thin (red) lines show the results obtained from the
  model. In spite of the randomness in the DNSs and in the CTRW model,
  all cases display a reasonable agreement.}}
\label{f:DZ2_MOD_LIN}
\end{figure}

\begin{figure}
\centering
\includegraphics[width=8cm]{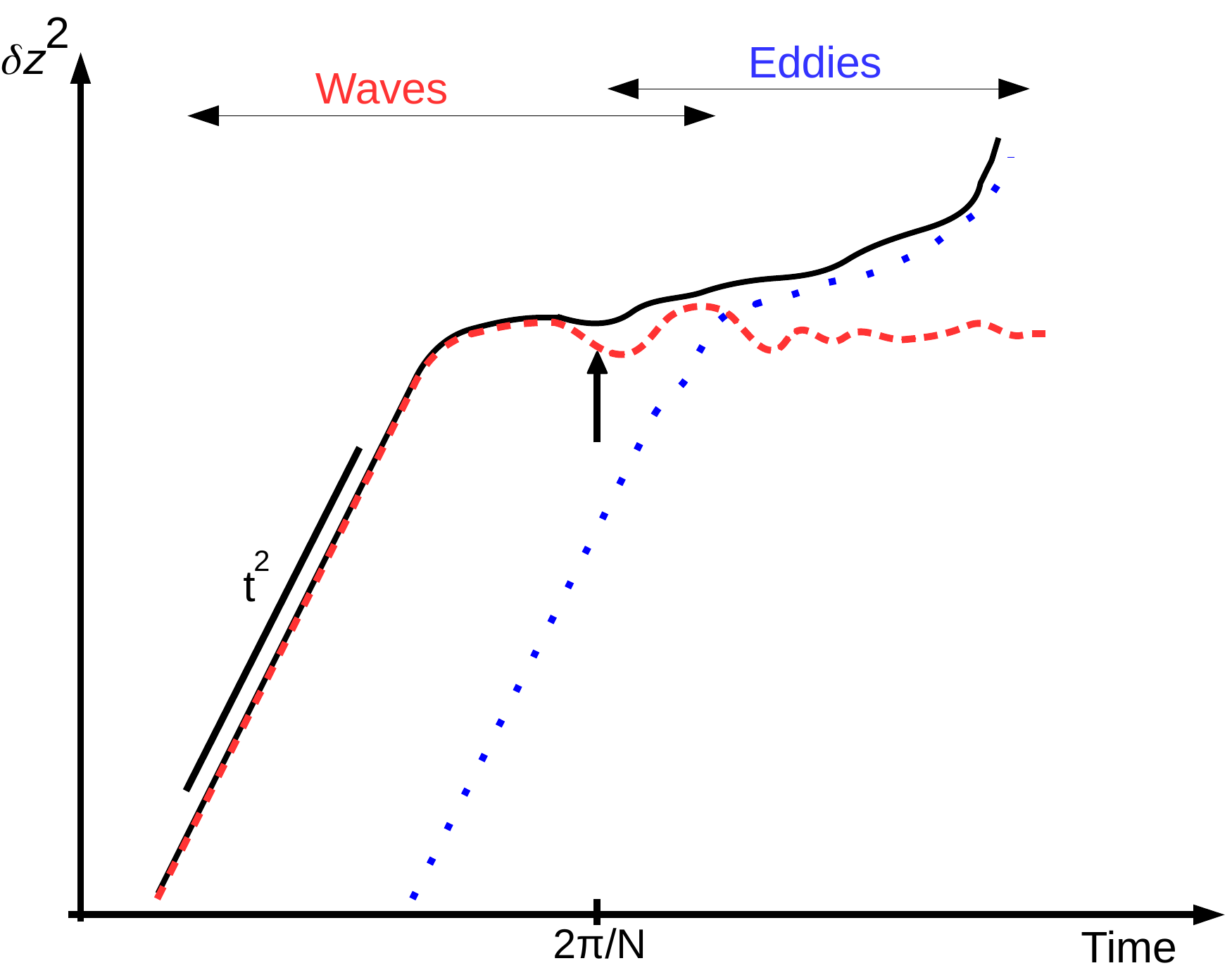}
\includegraphics[width=8cm]{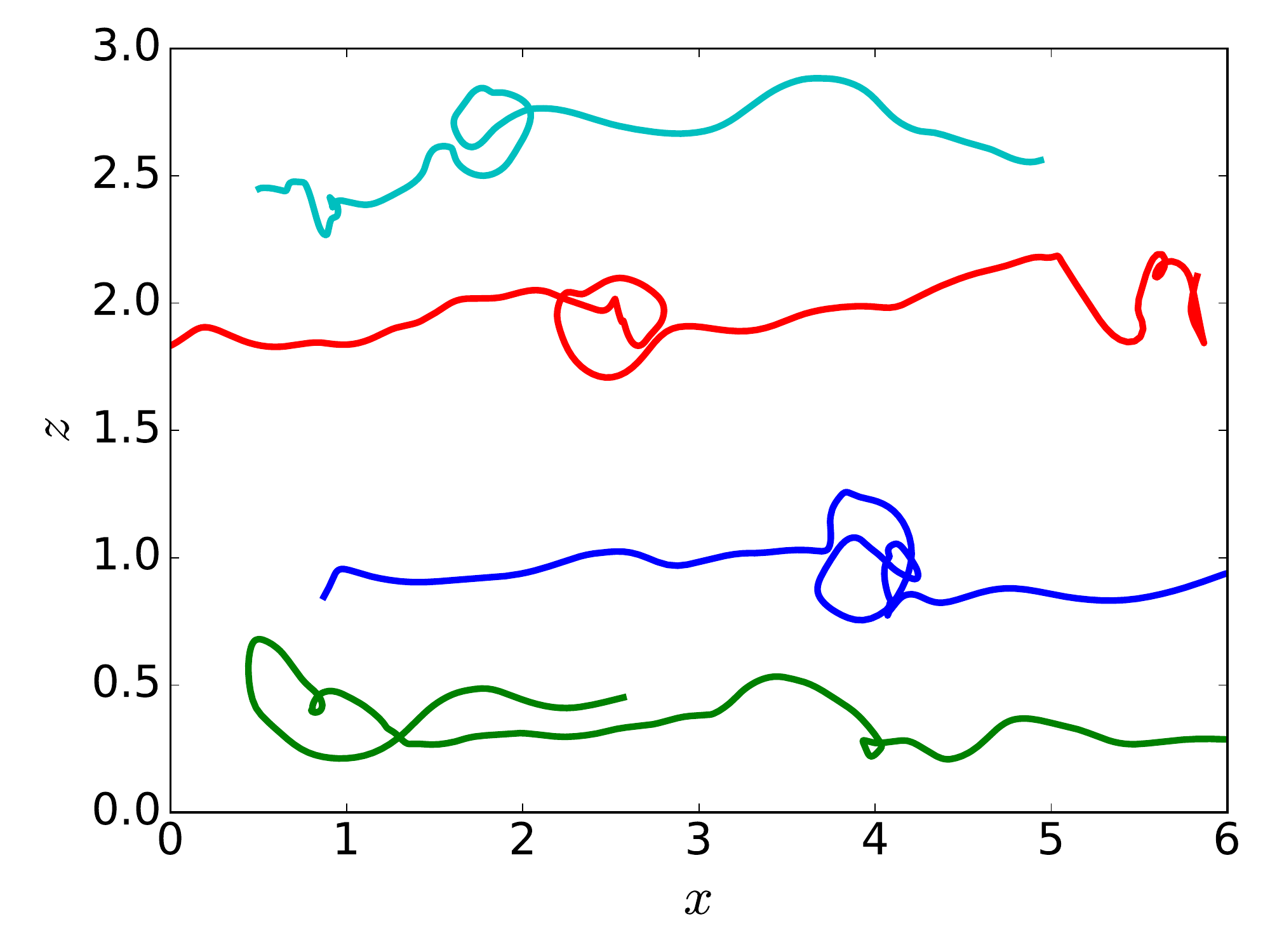}
\caption{({\it Color online}) {\it Left}: Sketch of the two
    contributions to vertical dispersion in the model. The random
    superposition of waves dominates at early times (red dashed
    curve), while the effect of turbulent eddies and overturning
    events become relevant at late times (blue dotted curve). The
    total vertical dispersion (black solid curve) results from the
    superposition of both effects. \AD{{\it Right}: A few particle
      trajectories from the TG$_{4}4$ run projected on the $x$-$z$
      plane. Note the small and wavy particles' displacements in the
      vertical directions, interrupted by sudden and close to circular
      trajectories that result in a larger vertical displacement,
      which we associate by trapping by overturning eddies (see
      Sec.~\ref{sec:MOD} for details).}}
\label{f:DZ2_EXP}
\end{figure}

\AD{We can summarize the computation of the entire model as follows:
\begin{enumerate}
\item Wave model: At any time, each particle vertical trajectory
  is computed as the sum of $N_{\omega}$ harmonic motions, each with
  amplitude $2U_{z}^{2}/(N_{w} \omega)$, and with uniformly
  distributed random frequencies $\omega$ between $N/2$ and $N$.
\item CTRW model:
	\begin{enumerate}
        \item  At each step $t$, a particle is trapped by and eddy
          with probability $R$ (or not trapped, with probability $1-R$).
 	\item If the particle is trapped:
		\begin{enumerate}
		\item The characteristic velocity of the eddy $U_{t}$
                  is given by $P(U_{t})$, a Rayleigh probability
                  distribution with its mean given by the mean
                  Eulerian vertical velocity of the flow.
		\item The eddy radius $r_{t}$ is chosen with a
                  probability distribution $P(r_{t})\sim r^{4/3}$
                  for $r_t\leq L_{oz}$, $P(r_{t}) = 0$ in other case.
		\item The trapping time $t_{t}$ is chosen from a
                  uniform probability distribution $P(t_{t})$ between
                  0 and $\tau_{oz}$.
		\item The central angle subtended by the arc traveled
                  by the particle during the trapping time is computed
                  as $\theta_{t}=U_{t} t_{t} / r_{t}$.
		\item The vertical displacement after the time $t_{t}$
                  is finally computed as $r_{t} \sin(\theta_{t})$.
		\end{enumerate}
	\item If the particle is not trapped, the particle is not
          displaced.
	\item The final vertical displacement resulting from the CTRW
          for any given particle and at a given time $t=\sum_{t=1}^{n}
          t_{t}$, is the sum over the $n$ vertical displacements.
	\end{enumerate}
\item Full model: At any given time, the total vertical displacement
  is obtained as the sum of the displacements generated by the waves
  and by the CTRW process.
\end{enumerate}
The key parameters of the full model then are $N$, $U_{z}$, $L_{oz}$,
and $R$, from which all other variables and probability distributions
(as well as the total displacements) can be computed.}

Figure \ref{f:DZ2_MOD} shows the mean vertical dispersion
obtained from several DNSs, and 
$\delta z^2(t) = [\delta z_{wav}(t) + \delta z_{CTRW}(t)]^2$ as
obtained from the wave and CTRW model (i.e., the full model). For the
runs in elongated domains (with TG forcing, or larger values of
$\textrm{Rb}$), as the dispersion is very similar for all runs, we
rescaled $\delta z^2$ using an arbitrary value (indicated in the
figure inset), so the curves can be distinguished more easily. In the
other cases, the same normalization as in Fig.~\ref{f:DZ2} was used
for $\delta z^2$ and time. The model is in good agreement with the DNS
data in all cases, and captures early and late time behavior
independently of the forcing function (TG or RND), as well as cases
with saturation of $\delta z^2(t)$ for $t>2\pi/N$ as cases in which
$\delta z^2(t)$ keeps growing at late times. \AD{The inset in
  Fig.~\ref{f:DZ2_MOD} shows a detail of the mean vertical
  dispersion for run $\textrm{RND}_{1}8$ (RND forcing with $N=8$), for
  which $\delta z^2$ almost completely saturates after $t(N/2\pi)
  \approx 1$, and grows only very slowly at late times. For this case,
  the inset also shows $\delta z^2$ obtained from the wave model alone
  (i.e., $\delta z_{wav}^2$), as well as $\delta z^2$ obtained from
  the full model). This case confirms that while the wave model can
  capture the saturation, the departure from this saturation and the
  growth observed at late times can only be captured if trapping by
  eddies is taken into account.}

\AD{As mentioned in Sec.~\ref{sec:SIN}, the aparent power laws
  observed at intermediate times in Fig.~\ref{f:DZ2} are the result of
  this competition between dispersion by waves and eddies, and for
  sufficiently long times $\delta z^2$ approaches a linear $\sim t$
  behavior if overturning is strong enough. To illustrate this, and to
  show the agreement between the model and the DNSs in more detail,
  Fig.~\ref{f:DZ2_MOD_LIN} shows $\delta z^2(t)$ in linear scale for
  four selected runs (corresponding to cases with TG or RND forcing,
  with different Brunt-V\"{a}is\"{a}l\"{a} frequencies, and with
  different domain aspect rations). Considering the inherent
  randomness of the DNSs results and of the CTRW process, a reasonable
  agreement is seen in all cases.}

The differences between the early and late time behavior can now be
further clarified using the model (see Fig.~\ref{f:DZ2_EXP}). At early
times, the waves dominate the dispersion resulting in the observed
ballistic regime up to the period of the slowest waves, $t \lesssim
2\pi/N$, for which the largest ``fast'' displacements (on the time
scale of the waves) can take place. If considered alone, trapping by
turbulent eddies in the CTRW model would also result in ballistic
growth of $\delta z^2$, but it has an initial value significantly
smaller, and as a result this process is subdominant to the dispersion
by the waves. At intermediate times ($t \approx 2\pi/N$) dispersion by
the waves saturates generating the plateau. If turbulence is moderate
(and thus $R$ is also moderate), trapping by eddies is inefficient,
resulting in a temporary saturation of the dispersion, or, in the most
extreme cases, in a complete saturation of $\delta z^2$. Depending on
how strong the turbulence is, at a certain time overturning events can
start enhancing the dispersion, and for longer times the turbulence
dominates the dynamic surpassing the effect of the waves. 
\AD{Remarkably, this simple picture is also compatible with the
  trajectories of individual Lagrangian particles. Figure
  \ref{f:DZ2_EXP} also shows as an example four trajectories
  projected into the $x$-$z$ plane, for four Eulerian eddy turn-over,
  and for run $\textrm{TG}_{4}4$. Small and wavy displacements can be
  seen in the vertical direction, interrupted by sudden and close to
  circular trajectories associated to trapping by overturning eddies,
  and which result in larger vertical displacements.}

\begin{figure}
\centering
\includegraphics[width=8cm]{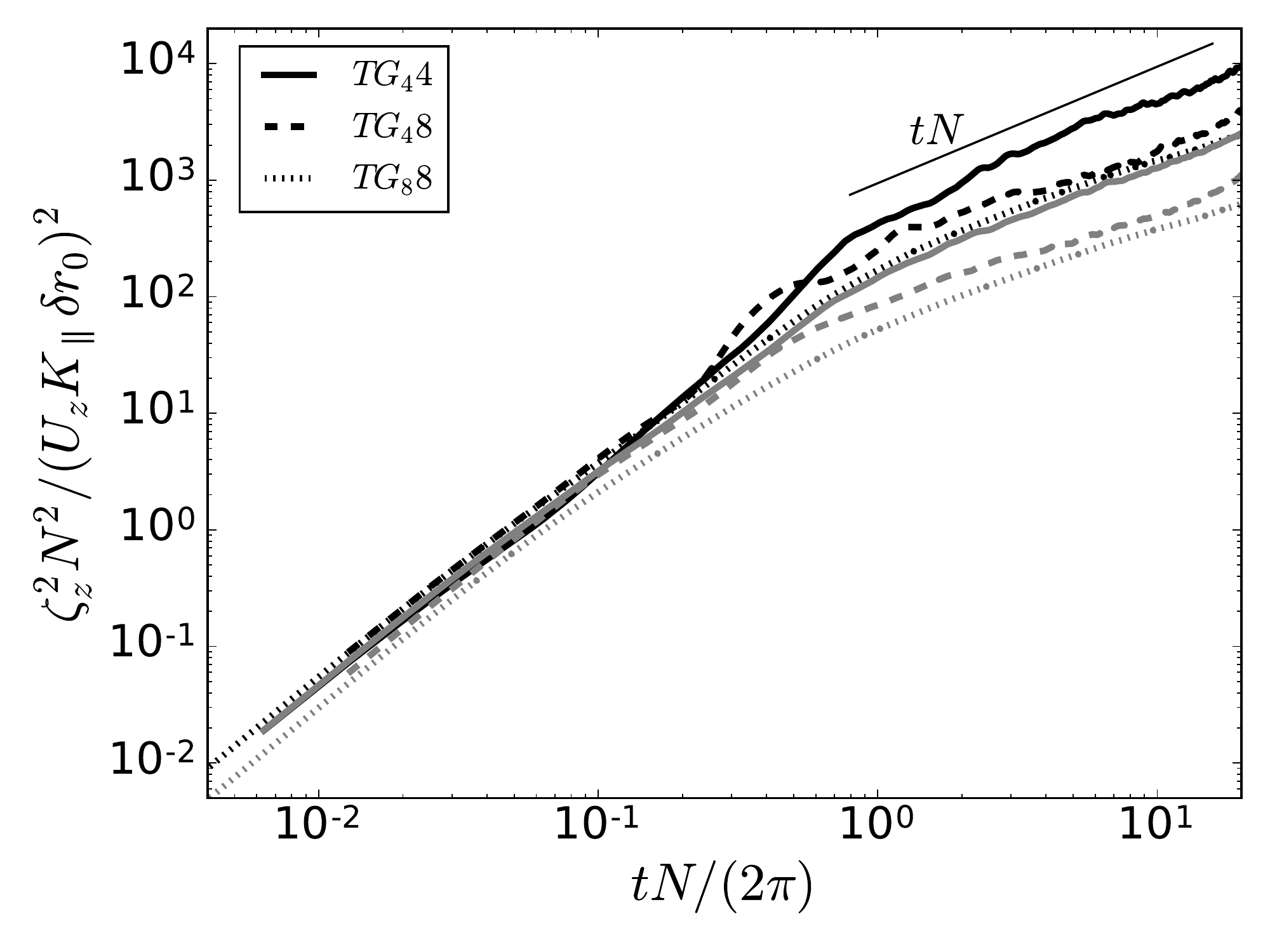}   
\includegraphics[width=8cm]{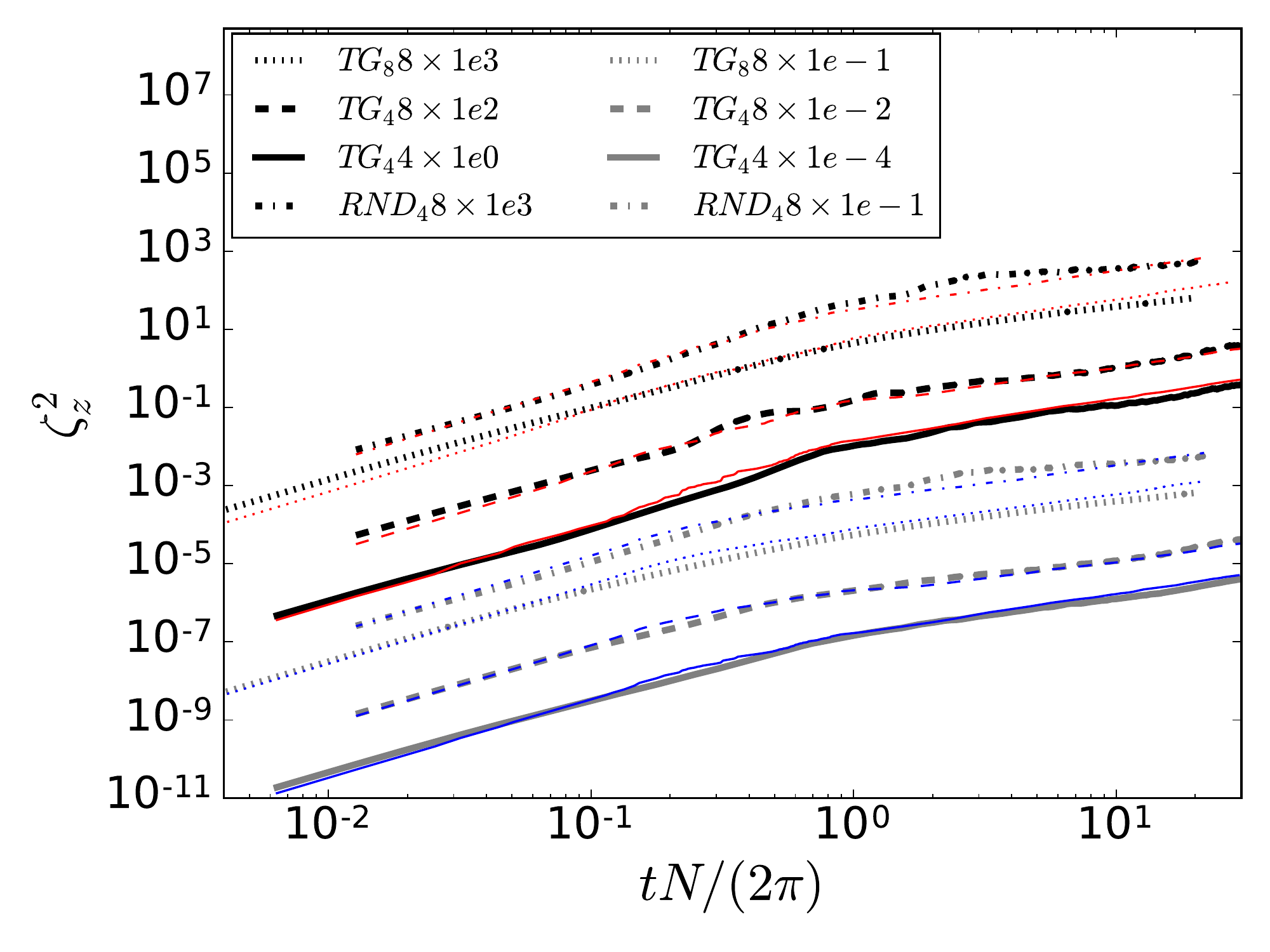}
\caption{({\it Color online}). {\it Left}: two-particle vertical
dispersion for particles with initial vertical separation 
$\delta z_0 \ll \eta$, and two initial horizontal separations: 
$\delta r_0 = \eta$ (black lines) $\delta r_0 = 2\eta$ (gray
lines). Results from three simulations with TG forcing are
shown. The vertical dispersion was normalized by 
$(w K_{\parallel} \delta r_0)^{2}/N^2$, and time by $1/N$. With this
normalization all curves collapse during the ballistic regime. A
power law at later times is indicated as a reference. {\it Right}:
two-particle vertical dispersion from DNSs with RND or TG forcing
(thick lines), and from the model (thin lines). The amplitude of the
curves have been rescaled for better visualization (see the
rescaling factor in the inset).}
\label{fig:D2P}
\end{figure}

\section{Two-particle vertical dispersion} \label{sec:TWO}

We can also study the two-particle vertical dispersion
$\zeta_z^2$, which the two simple processes presented above
(dispersion by a random superposition of waves, and a CTRW process to
capture the effect of turbulent overturning events) can also
model. The two-particle vertical dispersion is defined as 
$\zeta_z^2 = \langle[z_{i}(t)-z_{j}(t)]^{2}\rangle_{(i,j)}$, where 
$i \neq j$ are the labels of two particles that at the initial
time have a vertical separation $\delta z_{0}$ and a horizontal
separation $\delta r_{0}$, and where the subindex $(i,j)$ denotes that
the average is computed over pairs of particles. Figure \ref{fig:D2P}
shows the resulting two-particle vertical dispersion for runs TG$_44$,
TG$_48$, and TG$_88$. We consider pairs of particles which at time
$t=0$ have a vertical separation $\delta z_{0}\ll \eta$ (where $\eta$
is the Kolmogorov dissipation scale), and horizontal separations 
$\delta r_0 \approx \eta$ or $\delta r_0 \approx 2\eta$ (see
\cite{aartrijk_single-particle_2008} for a detailed study on different
choices of the initial separation in two-particle dispersion in
SST). Normalizing the vertical dispersion $\zeta_z^2$ by 
$(U_{z} K_{\parallel} \delta r_0)^2/N^2$ all curves approximately
collapse during the ballistic regime. As for the case of
single-particle dispersion, we see again a growth of the two-particle
dispersion at late times, which is linear or almost linear with $t$ in
all runs. Here we also see an effect of $\textrm{Rb}$: simulations
with larger $\textrm{Rb}$ display larger two-particle vertical
dispersions at late times. It is also worth pointing out that when
the initial horizontal separation $\delta r_0$  is increased (for a
given run), the short-time two-particle dispersion augments
proportionally, but the late-time two-particle dispersion remains
equal, indicating a decorrelation of the two particles at late times
as reported before in \cite{aartrijk_single-particle_2008} (note that
in Fig.~\ref{fig:D2P}, as $\zeta_z^2$ is normalized by 
$\delta r_{0}^{2}$, this late-time decorrelation results in different
amplitudes of $(\zeta_z/\delta r_{0})^{2}$ as $\delta r_{0}^{2}$ is
changed).

As mentioned above, the two-particle vertical dispersion can be
modeled using an extension of the single-particle model. As before, we
consider the effect of the waves and of the turbulent eddies
separately. First, if we have two particles which are initially very
close to each other (almost at the same height, but with a horizontal
displacement $\delta r_0$), we can assume they will be displaced by
the same waves but with a phase difference between the two (for each
wave with frequency $\omega$) given by
\begin{equation}
\phi'_{\omega} \approx \phi_{\omega} + k \, \delta r_0 .
\label{eq:phases}
\end{equation}
Here $\phi_{\omega}$ is the phase of the wave seen by one of the
particles, $\phi'_{\omega}$ is the phase seen by the other particle,
$k$ is the wavenumber, and we approximate the total separation between
the two particles by $\delta r_0$ (as $\delta z_0 \ll \eta$). Using the
expressions for the displacements of a particle in a superposition of
random waves given by Eqs.~(\ref{eq:traj}) and
(\ref{eq:displacement}), we can estimate the separation as a function
of time for a single pair as
\begin{equation}
\zeta_{ij,(wav)} (t) = \sum_{\omega} A_{\omega} \left\{  \left[
\cos(\omega t + \phi_{\omega}) - \cos(\phi_{\omega}) \right] -
\left[ \cos(\omega t + \phi'_{\omega}) - \cos(\phi'_{\omega})
\right]   \right\} + \delta z_{0},
\label{eq:dz2_2part}
\end{equation}
where the subindices $i$ and $j$ again label the pairs of particles
that at the initial time meet the condition $\delta r_0 \approx \eta$
(or $\approx 2\eta$). Equation (\ref{eq:dz2_2part}) is just the
difference between two single-particle vertical trajectories, to which
we added an initial vertical separation $\delta z_{0} \ll \eta$. As we
did in the previous section, the resulting dispersion, when averaged
over several pairs of particles and sets of random waves, can be
approximated as (see Appendix \ref{ap:B})
\begin{equation}
\label{eq:dz_2part_GM}
\left< \zeta_{z,(wav)}^2 \right>(t) \approx \left\{ \begin{array}{lcc}
U_{z}^{2} \left( 2.1 \delta r_0\dfrac{2\pi}{L_{\parallel}} 
   \right)^{2} t^{2}
   & \textrm{if}  & t < N^{-1} , \\
\\ Q_2(t) &  \textrm{if} & N^{-1} < t < 4 N^{-1} , \\
\\ \dfrac{16 U_{z}^{2} (2.1 \delta
  r_0 \pi)^{2}}{ \left( L_{\parallel} N \right)^2} &
  \textrm{if} & t \geq 4 N^{-1} , 
\end{array}
\right.
\end{equation} 
where, as in the single-particle approximation in
Eq.~(\ref{eq:dz_GM}), $Q_2(t)$ is a third order polynomial
interpolation for the two-particle case obtained by imposing the
function and its derivative to be continuous in time, and we neglected
all terms in $\delta{{z}_{0}}$ and $\delta{{z}_{0}}^{2}$ as they are
small when compared with the leading order terms.

As at late times the particles separate significantly from each other,
to take into account the effect of overturning we can assume the two
particles are uncorrelated, and as a result we can just consider two
independent CTRW processes with the same properties as the one
described in the previous section for single-particles, one for each
particle in the pair. This is compatible with observations of
two-particle dispersion in DNSs of SST
\cite{aartrijk_single-particle_2008}, and with the results from DNSs
shown above, indicating the late time dispersion becomes independent
of the original separation $\delta r_0$. The final result of combining
$\zeta_{z,(wav)}$ with the CTRW processes is shown in
Fig.~\ref{fig:D2P}. The model is in good agreement with the
two-particle vertical dispersion obtained from the DNSs, both in the
ballistic regime as well as for long times when dispersion becomes
dominated by the turbulent eddies, for both forcing functions
considered, different domain aspect ratios, and different values of
$\textrm{Fr}$ and $\textrm{Rb}$.

\section{Conclusions \label{sec:conclusions}}

In this paper we studied single- and two-particle vertical dispersion
for Lagrangian trajectories in forced stably stratified turbulence,
using two different forcing functions (Taylor-Green and random
forcing), domains with different aspect ratios, and different Froude
and Reynolds numbers. Using direct numerical simulations we showed
that late-time saturation of single-particle vertical dispersion,
reported in previous studies of this problem, is obtained only for
moderate values of the buoyancy Reynolds number, and that for larger
values of $\textrm{Rb}$, or even for moderate $\textrm{Rb}$ in the
case of the Taylor-Green flow that develops a vertical circulation,
the saturation does not take place. Instead, $\delta z^2$ keeps
growing in time after the ballistic regime, albeit at a slower rate
than in homogeneous and isotropic turbulence. 

We showed that the gradient Richardson number plays an important role
in the strength of the vertical transport of Lagrangian tracers, as
overturning fluid elements with $\textrm{Ri}_{g}<0$ give an important
contribution to vertical displacement of Lagrangian particles. In
particular, regions of the flows with higher vertical velocity present
a higher probability of having particles with 
$\textrm{Ri}_{g}< 0$ and vice versa. The joint probability (or
restricted PDFs) between $\textrm{Ri}_{g}$ and the Lagrangian vertical
velocity, temperature fluctuations and gradients were studied,
confirming this correlation.

Based on these results, we derived models for single- and two-particle
vertical dispersion that consist of a superposition of random waves
(to capture the early time ballistic regime), and an eddy-constrained
continuous-time random walk process (to capture the effect 
of turbulent eddies and overturning instabilities in the flow
at late times). \AD{These models, together with the model for
  single-particle horizontal dispersion in
  \cite{sujovolsky_single-particle_2017}, provide a description for
  the anisotropic dispersion in both the vertical and horizontal
  directions of stably stratified turbulence.} The vertical dispersion
obtained from \AD{the model presented here} is in good agreement with
the vertical dispersions obtained from the direct numerical
simulations. This agreement strengthens the observation that the waves
dominate the dynamic of particles at short times, resulting in the
initial ballistic regime, while at intermediate times 
($t\approx 2\pi/N$) linear and non-linear effects coexist in the
dynamics, giving rise to a transient that can develop (or not) a
plateau on the dispersion depending on how strong (or weak) is the
effect of overturning events. At later times, and if turbulence is 
sufficiently strong (as measured by $\textrm{Rb}$, or equivalently, by
the probability of a fluid element to suffer overturning, 
$R=P(\textrm{Ri}_g<0)$), turbulence (modeled here by the
continuous-time random walk process) dominates.
The superposition of linear and turbulent contributions to the
dispersion in the model thus allows clarification of the relevant time
and length scales involved in the dynamics of Lagrangian tracers in
stratified turbulence. Finally, as all parameters in the model can be
obtained from large-scale Eulerian properties of the flow, the model
opens the door to modeling turbulent dispersion of tracers in
Eulerian simulations of stratified flows that do not resolve the
smallest scales in the flow, as usually is the case in the study of
atmospheric and oceanic flows.

\appendix
\section{Derivation of the single-particle dispersion
wave model \label{ap:A}}

We want to derive averaged expressions for the dispersion as a
function of time resulting from a random superposition of waves as
that given by Eq.~(\ref{eq:displacement}). For short times, from
\begin{equation}
\delta z_{wav}(t) = \sum_{\omega} A_{w} \left[ \cos(\omega t +
\phi_{\omega}) - \cos(\phi_{\omega}) \right],
\end{equation}
we can take the square, use the trigonometric identity 
$\cos(\omega t+\phi_{\omega}) = \cos(\omega t) \cos(\phi_{\omega})-
\sin(\omega t) \sin(\phi_{\omega})$, 
the Taylor expansions to first order 
$\sin(\omega t)\approx \omega t$ and 
$\cos(\omega t)\approx 1$, and Eq.~(\ref{eq:AW}) with
$A_0=(2U_z^2/N_\omega)^{1/2}$, to get
\begin{equation}
\delta z^{2}_{wav} \approx t^{2} \dfrac{2
U_{z}^{2} }{N_{\omega}}
\left[ \sum_{\omega} \sin(\phi_{\omega}) \right]
\left[ \sum_{\omega'} \sin(\phi_{\omega'}) \right] =
t^{2} \dfrac{2 U_{z}^{2} }{N_{\omega}} \left[
\sum_{\omega} \sin^{2}(\phi_{\omega}) + \sum_{\omega, \omega' \neq
\omega} \sin(\phi_{\omega})  \sin(\phi_{\omega'})  \right] .
\label{eq:DZO2}
\end{equation}
As the average over random phases $\phi$ uniformly distributed between
$0$ and $2\pi$ is
\begin{eqnarray}
\left< \sin(\phi) \right>_{\phi} = \left< \cos(\phi) \right>_{\phi}
&=&0, \label{eq:prop_sin1} \\
\left< \sin^{2}(\phi) \right>_{\phi} = \left< \cos^{2}(\phi)
\right>_{\phi} &=& 1/2,
\label{eq:prop_sin2}
\end{eqnarray}
for $\phi_\omega$ and $\phi_{\omega'}$ two independent stochastic
variables, for short times and after averaging over an ensemble of
particles with different sets of random waves, we then have
\begin{equation}
\left< \delta z^{2}_{wav} \right>(t) \approx t^{2} U_{z}^{2}.
\label{eq:DZO3}
\end{equation}

For long times
\begin{equation}
\delta z_{wav}^{2}(t) = \dfrac{2
U_{z}^{2} }{N_{\omega}} \left\{ \sum_{\omega}
\dfrac{1}{\omega}\left[ \cos(\omega t + \phi_{\omega}) -
\cos(\phi_{\omega}) \right] \right\} \left\{ \sum_{\omega'}
\dfrac{1}{\omega'}\left[ \cos(\omega' t + \phi_{\omega'}) -
\cos(\phi_{\omega'}) \right] \right\},
\label{eq:DZO_long}
\end{equation}
which can be rewritten as
\begin{equation}
\delta z_{wav}^{2}(t) = \dfrac{2
U_{z}^{2}}{N_{\omega}} \left\{ \sum_{\omega}
\dfrac{1}{\omega^{2}} \left[ \cos(\omega t + \phi_{\omega}) -
\cos(\phi_{\omega}) \right]^{2} + 
\sum_{\omega, \omega' \neq \omega} 
\dfrac{1}{\omega \omega'}\left[ \cos(\omega' t + \phi_{\omega'}) -
\cos(\phi_{\omega'}) \right) \left( \cos(\omega t + \phi_{\omega})
- \cos(\phi_{\omega}) \right] \right\}.
\label{eq:DZO_long2}
\end{equation}
As the time average over several wave periods results in 
$\left< \cos(\omega t + \phi_{\omega})^2 \right>_{t}=1/2$, 
$\left< \cos(\omega t + \phi_{\omega}) \right>_{t}=0$, 
and 
$\left< \cos(\omega t + \phi_{\omega}) \cos(\omega' t +
\phi_{\omega'}) \right>_{t}=0$,
using Eqs.~(\ref{eq:prop_sin1}) and (\ref{eq:prop_sin2}) we obtain
the average of Eq.~(\ref{eq:DZO_long2}) over time and over an ensemble
of particles and random waves as
\begin{equation}
\label{eq:DZO_long3}
\left<\delta z_{wav}^{2} \right>_{t} \approx \dfrac{2
U_{z}^{2}}{N_{\omega}} \sum_{\omega=\omega_{min}}^{N}
\dfrac{1}{\omega^{2}}.
\end{equation}
Using $\Delta \omega = (N-\omega_{min})/N_{\omega}$, then
\begin{equation}
\label{eq:DZO_long4}
\sum_{\omega=\omega_{min}}^{N} \dfrac{1}{\omega^{2}} =
\sum_{\omega=\omega_{min}}^{N} \dfrac{1}{\omega^{2}} \dfrac{\Delta
\omega}{\Delta \omega} \approx \dfrac{1}{\Delta
\omega}\int_{\omega_{min}}^{N} \dfrac{1}{\omega^{2}} d\omega=
\dfrac{1}{\Delta \omega} \dfrac{N-\omega_{min}}{N
\omega_{min}}=\dfrac{N_{\omega}}{N\omega_{min}},
\end{equation} 
and finally, Eq.~(\ref{eq:DZO_long3}) can we rewritten as
\begin{equation}
\label{eq:DZO_long5}
\left<\delta z_{wav}^{2} \right>_{t} \approx \dfrac{2
U_{z}^{2}}{N \omega_{min}},
\end{equation} 
where we chose $\omega_{min} = N/2$.

This gives the early time ($t \le N^{-1}$) and late time 
($ t \ge 4N^{-1}$) expressions in Eq.~(\ref{eq:dz_GM}) (note the
choices of $N^{-1}$ and $4N^{-1}$ as the two limits for the validity
of the approximations are somewhat arbitrary). To obtain a smooth
(i.e., continuous in $\left<\delta z_{wav}^2\right>(t)$ and in its
time derivative) interpolation between these two expressions, we 
use a third order polynomial function to interpolate 
$\left<\delta z_{wav}^2 \right>(t)$  between $t_{a}=1/N$ and 
$t_{b}=4/N$. Writing a polynomial approximation
$\left<\delta z_{wav}^{2}\right>(t)=Q(t)=At^{3}+Bt^{2}+Ct+D$, then 
the coefficients after imposing the continuity conditions are 
\begin{eqnarray}
A &=&  \left[ Q'(t_{a}) (t_{a}-t_{b}) - 2 ( Q(t_{a}) - Q(t_{b}) )
\right] / (t_{a}-t_{b})^{3}, \\ 
B &=& -\left[ Q'(t_{a}) ( t_{a}^{2} + t_{a} t_{b} - 2 t_{b}^{2} ) +
3(t_{a}  + t_{b}) (Q(t_{b})-Q(t_{a}) ) \right]/(t_{a}-t_{b})^{3},
\\ 
C &=& t_{b} \left[ Q'(t_{a}) (2t_{a}^{2} - t_{a} t_{b} - t_{b}^{2} ) +
6t_{a} (Q(t_{b}) - Q(t_{a}))\right]/(t_{a}-t_{b})^{3}, \\
D &=& -\left[ t_{a} t_{b}^{2} Q'(t_{a}) (t_{a} -t_{b}) + t_{b}^{2}
Q(t_{a})(t_{b} -3 t_{a} ) + t_{a}^{2} Q(t_{b}) + (3t_{b}-t_{a})
\right] /(t_{a}-t_{b})^{3}, 
\end{eqnarray}
where the values $Q(t_{a})$ and $Q(t_{b})$ are given by the
expressions in Eqs.~(\ref{eq:dz_GM}) and (\ref{eq:dz_2part_GM})
evaluated at $t = N^{-1}$ or  $t = 4N^{-1}$. 

\section{Two-particle dispersion wave model \label{ap:B}}

To derive averaged expressions for the two-particle dispersion
resulting from a random superposition of waves, we start from 
Eq.~(\ref{eq:dz2_2part}),
\begin{equation}
\zeta_{ij,(wav)} (t) = z_{i,(wav)}(t)-z_{j,(wav)}(t) = \sum_{\omega} A_{\omega}
\left\{  \left[ \cos(\omega t + \phi_{\omega}) - \cos(\phi_{\omega})
\right] - \left[ \cos(\omega t + \phi'_{\omega}) -
\cos(\phi'_{\omega}) \right]   \right\} + \delta z_{0},
\label{eq:dz2_appendix}
\end{equation}
where $\phi'_\omega$ is the phase of the wave with frequency $\omega$
as seen by the particle $j$, which is displaced a distance 
$\approx \delta r_0$ (as  $\delta z_0 \ll \delta r_0$) from the
particle $i$ (with phase $\phi_\omega$). Thus, 
$\phi'_{\omega} \approx \phi_{\omega} + k \, \delta r_0 $ (note we
ignore any increase in time of the horizontal distance between the two
particles, and in the following we consider only the increase in the
vertical distance between them). We can approximate 
$k \approx K_{\parallel}/\cos \alpha$, where $\alpha$ is the angle of
propagation of the waves with respect to the vertical direction, and
$K_{\parallel}$ is the parallel integral wave number as introduced in
Sec.~\ref{sec:NSIM}. From the dispersion relation of internal gravity
waves we have 
$\omega = N \sin \alpha $ (or $\sin \alpha = \omega/N$), and using 
$\cos \alpha =(1-\sin^{2} \alpha)^{1/2}$ and that $\omega/N$ in the
single-particle model is a random variable uniformly distributed 
between 1/2 and 1, we can estimate the mean value of the wavenumber
$k$ for an ensemble of waves propagating in all available directions as 
\begin{equation}
\left<k\right> = \left< \dfrac{K_{\parallel}}{\cos \alpha}
\right>_{\omega / N} = 2K_{\parallel} \int_{1/2}^{1}
\dfrac{d(\omega / N)}{\sqrt{1-(\omega/N)^{2}}}  \approx 2.1
K_\parallel,
\label{eq:cos}
\end{equation}
where the factor $2$ multiplying the integral comes from computing the
mean of $\omega / N$ in the interval $[1/2,1]$. Thus, the mean phase
shift results to be 
$\left<\delta\phi\right>\approx 2.1K_\parallel \, \delta r_0$. 

The two-particle mean squared vertical displacement caused by the
waves $\langle \zeta_{z,(wav)}^2 \rangle$ is the average over an
ensemble of particle pairs of the square of the vertical two-particle
displacements for a single pair $\zeta_{ij,(wav)}$. For a very small
initial separation between the particle pairs $\delta \phi$ is also
small and we can use in Eq.~(\ref{eq:dz2_appendix}) the approximation 
$\cos(\omega t + \phi +\delta \phi) \approx \cos(\omega t + \phi) -
\delta \phi \sin(\omega t + \phi)$ to get
\begin{equation}
\zeta_{ij,(wav)} \approx \sum_{\omega} -k \delta r_0 A_{\omega} \left[
\sin(\omega t + \phi_{\omega}) - \sin(\phi_{\omega}) \right] +
\delta z_{0} .
\label{eq:dz2_2part_2}
\end{equation}
For short times we can take the first order Taylor approximations 
$\sin(\omega t)\approx -\omega t$ and 
$\cos(\omega t) \approx 1$. Also, using the trigonometrical identity
$\sin(\omega t + \phi_{\omega})=\sin(\omega t )\cos(\phi_{\omega}) +
\cos(\omega t)\sin(\phi_{\omega})$ we obtain
\begin{equation}
\zeta_{ij,(wav)} \approx t \delta r_0 \sum_{\omega} \left[ - \omega k
A_{\omega} \cos(\phi_{\omega}) \right] + \delta z_{0}.
\label{eq:dz2_2part_3}
\end{equation}
Finally, we take the ensemble average of the square of
$\zeta_{ij,(wav)}$, we use that $\alpha$ and $\omega$ are stochastic
variables, we use Eq.~(\ref{eq:AW}) for $A_{\omega}$, and we use
Eqs.~(\ref{eq:cos}), (\ref{eq:prop_sin1}), and (\ref{eq:prop_sin2}) to
get
\begin{equation}
\left< \zeta_{wav}^2 \right>(t) \approx U_{z}^{2} \left(2.1
K_{\parallel} \delta r_0 \right)^{2} t^{2},
\end{equation}
where terms in $\delta z_{0}$ and $\delta z_{0}^{2}$ are neglected
for being much smaller than the leading order term.

To obtain the long time approximation we start by neglecting the term
$\delta z_{0}$ and taking the mean squared value of
Eq.~(\ref{eq:dz2_appendix}), 
\begin{equation}
\begin{split}
\left< \zeta_{ij,(wav)}^2 \right> = \left< \sum_{\omega} A_{\omega}^{2}
\left\{  \left[ \cos(\omega t + \phi_{\omega}) - \cos(\phi_{\omega})
\right]^{2} + \left[ \cos(\omega t + \phi'_{\omega}) -
\cos(\phi'_{\omega}) \right]^{2} -
\right. \right. \\\left. \left. 2\left[ \cos(\omega t +
\phi_{\omega}) - \cos(\phi_{\omega}) \right] \left[ \cos(\omega
t + \phi'_{\omega}) - \cos(\phi'_{\omega}) \right] \right\} +
\sum_{\omega,\omega'\neq\omega} (...) \right>, 
\end{split}
\label{eq:dz2_2part_1}
\end{equation}
where the mean is taken both over time and over particle pairs. The
cross-product terms in Eq.~(\ref{eq:dz2_2part_1}) have mean value
$\langle \sum_{\omega,\omega'\neq\omega} (...) \rangle = 0$, as
discussed in Appendix \ref{ap:A}. Using again the approximation 
$\cos(\omega t+ \phi_{\omega} + \delta \phi) \approx \cos(\omega t +
\phi_{\omega})-\delta  \phi \sin(\omega t + \phi)$ we get  
\begin{equation}
\begin{split}
\left< \zeta_{ij,(wav)}^2 \right> \approx \left<\sum_{\omega}
A_{\omega}^{2} \left\{  \left[ \cos(\omega t + \phi_{\omega}) -
\cos(\phi_{\omega}) \right]^{2} + \left[ \cos(\omega t +
\phi_{\omega}) - \cos(\phi_{\omega}) - \delta \phi
\left(\sin(\omega t + \phi_{\omega}) - \sin(\phi) \right)
\right]^{2} - \right. \right. \\ \left. \left. 2\left[ \cos(\omega
t + \phi_{\omega}) - \cos(\phi_{\omega}) \right] \left[
\cos(\omega t + \phi_{\omega}) - \cos(\phi_{\omega}) - \delta
\phi \left(\sin(\omega t + \phi_{\omega}) - \sin(\phi) \right)
\right] \right\} \right>.
\end{split}
\label{eq:dz2_2part_2}
\end{equation}
Finally, using Eqs.~(\ref{eq:prop_sin1}), (\ref{eq:prop_sin2}),
(\ref{eq:DZO_long2}), (\ref{eq:DZO_long3}) and (\ref{eq:DZO_long5}),
and given that $\left< \sin(\phi) \cos(\phi) \right>_{\phi}=0$ with
$\phi$ uniformly distributed between $0$ and $2\pi$, we have 
\begin{equation}
\left< \zeta_{z,(wav)}^2 \right>_{t} \approx \left< (k \delta r_0)^{2}
\right> \left< \sum_{\omega} A_{\omega}^{2} \left[ \sin(\omega t +
\phi)- \sin(\phi) \right] \right>_{t} \approx \left< (k \delta
r_0)^{2}\right> \left< \delta z_{wav}^{2}
\right>_{t} \approx \dfrac{2
(2.1 K_{\parallel} \delta r_0)^{2} U_{z}^{2} }{N \omega_{min}}.
\label{eq:dz2_2part_3}
\end{equation}

\begin{acknowledgments}
The authors acknowledge support from PICT Grant
No.~2015-3530. 
\end{acknowledgments}
\bibliography{ms}

\begin{thebibliography}{41}%
\makeatletter
\providecommand \@ifxundefined [1]{%
 \@ifx{#1\undefined}
}%
\providecommand \@ifnum [1]{%
 \ifnum #1\expandafter \@firstoftwo
 \else \expandafter \@secondoftwo
 \fi
}%
\providecommand \@ifx [1]{%
 \ifx #1\expandafter \@firstoftwo
 \else \expandafter \@secondoftwo
 \fi
}%
\providecommand \natexlab [1]{#1}%
\providecommand \enquote  [1]{``#1''}%
\providecommand \bibnamefont  [1]{#1}%
\providecommand \bibfnamefont [1]{#1}%
\providecommand \citenamefont [1]{#1}%
\providecommand \href@noop [0]{\@secondoftwo}%
\providecommand \href [0]{\begingroup \@sanitize@url \@href}%
\providecommand \@href[1]{\@@startlink{#1}\@@href}%
\providecommand \@@href[1]{\endgroup#1\@@endlink}%
\providecommand \@sanitize@url [0]{\catcode `\\12\catcode `\$12\catcode
  `\&12\catcode `\#12\catcode `\^12\catcode `\_12\catcode `\%12\relax}%
\providecommand \@@startlink[1]{}%
\providecommand \@@endlink[0]{}%
\providecommand \url  [0]{\begingroup\@sanitize@url \@url }%
\providecommand \@url [1]{\endgroup\@href {#1}{\urlprefix }}%
\providecommand \urlprefix  [0]{URL }%
\providecommand \Eprint [0]{\href }%
\providecommand \doibase [0]{http://dx.doi.org/}%
\providecommand \selectlanguage [0]{\@gobble}%
\providecommand \bibinfo  [0]{\@secondoftwo}%
\providecommand \bibfield  [0]{\@secondoftwo}%
\providecommand \translation [1]{[#1]}%
\providecommand \BibitemOpen [0]{}%
\providecommand \bibitemStop [0]{}%
\providecommand \bibitemNoStop [0]{.\EOS\space}%
\providecommand \EOS [0]{\spacefactor3000\relax}%
\providecommand \BibitemShut  [1]{\csname bibitem#1\endcsname}%
\let\auto@bib@innerbib\@empty
\bibitem [{\citenamefont {Wyngaard}(1992)}]{wyngaard_atmospheric_1992}%
  \BibitemOpen
  \bibfield  {author} {\bibinfo {author} {\bibfnamefont {J.~C.}\ \bibnamefont
  {Wyngaard}},\ }\bibfield  {title} {\enquote {\bibinfo {title} {Atmospheric
  {Turbulence}},}\ }\href {\doibase 10.1146/annurev.fl.24.010192.001225}
  {\bibfield  {journal} {\bibinfo  {journal} {Annu. Rev. Fluid Mech.}\ }\textbf
  {\bibinfo {volume} {24}},\ \bibinfo {pages} {205} (\bibinfo {year}
  {1992})}\BibitemShut {NoStop}%
\bibitem [{\citenamefont {D'Asaro}\ and\ \citenamefont
  {Lien}(2000)}]{dasaro_lagrangian_2000}%
  \BibitemOpen
  \bibfield  {author} {\bibinfo {author} {\bibfnamefont {E.~A.}\ \bibnamefont
  {D'Asaro}}\ and\ \bibinfo {author} {\bibfnamefont {R.-C.}\ \bibnamefont
  {Lien}},\ }\bibfield  {title} {\enquote {\bibinfo {title} {Lagrangian
  {Measurements} of {Waves} and {Turbulence} in {Stratified} {Flows}},}\ }\href
  {\doibase 10.1175/1520-0485(2000)030<0641:LMOWAT>2.0.CO;2} {\bibfield
  {journal} {\bibinfo  {journal} {J. Phys. Oceanogr.}\ }\textbf {\bibinfo
  {volume} {30}},\ \bibinfo {pages} {641} (\bibinfo {year} {2000})}\BibitemShut
  {NoStop}%
\bibitem [{\citenamefont {Watanabe}\ \emph {et~al.}(2017)\citenamefont
  {Watanabe}, \citenamefont {Riley},\ and\ \citenamefont
  {Nagata}}]{watanabe_2017}%
  \BibitemOpen
  \bibfield  {author} {\bibinfo {author} {\bibfnamefont {T.}~\bibnamefont
  {Watanabe}}, \bibinfo {author} {\bibfnamefont {J.J.}\ \bibnamefont {Riley}},
  \ and\ \bibinfo {author} {\bibfnamefont {K.}~\bibnamefont {Nagata}},\
  }\bibfield  {title} {\enquote {\bibinfo {title} {Turbulent entrainment across
  turbulent-nonturbulent interfaces in stably stratified mixing layers},}\
  }\href@noop {} {\bibfield  {journal} {\bibinfo  {journal} {Phys. Rev.
  Fluids}\ }\textbf {\bibinfo {volume} {2}},\ \bibinfo {pages} {104803}
  (\bibinfo {year} {2017})}\BibitemShut {NoStop}%
\bibitem [{\citenamefont {Amir}\ \emph {et~al.}(2017)\citenamefont {Amir},
  \citenamefont {Bar}, \citenamefont {Eidelman}, \citenamefont {Elperin},
  \citenamefont {Kleeorin},\ and\ \citenamefont {Rogachevskii}}]{amir_2017}%
  \BibitemOpen
  \bibfield  {author} {\bibinfo {author} {\bibfnamefont {G.}~\bibnamefont
  {Amir}}, \bibinfo {author} {\bibfnamefont {N.}~\bibnamefont {Bar}}, \bibinfo
  {author} {\bibfnamefont {A.}~\bibnamefont {Eidelman}}, \bibinfo {author}
  {\bibfnamefont {T.}~\bibnamefont {Elperin}}, \bibinfo {author} {\bibfnamefont
  {N.}~\bibnamefont {Kleeorin}}, \ and\ \bibinfo {author} {\bibfnamefont
  {I.}~\bibnamefont {Rogachevskii}},\ }\bibfield  {title} {\enquote {\bibinfo
  {title} {Turbulent thermal diffusion in strongly stratified turbulence:
  Theory and experiments},}\ }\href@noop {} {\bibfield  {journal} {\bibinfo
  {journal} {Phys. Rev. Fluids}\ }\textbf {\bibinfo {volume} {2}},\ \bibinfo
  {pages} {064605} (\bibinfo {year} {2017})}\BibitemShut {NoStop}%
\bibitem [{\citenamefont {Lindborg}\ and\ \citenamefont
  {Brethouwer}(2008)}]{lindborg_vertical_2008}%
  \BibitemOpen
  \bibfield  {author} {\bibinfo {author} {\bibfnamefont {E.}~\bibnamefont
  {Lindborg}}\ and\ \bibinfo {author} {\bibfnamefont {G.}~\bibnamefont
  {Brethouwer}},\ }\bibfield  {title} {\enquote {\bibinfo {title} {Vertical
  dispersion by stratified turbulence},}\ }\href {\doibase
  10.1017/S0022112008003595} {\bibfield  {journal} {\bibinfo  {journal} {J.
  Fluid Mech.}\ }\textbf {\bibinfo {volume} {614}},\ \bibinfo {pages} {303}
  (\bibinfo {year} {2008})}\BibitemShut {NoStop}%
\bibitem [{\citenamefont {Waite}(2011)}]{waite_2011}%
  \BibitemOpen
  \bibfield  {author} {\bibinfo {author} {\bibfnamefont {M.~L.}\ \bibnamefont
  {Waite}},\ }\bibfield  {title} {\enquote {\bibinfo {title} {Stratified
  turbulence at the buoyancy scale},}\ }\href@noop {} {\bibfield  {journal}
  {\bibinfo  {journal} {Phys. Fluids}\ }\textbf {\bibinfo {volume} {23}},\
  \bibinfo {pages} {066602} (\bibinfo {year} {2011})}\BibitemShut {NoStop}%
\bibitem [{\citenamefont {Marino}\ \emph {et~al.}(2014)\citenamefont {Marino},
  \citenamefont {Mininni}, \citenamefont {Rosenberg},\ and\ \citenamefont
  {Pouquet}}]{marino_large-scale_2014}%
  \BibitemOpen
  \bibfield  {author} {\bibinfo {author} {\bibfnamefont {R.}~\bibnamefont
  {Marino}}, \bibinfo {author} {\bibfnamefont {P.~D.}\ \bibnamefont {Mininni}},
  \bibinfo {author} {\bibfnamefont {D.~L.}\ \bibnamefont {Rosenberg}}, \ and\
  \bibinfo {author} {\bibfnamefont {A.}~\bibnamefont {Pouquet}},\ }\bibfield
  {title} {\enquote {\bibinfo {title} {Large-scale anisotropy in stably
  stratified rotating flows},}\ }\href {\doibase 10.1103/PhysRevE.90.023018}
  {\bibfield  {journal} {\bibinfo  {journal} {Phys. Rev. E}\ }\textbf {\bibinfo
  {volume} {90}},\ \bibinfo {pages} {023018} (\bibinfo {year}
  {2014})}\BibitemShut {NoStop}%
\bibitem [{\citenamefont {Maffioli}(2017)}]{maffioli_2017}%
  \BibitemOpen
  \bibfield  {author} {\bibinfo {author} {\bibfnamefont {A.}~\bibnamefont
  {Maffioli}},\ }\bibfield  {title} {\enquote {\bibinfo {title} {Vertical
  spectra of stratified turbulence at large horizontal scales},}\ }\href@noop
  {} {\bibfield  {journal} {\bibinfo  {journal} {Phys. Rev. Fluids}\ }\textbf
  {\bibinfo {volume} {2}},\ \bibinfo {pages} {104802} (\bibinfo {year}
  {2017})}\BibitemShut {NoStop}%
\bibitem [{\citenamefont {Smith}\ and\ \citenamefont
  {Waleffe}(2002)}]{smith_generation_2002}%
  \BibitemOpen
  \bibfield  {author} {\bibinfo {author} {\bibfnamefont {L.~M.}\ \bibnamefont
  {Smith}}\ and\ \bibinfo {author} {\bibfnamefont {F.}~\bibnamefont
  {Waleffe}},\ }\bibfield  {title} {\enquote {\bibinfo {title} {Generation of
  slow large scales in forced rotating stratified turbulence},}\ }\href
  {\doibase 10.1017/S0022112001006309} {\bibfield  {journal} {\bibinfo
  {journal} {J. Fluid Mech.}\ }\textbf {\bibinfo {volume} {451}},\ \bibinfo
  {pages} {145} (\bibinfo {year} {2002})}\BibitemShut {NoStop}%
\bibitem [{\citenamefont {Clark~di Leoni}\ and\ \citenamefont
  {Mininni}(2015)}]{clark_di_leoni_absorption_2015}%
  \BibitemOpen
  \bibfield  {author} {\bibinfo {author} {\bibfnamefont {P.}~\bibnamefont
  {Clark~di Leoni}}\ and\ \bibinfo {author} {\bibfnamefont {P.~D.}\
  \bibnamefont {Mininni}},\ }\bibfield  {title} {\enquote {\bibinfo {title}
  {Absorption of waves by large-scale winds in stratified turbulence},}\ }\href
  {\doibase 10.1103/PhysRevE.91.033015} {\bibfield  {journal} {\bibinfo
  {journal} {Phys. Rev. E}\ }\textbf {\bibinfo {volume} {91}},\ \bibinfo
  {pages} {033015} (\bibinfo {year} {2015})}\BibitemShut {NoStop}%
\bibitem [{\citenamefont {Sujovolsky}\ \emph
  {et~al.}(2018{\natexlab{a}})\citenamefont {Sujovolsky}, \citenamefont
  {Mininni},\ and\ \citenamefont {Rast}}]{sujovolsky_single-particle_2017}%
  \BibitemOpen
  \bibfield  {author} {\bibinfo {author} {\bibfnamefont {N.~E.}\ \bibnamefont
  {Sujovolsky}}, \bibinfo {author} {\bibfnamefont {P.~D.}\ \bibnamefont
  {Mininni}}, \ and\ \bibinfo {author} {\bibfnamefont {M.~P.}\ \bibnamefont
  {Rast}},\ }\bibfield  {title} {\enquote {\bibinfo {title} {Single-particle
  dispersion in stably stratified turbulence},}\ }\href {\doibase
  10.1103/PhysRevFluids.3.034603} {\bibfield  {journal} {\bibinfo  {journal}
  {Phys. Rev. Fluids}\ }\textbf {\bibinfo {volume} {3}},\ \bibinfo {pages}
  {034603} (\bibinfo {year} {2018}{\natexlab{a}})}\BibitemShut {NoStop}%
\bibitem [{\citenamefont {Kimura}\ and\ \citenamefont
  {Herring}(1996)}]{kimura_diffusion_1996}%
  \BibitemOpen
  \bibfield  {author} {\bibinfo {author} {\bibfnamefont {Y.}~\bibnamefont
  {Kimura}}\ and\ \bibinfo {author} {\bibfnamefont {J.~R.}\ \bibnamefont
  {Herring}},\ }\bibfield  {title} {\enquote {\bibinfo {title} {Diffusion in
  stably stratified turbulence},}\ }\href {\doibase 10.1017/S0022112096008713}
  {\bibfield  {journal} {\bibinfo  {journal} {J. Fluid Mech.}\ }\textbf
  {\bibinfo {volume} {328}},\ \bibinfo {pages} {253} (\bibinfo {year}
  {1996})}\BibitemShut {NoStop}%
\bibitem [{\citenamefont {Kaneda}\ and\ \citenamefont
  {Ishida}(2000)}]{kaneda_suppression_2000}%
  \BibitemOpen
  \bibfield  {author} {\bibinfo {author} {\bibfnamefont {Y.}~\bibnamefont
  {Kaneda}}\ and\ \bibinfo {author} {\bibfnamefont {T.}~\bibnamefont
  {Ishida}},\ }\bibfield  {title} {\enquote {\bibinfo {title} {Suppression of
  vertical diffusion in strongly stratified turbulence},}\ }\href {\doibase
  10.1017/S0022112099007041} {\bibfield  {journal} {\bibinfo  {journal} {J.
  Fluid Mech.}\ }\textbf {\bibinfo {volume} {402}},\ \bibinfo {pages}
  {311--327} (\bibinfo {year} {2000})}\BibitemShut {NoStop}%
\bibitem [{\citenamefont {Liechtenstein}\ \emph {et~al.}(2006)\citenamefont
  {Liechtenstein}, \citenamefont {Godeferd},\ and\ \citenamefont
  {Cambon}}]{liechtenstein_role_2006}%
  \BibitemOpen
  \bibfield  {author} {\bibinfo {author} {\bibfnamefont {L.}~\bibnamefont
  {Liechtenstein}}, \bibinfo {author} {\bibfnamefont {F.~S.}\ \bibnamefont
  {Godeferd}}, \ and\ \bibinfo {author} {\bibfnamefont {C.}~\bibnamefont
  {Cambon}},\ }\bibfield  {title} {\enquote {\bibinfo {title} {The role of
  nonlinearity in turbulent diffusion models for stably stratified and rotating
  turbulence},}\ }\href {\doibase 10.1016/j.ijheatfluidflow.2006.02.010}
  {\bibfield  {journal} {\bibinfo  {journal} {Int. J. Heat Fluid. Fl.}\
  }\textbf {\bibinfo {volume} {27}},\ \bibinfo {pages} {644} (\bibinfo {year}
  {2006})}\BibitemShut {NoStop}%
\bibitem [{\citenamefont {Rorai}\ \emph {et~al.}(2014)\citenamefont {Rorai},
  \citenamefont {Mininni},\ and\ \citenamefont
  {Pouquet}}]{rorai_turbulence_2014}%
  \BibitemOpen
  \bibfield  {author} {\bibinfo {author} {\bibfnamefont {C.}~\bibnamefont
  {Rorai}}, \bibinfo {author} {\bibfnamefont {P.~D.}\ \bibnamefont {Mininni}},
  \ and\ \bibinfo {author} {\bibfnamefont {A.}~\bibnamefont {Pouquet}},\
  }\bibfield  {title} {\enquote {\bibinfo {title} {Turbulence comes in bursts
  in stably stratified flows},}\ }\href {\doibase 10.1103/PhysRevE.89.043002}
  {\bibfield  {journal} {\bibinfo  {journal} {Phys. Rev. E}\ }\textbf {\bibinfo
  {volume} {89}},\ \bibinfo {pages} {043002} (\bibinfo {year}
  {2014})}\BibitemShut {NoStop}%
\bibitem [{\citenamefont {Feraco}\ \emph {et~al.}(2018)\citenamefont {Feraco},
  \citenamefont {Marino}, \citenamefont {Pumir}, \citenamefont {Primavera},
  \citenamefont {Mininni}, \citenamefont {Pouquet},\ and\ \citenamefont
  {Rosenberg}}]{feraco_vertical_2018}%
  \BibitemOpen
  \bibfield  {author} {\bibinfo {author} {\bibfnamefont {F.}~\bibnamefont
  {Feraco}}, \bibinfo {author} {\bibfnamefont {R.}~\bibnamefont {Marino}},
  \bibinfo {author} {\bibfnamefont {A.}~\bibnamefont {Pumir}}, \bibinfo
  {author} {\bibfnamefont {L.}~\bibnamefont {Primavera}}, \bibinfo {author}
  {\bibfnamefont {P.~D.}\ \bibnamefont {Mininni}}, \bibinfo {author}
  {\bibfnamefont {A.}~\bibnamefont {Pouquet}}, \ and\ \bibinfo {author}
  {\bibfnamefont {D.}~\bibnamefont {Rosenberg}},\ }\bibfield  {title} {\enquote
  {\bibinfo {title} {Vertical drafts and mixing in stratified turbulence: sharp
  transition with {Froude} number},}\ }\href {http://arxiv.org/abs/1806.00342}
  {\bibfield  {journal} {\bibinfo  {journal} {EPL in press}\ } (\bibinfo {year}
  {2018})},\ \bibinfo {note} {arXiv: 1806.00342}\BibitemShut {NoStop}%
\bibitem [{\citenamefont {Billant}\ and\ \citenamefont
  {Chomaz}(2001)}]{billant_self-similarity_2001}%
  \BibitemOpen
  \bibfield  {author} {\bibinfo {author} {\bibfnamefont {P.}~\bibnamefont
  {Billant}}\ and\ \bibinfo {author} {\bibfnamefont {J.-M.}\ \bibnamefont
  {Chomaz}},\ }\bibfield  {title} {\enquote {\bibinfo {title} {Self-similarity
  of strongly stratified inviscid flows},}\ }\href {\doibase 10.1063/1.1369125}
  {\bibfield  {journal} {\bibinfo  {journal} {Phys. Fluids}\ }\textbf {\bibinfo
  {volume} {13}},\ \bibinfo {pages} {1645} (\bibinfo {year}
  {2001})}\BibitemShut {NoStop}%
\bibitem [{\citenamefont {de~Bruyn~Kops}\ \emph {et~al.}(2004)\citenamefont
  {de~Bruyn~Kops}, \citenamefont {Riley},\ and\ \citenamefont
  {Winters}}]{deBruyn_2004_reynolds}%
  \BibitemOpen
  \bibfield  {author} {\bibinfo {author} {\bibfnamefont {S.~M.}\ \bibnamefont
  {de~Bruyn~Kops}}, \bibinfo {author} {\bibfnamefont {J.~J.}\ \bibnamefont
  {Riley}}, \ and\ \bibinfo {author} {\bibfnamefont {K.~B.}\ \bibnamefont
  {Winters}},\ }\bibfield  {title} {\enquote {\bibinfo {title} {Reynolds and
  froude number scaling in stably-stratified flows},}\ }\href@noop {}
  {\bibfield  {journal} {\bibinfo  {journal} {Fluid Mech. Appl.}\ }\textbf
  {\bibinfo {volume} {74}},\ \bibinfo {pages} {71} (\bibinfo {year}
  {2004})}\BibitemShut {NoStop}%
\bibitem [{\citenamefont {Bauer}(1974)}]{bauer_dispersion_1974}%
  \BibitemOpen
  \bibfield  {author} {\bibinfo {author} {\bibfnamefont {E.}~\bibnamefont
  {Bauer}},\ }\bibfield  {title} {\enquote {\bibinfo {title} {Dispersion of
  tracers in the atmosphere and ocean: {Survey} and comparison of experimental
  data},}\ }\href@noop {} {\bibfield  {journal} {\bibinfo  {journal} {J.
  Geophys. Res.}\ }\textbf {\bibinfo {volume} {79}},\ \bibinfo {pages} {789}
  (\bibinfo {year} {1974})}\BibitemShut {NoStop}%
\bibitem [{\citenamefont {Fernando}(1991)}]{fernando_turbulent_1991}%
  \BibitemOpen
  \bibfield  {author} {\bibinfo {author} {\bibfnamefont {H.~J.~S.}\
  \bibnamefont {Fernando}},\ }\bibfield  {title} {\enquote {\bibinfo {title}
  {Turbulent {Mixing} in {Stratified} {Fluids}},}\ }\href {\doibase
  10.1146/annurev.fl.23.010191.002323} {\bibfield  {journal} {\bibinfo
  {journal} {Annu. Rev. Fluid Mech.}\ }\textbf {\bibinfo {volume} {23}},\
  \bibinfo {pages} {455} (\bibinfo {year} {1991})}\BibitemShut {NoStop}%
\bibitem [{\citenamefont {Polzin}\ \emph {et~al.}(1997)\citenamefont {Polzin},
  \citenamefont {Toole}, \citenamefont {Ledwell},\ and\ \citenamefont
  {Schmitt}}]{polzin_spatial_1997}%
  \BibitemOpen
  \bibfield  {author} {\bibinfo {author} {\bibfnamefont {K.~L.}\ \bibnamefont
  {Polzin}}, \bibinfo {author} {\bibfnamefont {J.~M.}\ \bibnamefont {Toole}},
  \bibinfo {author} {\bibfnamefont {J.~R.}\ \bibnamefont {Ledwell}}, \ and\
  \bibinfo {author} {\bibfnamefont {R.~W.}\ \bibnamefont {Schmitt}},\
  }\bibfield  {title} {\enquote {\bibinfo {title} {Spatial {Variability} of
  {Turbulent} {Mixing} in the {Abyssal} {Ocean}},}\ }\href {\doibase
  10.1126/science.276.5309.93} {\bibfield  {journal} {\bibinfo  {journal}
  {Science}\ }\textbf {\bibinfo {volume} {276}},\ \bibinfo {pages} {93}
  (\bibinfo {year} {1997})}\BibitemShut {NoStop}%
\bibitem [{\citenamefont {Wunsch}\ and\ \citenamefont
  {Ferrari}(2004)}]{wunsch_vertical_2004}%
  \BibitemOpen
  \bibfield  {author} {\bibinfo {author} {\bibfnamefont {C.}~\bibnamefont
  {Wunsch}}\ and\ \bibinfo {author} {\bibfnamefont {R.}~\bibnamefont
  {Ferrari}},\ }\bibfield  {title} {\enquote {\bibinfo {title} {Vertical
  {Mixing}, {Energy}, and the {General} {Circulation} of the {Oceans}},}\
  }\href {\doibase 10.1146/annurev.fluid.36.050802.122121} {\bibfield
  {journal} {\bibinfo  {journal} {Annu. Rev. Fluid Mech.}\ }\textbf {\bibinfo
  {volume} {36}},\ \bibinfo {pages} {281} (\bibinfo {year} {2004})}\BibitemShut
  {NoStop}%
\bibitem [{\citenamefont {Ivey}\ \emph {et~al.}(2008)\citenamefont {Ivey},
  \citenamefont {Winters},\ and\ \citenamefont {Koseff}}]{ivey_density_2008}%
  \BibitemOpen
  \bibfield  {author} {\bibinfo {author} {\bibfnamefont {G.N.}\ \bibnamefont
  {Ivey}}, \bibinfo {author} {\bibfnamefont {K.B.}\ \bibnamefont {Winters}}, \
  and\ \bibinfo {author} {\bibfnamefont {J.R.}\ \bibnamefont {Koseff}},\
  }\bibfield  {title} {\enquote {\bibinfo {title} {Density {Stratification},
  {Turbulence}, but {How} {Much} {Mixing}?}}\ }\href {\doibase
  10.1146/annurev.fluid.39.050905.110314} {\bibfield  {journal} {\bibinfo
  {journal} {Annu. Rev. Fluid Mech.}\ }\textbf {\bibinfo {volume} {40}},\
  \bibinfo {pages} {169--184} (\bibinfo {year} {2008})}\BibitemShut {NoStop}%
\bibitem [{\citenamefont {Klein}\ and\ \citenamefont
  {Lapeyre}(2009)}]{klein_oceanic_2009}%
  \BibitemOpen
  \bibfield  {author} {\bibinfo {author} {\bibfnamefont {P.}~\bibnamefont
  {Klein}}\ and\ \bibinfo {author} {\bibfnamefont {G.}~\bibnamefont
  {Lapeyre}},\ }\bibfield  {title} {\enquote {\bibinfo {title} {The {Oceanic}
  {Vertical} {Pump} {Induced} by {Mesoscale} and {Submesoscale}
  {Turbulence}},}\ }\href {\doibase 10.1146/annurev.marine.010908.163704}
  {\bibfield  {journal} {\bibinfo  {journal} {Annu. Rev. Mar. Sci.}\ }\textbf
  {\bibinfo {volume} {1}},\ \bibinfo {pages} {351} (\bibinfo {year}
  {2009})}\BibitemShut {NoStop}%
\bibitem [{\citenamefont {Mingari}\ \emph {et~al.}(2017)\citenamefont
  {Mingari}, \citenamefont {Collini}, \citenamefont {Folch}, \citenamefont
  {B{\'a}ez}, \citenamefont {Bustos}, \citenamefont {Osores}, \citenamefont
  {Reckziegel}, \citenamefont {Alexander},\ and\ \citenamefont
  {Viramonte}}]{mingari_2017}%
  \BibitemOpen
  \bibfield  {author} {\bibinfo {author} {\bibfnamefont {L.A.}\ \bibnamefont
  {Mingari}}, \bibinfo {author} {\bibfnamefont {E.A.}\ \bibnamefont {Collini}},
  \bibinfo {author} {\bibfnamefont {A.}~\bibnamefont {Folch}}, \bibinfo
  {author} {\bibfnamefont {W.}~\bibnamefont {B{\'a}ez}}, \bibinfo {author}
  {\bibfnamefont {E.}~\bibnamefont {Bustos}}, \bibinfo {author} {\bibfnamefont
  {M.S.}\ \bibnamefont {Osores}}, \bibinfo {author} {\bibfnamefont
  {F.}~\bibnamefont {Reckziegel}}, \bibinfo {author} {\bibfnamefont
  {P.}~\bibnamefont {Alexander}}, \ and\ \bibinfo {author} {\bibfnamefont
  {J.G.}\ \bibnamefont {Viramonte}},\ }\bibfield  {title} {\enquote {\bibinfo
  {title} {Numerical simulations of windblown dust over complex terrain: the
  {F}iambal{\'a} {B}asin episode in {J}une 2015},}\ }\href@noop {} {\bibfield
  {journal} {\bibinfo  {journal} {Atmos. Chem. Phys.}\ }\textbf {\bibinfo
  {volume} {17}},\ \bibinfo {pages} {6759} (\bibinfo {year}
  {2017})}\BibitemShut {NoStop}%
\bibitem [{\citenamefont {Jones}\ \emph {et~al.}(2007)\citenamefont {Jones},
  \citenamefont {Thomson}, \citenamefont {Hort},\ and\ \citenamefont
  {Devenish}}]{jones_2007}%
  \BibitemOpen
  \bibfield  {author} {\bibinfo {author} {\bibfnamefont {A.}~\bibnamefont
  {Jones}}, \bibinfo {author} {\bibfnamefont {D.}~\bibnamefont {Thomson}},
  \bibinfo {author} {\bibfnamefont {M.}~\bibnamefont {Hort}}, \ and\ \bibinfo
  {author} {\bibfnamefont {B.}~\bibnamefont {Devenish}},\ }\bibfield  {title}
  {\enquote {\bibinfo {title} {The {UK} {M}et {O}ffice's next-generation
  atmospheric dispersion model, {NAME} {III}},}\ }in\ \href@noop {} {\emph
  {\bibinfo {booktitle} {Air pollution modeling and its application XVII}}}\
  (\bibinfo  {publisher} {Springer},\ \bibinfo {year} {2007})\ pp.\ \bibinfo
  {pages} {580--589}\BibitemShut {NoStop}%
\bibitem [{\citenamefont {Godeferd}\ \emph {et~al.}(1996)\citenamefont
  {Godeferd}, \citenamefont {Malik}, \citenamefont {Cambon},\ and\
  \citenamefont {Nicolleau}}]{godeferd_eulerian_1996}%
  \BibitemOpen
  \bibfield  {author} {\bibinfo {author} {\bibfnamefont {F.~S.}\ \bibnamefont
  {Godeferd}}, \bibinfo {author} {\bibfnamefont {N.~A.}\ \bibnamefont {Malik}},
  \bibinfo {author} {\bibfnamefont {C.}~\bibnamefont {Cambon}}, \ and\ \bibinfo
  {author} {\bibfnamefont {F.}~\bibnamefont {Nicolleau}},\ }\bibfield  {title}
  {\enquote {\bibinfo {title} {Eulerian and {Lagrangian} statistics in
  homogeneous stratified flows},}\ }\href@noop {} {\bibfield  {journal}
  {\bibinfo  {journal} {Appl- Sci. Res.}\ }\textbf {\bibinfo {volume} {57}},\
  \bibinfo {pages} {319} (\bibinfo {year} {1996})}\BibitemShut {NoStop}%
\bibitem [{\citenamefont {Nicolleau}\ and\ \citenamefont
  {Vassilicos}(2000)}]{nicolleau_turbulent_2000}%
  \BibitemOpen
  \bibfield  {author} {\bibinfo {author} {\bibfnamefont {F.}~\bibnamefont
  {Nicolleau}}\ and\ \bibinfo {author} {\bibfnamefont {J.~C.}\ \bibnamefont
  {Vassilicos}},\ }\bibfield  {title} {\enquote {\bibinfo {title} {Turbulent
  diffusion in stably stratified non-decaying turbulence},}\ }\href {\doibase
  10.1017/S0022112099008113} {\bibfield  {journal} {\bibinfo  {journal} {J.
  Fluid Mech.}\ }\textbf {\bibinfo {volume} {410}},\ \bibinfo {pages} {123}
  (\bibinfo {year} {2000})}\BibitemShut {NoStop}%
\bibitem [{\citenamefont {Aartrijk}\ \emph {et~al.}(2008)\citenamefont
  {Aartrijk}, \citenamefont {Clercx},\ and\ \citenamefont
  {Winters}}]{aartrijk_single-particle_2008}%
  \BibitemOpen
  \bibfield  {author} {\bibinfo {author} {\bibfnamefont {M.~van}\ \bibnamefont
  {Aartrijk}}, \bibinfo {author} {\bibfnamefont {H.~J.~H.}\ \bibnamefont
  {Clercx}}, \ and\ \bibinfo {author} {\bibfnamefont {K.~B.}\ \bibnamefont
  {Winters}},\ }\bibfield  {title} {\enquote {\bibinfo {title}
  {Single-particle, particle-pair, and multiparticle dispersion of fluid
  particles in forced stably stratified turbulence},}\ }\href {\doibase
  10.1063/1.2838593} {\bibfield  {journal} {\bibinfo  {journal} {Phys. Fluids
  (1994-present)}\ }\textbf {\bibinfo {volume} {20}},\ \bibinfo {pages}
  {025104} (\bibinfo {year} {2008})}\BibitemShut {NoStop}%
\bibitem [{\citenamefont {Bec}\ \emph {et~al.}(2007)\citenamefont {Bec},
  \citenamefont {Biferale}, \citenamefont {Cencini}, \citenamefont {Lanotte},
  \citenamefont {Musacchio},\ and\ \citenamefont {Toschi}}]{bec_heavy_2007}%
  \BibitemOpen
  \bibfield  {author} {\bibinfo {author} {\bibfnamefont {J.}~\bibnamefont
  {Bec}}, \bibinfo {author} {\bibfnamefont {L.}~\bibnamefont {Biferale}},
  \bibinfo {author} {\bibfnamefont {M.}~\bibnamefont {Cencini}}, \bibinfo
  {author} {\bibfnamefont {A.}~\bibnamefont {Lanotte}}, \bibinfo {author}
  {\bibfnamefont {S.}~\bibnamefont {Musacchio}}, \ and\ \bibinfo {author}
  {\bibfnamefont {F.}~\bibnamefont {Toschi}},\ }\bibfield  {title} {\enquote
  {\bibinfo {title} {Heavy {Particle} {Concentration} in {Turbulence} at
  {Dissipative} and {Inertial} {Scales}},}\ }\href {\doibase
  10.1103/PhysRevLett.98.084502} {\bibfield  {journal} {\bibinfo  {journal}
  {Phys. Rev. Lett.}\ }\textbf {\bibinfo {volume} {98}},\ \bibinfo {pages}
  {084502} (\bibinfo {year} {2007})}\BibitemShut {NoStop}%
\bibitem [{\citenamefont {Sumbekova}\ \emph {et~al.}(2017)\citenamefont
  {Sumbekova}, \citenamefont {Cartellier}, \citenamefont {Aliseda},\ and\
  \citenamefont {Bourgoin}}]{sumbekova_2017}%
  \BibitemOpen
  \bibfield  {author} {\bibinfo {author} {\bibfnamefont {S.}~\bibnamefont
  {Sumbekova}}, \bibinfo {author} {\bibfnamefont {A.}~\bibnamefont
  {Cartellier}}, \bibinfo {author} {\bibfnamefont {A.}~\bibnamefont {Aliseda}},
  \ and\ \bibinfo {author} {\bibfnamefont {M.}~\bibnamefont {Bourgoin}},\
  }\bibfield  {title} {\enquote {\bibinfo {title} {Preferential concentration
  of inertial sub-kolmogorov particles: The roles of mass loading of particles,
  stokes numbers, and reynolds numbers},}\ }\href@noop {} {\bibfield  {journal}
  {\bibinfo  {journal} {Phys. Rev. Fluids}\ }\textbf {\bibinfo {volume} {2}},\
  \bibinfo {pages} {024302} (\bibinfo {year} {2017})}\BibitemShut {NoStop}%
\bibitem [{\citenamefont {van Aartrijk}\ and\ \citenamefont
  {Clercx}(2010)}]{van_aartrijk_vertical_2010}%
  \BibitemOpen
  \bibfield  {author} {\bibinfo {author} {\bibfnamefont {M.}~\bibnamefont {van
  Aartrijk}}\ and\ \bibinfo {author} {\bibfnamefont {H.~J.~H.}\ \bibnamefont
  {Clercx}},\ }\bibfield  {title} {\enquote {\bibinfo {title} {Vertical
  dispersion of light inertial particles in stably stratified turbulence: {The}
  influence of the {Basset} force},}\ }\href {\doibase 10.1063/1.3291678}
  {\bibfield  {journal} {\bibinfo  {journal} {Phys. Fluids}\ }\textbf {\bibinfo
  {volume} {22}},\ \bibinfo {pages} {013301} (\bibinfo {year}
  {2010})}\BibitemShut {NoStop}%
\bibitem [{\citenamefont {Sozza}\ \emph {et~al.}(2016)\citenamefont {Sozza},
  \citenamefont {De~Lillo}, \citenamefont {Musacchio},\ and\ \citenamefont
  {Boffetta}}]{sozza_large-scale_2016}%
  \BibitemOpen
  \bibfield  {author} {\bibinfo {author} {\bibfnamefont {A.}~\bibnamefont
  {Sozza}}, \bibinfo {author} {\bibfnamefont {F.}~\bibnamefont {De~Lillo}},
  \bibinfo {author} {\bibfnamefont {S.}~\bibnamefont {Musacchio}}, \ and\
  \bibinfo {author} {\bibfnamefont {G.}~\bibnamefont {Boffetta}},\ }\bibfield
  {title} {\enquote {\bibinfo {title} {Large-scale confinement and small-scale
  clustering of floating particles in stratified turbulence},}\ }\href
  {\doibase 10.1103/PhysRevFluids.1.052401} {\bibfield  {journal} {\bibinfo
  {journal} {Phys. Rev. Fluids}\ }\textbf {\bibinfo {volume} {1}},\ \bibinfo
  {pages} {052401(R)} (\bibinfo {year} {2016})}\BibitemShut {NoStop}%
\bibitem [{\citenamefont {Sozza}\ \emph {et~al.}(2018)\citenamefont {Sozza},
  \citenamefont {De~Lillo},\ and\ \citenamefont
  {Boffetta}}]{sozza2018inertial}%
  \BibitemOpen
  \bibfield  {author} {\bibinfo {author} {\bibfnamefont {A.}~\bibnamefont
  {Sozza}}, \bibinfo {author} {\bibfnamefont {F.}~\bibnamefont {De~Lillo}}, \
  and\ \bibinfo {author} {\bibfnamefont {G.}~\bibnamefont {Boffetta}},\
  }\bibfield  {title} {\enquote {\bibinfo {title} {Inertial floaters in
  stratified turbulence},}\ }\href@noop {} {\bibfield  {journal} {\bibinfo
  {journal} {EPL (Europhys. Letters)}\ }\textbf {\bibinfo {volume} {121}},\
  \bibinfo {pages} {14002} (\bibinfo {year} {2018})}\BibitemShut {NoStop}%
\bibitem [{\citenamefont {Mininni}\ \emph {et~al.}(2011)\citenamefont
  {Mininni}, \citenamefont {Rosenberg}, \citenamefont {Reddy},\ and\
  \citenamefont {Pouquet}}]{mininni_hybrid_2011}%
  \BibitemOpen
  \bibfield  {author} {\bibinfo {author} {\bibfnamefont {P.~D.}\ \bibnamefont
  {Mininni}}, \bibinfo {author} {\bibfnamefont {D.}~\bibnamefont {Rosenberg}},
  \bibinfo {author} {\bibfnamefont {R.}~\bibnamefont {Reddy}}, \ and\ \bibinfo
  {author} {\bibfnamefont {A.}~\bibnamefont {Pouquet}},\ }\bibfield  {title}
  {\enquote {\bibinfo {title} {A hybrid {MPI}–{OpenMP} scheme for scalable
  parallel pseudospectral computations for fluid turbulence},}\ }\href
  {\doibase 10.1016/j.parco.2011.05.004} {\bibfield  {journal} {\bibinfo
  {journal} {Parallel Comput.}\ }\textbf {\bibinfo {volume} {37}},\ \bibinfo
  {pages} {316} (\bibinfo {year} {2011})}\BibitemShut {NoStop}%
\bibitem [{\citenamefont {Yeung}\ and\ \citenamefont
  {Pope}(1988)}]{yeung_1988}%
  \BibitemOpen
  \bibfield  {author} {\bibinfo {author} {\bibfnamefont {P.K.}\ \bibnamefont
  {Yeung}}\ and\ \bibinfo {author} {\bibfnamefont {S.B.}\ \bibnamefont
  {Pope}},\ }\bibfield  {title} {\enquote {\bibinfo {title} {An algorithm for
  tracking fluid particles in numerical simulations of homogeneous
  turbulence},}\ }\href@noop {} {\bibfield  {journal} {\bibinfo  {journal} {J.
  Comp. Phys.}\ }\textbf {\bibinfo {volume} {79}},\ \bibinfo {pages} {373}
  (\bibinfo {year} {1988})}\BibitemShut {NoStop}%
\bibitem [{\citenamefont {Sujovolsky}\ \emph
  {et~al.}(2018{\natexlab{b}})\citenamefont {Sujovolsky}, \citenamefont
  {Mininni},\ and\ \citenamefont {Pouquet}}]{mininni_generation_2017}%
  \BibitemOpen
  \bibfield  {author} {\bibinfo {author} {\bibfnamefont {N.~E.}\ \bibnamefont
  {Sujovolsky}}, \bibinfo {author} {\bibfnamefont {P.~D.}\ \bibnamefont
  {Mininni}}, \ and\ \bibinfo {author} {\bibfnamefont {A.}~\bibnamefont
  {Pouquet}},\ }\bibfield  {title} {\enquote {\bibinfo {title} {Generation of
  turbulence through frontogenesis in sheared stratified flows},}\ }\href@noop
  {} {\bibfield  {journal} {\bibinfo  {journal} {Phys. Fluids}\ }\textbf
  {\bibinfo {volume} {30}},\ \bibinfo {pages} {086601} (\bibinfo {year}
  {2018}{\natexlab{b}})}\BibitemShut {NoStop}%
\bibitem [{\citenamefont {Billant}\ and\ \citenamefont
  {Chomaz}(2000)}]{billant_theoretical_2000}%
  \BibitemOpen
  \bibfield  {author} {\bibinfo {author} {\bibfnamefont {P.}~\bibnamefont
  {Billant}}\ and\ \bibinfo {author} {\bibfnamefont {J.-M.}\ \bibnamefont
  {Chomaz}},\ }\bibfield  {title} {\enquote {\bibinfo {title} {Theoretical
  analysis of the zigzag instability of a vertical columnar vortex pair in a
  strongly stratified fluid},}\ }\href@noop {} {\bibfield  {journal} {\bibinfo
  {journal} {J. Fluid Mech.}\ }\textbf {\bibinfo {volume} {419}},\ \bibinfo
  {pages} {29} (\bibinfo {year} {2000})}\BibitemShut {NoStop}%
\bibitem [{\citenamefont {Riley}\ and\ \citenamefont
  {deBruynKops}(2003)}]{riley_2003}%
  \BibitemOpen
  \bibfield  {author} {\bibinfo {author} {\bibfnamefont {J.J.}\ \bibnamefont
  {Riley}}\ and\ \bibinfo {author} {\bibfnamefont {S.M.}\ \bibnamefont
  {deBruynKops}},\ }\bibfield  {title} {\enquote {\bibinfo {title} {Dynamics of
  turbulence strongly influenced by buoyancy},}\ }\href@noop {} {\bibfield
  {journal} {\bibinfo  {journal} {Phys. Fluids}\ }\textbf {\bibinfo {volume}
  {15}},\ \bibinfo {pages} {2047} (\bibinfo {year} {2003})}\BibitemShut
  {NoStop}%
\bibitem [{\citenamefont {Davidson}(2013)}]{davidson_turbulence_2013}%
  \BibitemOpen
  \bibfield  {author} {\bibinfo {author} {\bibfnamefont {P.~A.}\ \bibnamefont
  {Davidson}},\ }\href@noop {} {\emph {\bibinfo {title} {Turbulence in
  {Rotating}, {Stratified} and {Electrically} {Conducting} {Fluids}}}}\
  (\bibinfo  {publisher} {Cambridge University Press},\ \bibinfo {year}
  {2013})\BibitemShut {NoStop}%
\bibitem [{\citenamefont {Rast}\ \emph {et~al.}(2016)\citenamefont {Rast},
  \citenamefont {Pinton},\ and\ \citenamefont {Mininni}}]{rast_turbulent_2016}%
  \BibitemOpen
  \bibfield  {author} {\bibinfo {author} {\bibfnamefont {M.P.}\ \bibnamefont
  {Rast}}, \bibinfo {author} {\bibfnamefont {J.-F.}\ \bibnamefont {Pinton}}, \
  and\ \bibinfo {author} {\bibfnamefont {P.D.}\ \bibnamefont {Mininni}},\
  }\bibfield  {title} {\enquote {\bibinfo {title} {Turbulent transport with
  intermittency: {Expectation} of a scalar concentration},}\ }\href {\doibase
  10.1103/PhysRevE.93.043120} {\bibfield  {journal} {\bibinfo  {journal} {Phys.
  Rev. E}\ }\textbf {\bibinfo {volume} {93}},\ \bibinfo {pages} {043120}
  (\bibinfo {year} {2016})}\BibitemShut {NoStop}%
\end{thebibliography}%

\end{document}